\documentclass[11pt,a4paper]{article}
\pdfoutput=1
\usepackage{jheppub}
\usepackage[usenames,dvipsnames]{xcolor}
\usepackage{graphicx}
\usepackage{amsmath,amssymb,mathrsfs}

\newcommand{\be}{\begin{equation}}
\newcommand{\ee}{\end{equation}}
\newcommand{\bea}{\begin{eqnarray}}
\newcommand{\eea}{\end{eqnarray}}

\newcommand{\eqn}{\begin{eqnarray}}
\newcommand{\eqnx}{\end{eqnarray}}

\begin{document}

\title{The $\phi^4$ model with the BPS preserving defect}

%%%   JHEP    %%%%%%

\author[a]{C. Adam}
\emailAdd{adam@fpaxp1.usc.es}
\author[b]{T. Romanczukiewicz}
\emailAdd{trom@th.if.uj.edu.pl}
\author[c]{A. Wereszczynski}
\emailAdd{andrzej.wereszczynski@uj.edu.pl}

\affiliation[a]{Departamento de F\'isica de Part\'iculas, Universidad de Santiago de Compostela and Instituto Galego de F\'isica de Altas Enerxias (IGFAE) E-15782 Santiago de Compostela, Spain}

\affiliation[b,c]{Institute of Physics,  Jagiellonian University, Lojasiewicza 11, Krak\'{o}w, Poland}

\abstract{
The $\phi^4$ model is coupled to an impurity in a way that preserves one-half of the BPS property. This means that the antikink-impurity bound state is still a BPS solution, i.e., a zero-pressure solution saturating the topological energy bound. The kink-impurity bound state, on the other hand, does not saturate the bound, in general. 

We found that, although the impurity breaks translational invariance, it is, in some sense, restored in the BPS sector where the energy of the antikink-impurity solution does not depend on their mutual distance. This is reflected in the existence of a generalised translational symmetry and a zero mode.

We also investigate scattering processes. In particular, we compare the antikink-impurity interaction close to the BPS regime, which presents a rather smooth, elastic like nature, with other scattering processes. However, even in this case, after exciting a sufficiently large linear mode on the incoming antikink, we can depart from the close-to-BPS regime. This results, for example, in a backward scattering.
}

\maketitle 

%%%%%%%%%%%%%%%%%%%%%%%%%%%%%%%%%%%%%%%%%
\section{Introduction }
%%%%%%%%%%%%%%%%%%%%%%%%%%%%%%%%%%%%%%%%%
Scalar field theories in (1+1) dimensions with double-vacuum potentials are known to support  topological solitons, called {\it kinks} or {\it antikinks}, depending on the value of the topological charge. Here the topological charge is proportional to the difference of the asymptotic values of the scalar field at plus and minus infinity. Further, static solitons are zero-pressure configurations and satisfy the so-called Bogomolny equation \cite{Bogomol} which, in contrast to the full Euler-Lagrange (EL) equation, is a first order equation. This guarantees that the pertinent energy bound is saturated, leading to the topological stability of the solitons. A model which possesses these properties is usually referred to as a Bogomol'nyi-Prasad-Sommerfield (BPS) theory. 

In fact, it has been shown recently that {\it all} scalar soliton models in (1+1) dimensions with at least two vacua enjoy the BPS property. This comprises both the inclusion of an arbitrary number of scalar fields and any dependence of the energy density on the first spatial derivative \cite{FOEL} as well as on higher derivatives \cite{Adam:2018pvd, Bazeia:2007df, Bazeia:2008tj, Liu:2009ega}. In other words, the BPS property in (1+1) dimensions is a completely generic feature which does not imply any specific requirements on translationally invariant models. This differs completely from higher-dimensional generalizations, where only very particular models enjoy the BPS property - for example the $O(3)$ $\sigma$-model and the Abelian Higgs model in (2+1) dimensions, the t'Hooft-Polyakov monopole (see \cite{Manton} for a review) and the BPS Skyrme model in (3+1) dimensions \cite{Adam:2010fg}, as well as the self-dual sector of the $SU(2)$ Yang-Mills theory in (4+0) dimensions. In some sense, this drastic difference limits the application of (1+1) dimensional theories for the modeling of dynamical processes of higher dimensional solitons. In addition, the BPS sector in (1+1) dimensions contains only one-particle states, i.e., a kink or antikink separately. For models supporting compactons, i.e., solitons which differ from the vacuum only on a finite segment, one may easily construct a multi-particle state by a simple collection of non-overlapping compactons. However, by construction, all constituents do not interact with each other and can in fact be treated independently. In higher dimensions, the BPS sector is significantly more complicated and usually allows for multi-soliton configurations.

One way to destroy the BPS property is to break the translational invariance by adding an impurity  located in space \cite{Adam:2018pvd}, \cite{Malomed:1992hp}, \cite{Fei:1992dk}. 
However, quite interestingly, among the infinitely many couplings to defects (see for example \cite{Malomed:1992hp}, \cite{Fei:1992dk}), there is a specific way of introducing an impurity where one-half of the BPS-ness is preserved \cite{Adam:2018pvd}. 
This means that only one topological sector is a BPS sector saturating the corresponding topological bound. Hence, we have a rather unique chance to investigate, within the same model, both BPS and non-BPS solitons, which can allow us to analyze the impact of the BPS property on dynamics of solitons. 
Such a BPS preserving coupling of defects exists for any 1+1 dimensional scalar soliton model without any restriction on the potential. 
Also the specific spatial distribution of the impurity is arbitrary, except for the finiteness of the $L^2$ norm, which physically means that the impurity is sufficiently well localized. This large freedom may give the chance for an experimental realization of this class of models, especially if one extends the impurity to a periodical lattice of BPS preserving impurities. 
 This can be compared with the integrability preserving defect \cite{MacIntyre:1994sr}, \cite{Bowcock:2003dr}. It is not surprising that the integrability puts much more constraints on the defect, as there are infinitely many currents which must be conserved. Therefore, it strongly restricts its form.

To study properties of solitons in a model with the BPS preserving defect, we choose the $\phi^4$ theory, which together with the sine-Gordon model, is perhaps one of the most studied examples of solitonic field theories in (1+1) dimensions. 
It is a prototypical field theory with solitons (kinks), which found applications in various physical systems from condensed matter  \cite{Yamaletdinov:2017dlz} and biophysics to cosmology. 
It has been widely studied, especially in the context of time dependent solutions describing a collision of kinks, annihilation and creation processes as well as an interaction with radiation. 
Unlike the sine-Gordon model, where the collisions between defects are perfectly elastic, the $\phi^4$ model reveals a very interesting resonant structure. 
The presence of this structure is related to the existence of the internal mode of the kink which can store the energy and give it back to the translational degrees of freedom in a resonant way \cite{Sugiyama:1979mi, Peyrard:1984qn, Anninos:1991un}. Later, the effect with some improvement was  discussed in \cite{Weigel:2013kwa, Takyi:2016tnc} where a typographic error was corrected and the role of the shape mode was questioned. 
Similar structures were later observed in many other models, including the nonlinear Schr\"odinger equation \cite{Goodman, Yang}, double sine-Gordone model \cite{Zhang1992}, multicomponent models \cite{Halavanau:2012dv, Ashcroft:2016tgj, Alonso-Izquierdo:2018uuj, Alonso-Izquierdo:2017gns, Alonso-Izquierdo:2017zta} or collisions with a dynamical boundary \cite{Dorey:2015sha} or impurity \cite{Goodman:2002dfct}. 
A different mechanism was found in the case of non-symmetric kinks when the bound modes were formed between the kinks \cite{Dorey:2011yw}. 
Recently, it was shown that also quasi-normal modes can be responsible for the creation of such a structure \cite{Dorey:2017dsn}.

The main aim of the present paper is to better understand the role of the BPS property in the dynamics of solitonic models. 
Specifically, we begin with a systematic description of the static spectrum of the model in the topologically trivial sector (lump) as well as in the topologically nontrivial sectors ($Q=\pm1$) where kink and antikink exist. 
As only one of the solitons is a BPS state, we carefully examine their asymmetry as a function of a parameter which measures the strength of the impurity. 
Next, we analyze spectral properties of the static solutions which means the linear perturbation theory. Finally, we consider scattering processes between these objects. 
 
However, as one may chose the form of the impurity arbitrarily, we also want to investigate which properties are generic, i.e., independent on a particular choice of the defect. Here, the main finding is that a moduli space for the topologically nontrivial BPS soliton exists. As a consequence, there is a generalized translation symmetry which is reflected by the existence of  a zero mode. Furthermore, we want to disentangle dynamical features which are inherited from the original pure $\phi^4$ model from those which are due to the (specific) BPS preserving impurity. 

The last motivation comes from the fact that, in many aspects, our model is very similar to the Abelian Higgs model with the (one-half) BPS preserving defect introduced in \cite{Tong:2013iqa}. In fact, the coupling to the impurity is given by exactly the same two terms. In the first one, the impurity couples to a topological quantity (the topological charge density and the magnetic field, respectively) while in the second one it multiplies the square root of the potential. As numerical studies in (1+1) dimensions are much simpler, our paper may also be viewed as a guideline for numerics for this vortex model \cite{Cockburn:2015huk}, \cite{Ashcroft:2018gkp}. In addition, we prove the existence of the zero mode (moduli space), which has not yet been completely achieved for the vortex model. 

%%%%%%%%%%%%%%%%%%%%%%%%%%%%%%%%%%%%%%%%%
\section{The BPS $\phi^4$ impurity model}
%%%%%%%%%%%%%%%%%%%%%%%%%%%%%%%%%%%%%%%%%
%%%%%%%%%%%%%%%%%%%%%%%%%%%%%%%%%%%%%%%%%
\subsection{The BPS soliton-impurity models}
%%%%%%%%%%%%%%%%%%%%%%%%%%%%%%%%%%%%%%%%%
Here we briefly summarize some recent findings \cite{Adam:2018pvd} on BPS soliton-impurity models. Let us begin with a general Lagrange density 
 \be
\mathcal{L}= \frac{1}{2} \phi_t^2 - \frac{1}{2} \phi_x^2 - U - 2\sigma \sqrt{U} - \sqrt{2} \sigma \phi_x - \sigma^2 
\ee
where $\phi (t,x)$ is a real scalar field in one spatial dimension, $\sigma (x)$ is a space-located impurity and $U(\phi)$ is a potential with at least two isolated global minima $\phi^v_+>\phi^v_-$. For static configurations the energy reads 
 \be
E=\int_{-\infty}^{\infty} dx \left[\frac{1}{2} \phi_x^2 + U(\phi) + 2\sigma \sqrt{U} + \sqrt{2} \sigma \phi_x\right] +\int_{-\infty}^{\infty} dx \sigma^2 \label{BPS-imp-stat}
\ee
where the last term, obviously, does not contribute to the field equations but sets the zero of the energy scale. 
It leads to the following Euler-Lagrange equation
\be
\phi_{xx}-U_\phi-\sigma \frac{U_\phi}{\sqrt{U}} +\sqrt{2}\sigma_x=0
\ee
where $U_\phi$ denotes the derivative w.r.t. the target space variable. Now, using the standard completing of the square trick, we can compute a lower bound on the energy 
\bea
 E &=&\int_{-\infty}^{\infty} dx \left( \frac{1}{\sqrt{2}} \phi_x + (\sigma + \sqrt{U})\right)^2  -\sqrt{2} \int_{-\infty}^{\infty} dx \phi_x \sqrt{U} \geq   - \sqrt{2} \int_{\phi (-\infty)}^{\phi (+\infty)} d\phi \sqrt{U}  \\ &=&  
 - Q \sqrt{2} \int_{\phi^v_-}^{\phi^v_+} d\phi \sqrt{U} 
 \eea
 which is saturated by solutions of the following first order equation (the so-called Bogomolny equation)
 \be
  \frac{1}{\sqrt{2}} \phi_x + \sigma + \sqrt{U}=0.
 \ee
This is equivalent to the zero pressure condition $T^{11}=0$. Here the topological charge 
 \be
 Q=\frac{\phi(+\infty) - \phi(-\infty)}{\phi^v_+-\phi^v_-}
 \ee
 is $+1$ for the kink and $-1$ for the antikink. Obviously, the topological charge is just a spatial integral of the temporal component of the topological current which is $j^\mu=(\phi^v_+-\phi^v_-)^{-1} \epsilon^{\mu \nu} \partial_\nu \phi$.   Note that, contrary to the usual solitonic models, there is a fixed sign in front of the potential term in the Bogomolny equation. Therefore, only the antikink (or the kink if we change some signs in the model) is a BPS solution in such an impurity extended system, while the kink is a genuine non-BPS solution which generally does not saturate the bound. Furthermore, the energy  bound suggests an asymmetry in the energies of kink and antikink. We also remark that, although the energy can be negative, it is bounded from below in each topological sector. Hence, the model is well defined and has a true vacuum.

Let us remark that in the case of the BPS preserving impurity model a natural object is the prepotential $W$ (also called superpotential), where $U=W^2(\phi)$. The energy then reads
 \be
 E=\int_{-\infty}^{\infty} dx \left[\frac{1}{2} \phi_x^2 + W^2(\phi) + 2\sigma W + \sqrt{2} \sigma \phi_x\right] +\int_{-\infty}^{\infty} dx \sigma^2 .  \label{BPS-imp}
\ee

Although the BPS soliton-impurity model seems to contain quite nontrivial couplings between the scalar field and the impurity via a potential as well as a derivative term, it may be written in a sort of effective potential way. Indeed, the term which is linear in $\phi_x$ can be replaced, up to a total derivative, by a new potential-like term. Then we get an equivalent formulation of the model
 \be
  E=\int_{-\infty}^{\infty} dx \left[\frac{1}{2} \phi_x^2 + U(\phi) + 2\sigma \sqrt{U} - \sqrt{2} \sigma_x \phi \right] +\int_{-\infty}^{\infty} dx \sigma^2 
 \ee
 or simply
 \be
 E=\int_{-\infty}^{\infty} dx \left[\frac{1}{2} \phi_x^2 + U_{eff}(\phi, \sigma)  \right] 
 \ee
 where the effective potential is
  \be
U_{eff} (\phi, \sigma)= U(\phi) + 2\sigma \sqrt{U} - \sqrt{2} \sigma_x \phi + \sigma^2 .
\ee
Observe that, since the effective potential explicitly depends on the spatial coordinates, we cannot apply the former trick to derive the pertinent  Bogomolny equation. 
We believe that this formulation might open a way for an experimental realization of this soliton-impurity model.
%%%%%%%%%%%%%%%%%%%%%%%%%%%%%%%%%%%%%%%%%
\subsection{Kink-form preserving impurity}
%%%%%%%%%%%%%%%%%%%%%%%%%%%%%%%%%%%%%%%%%
It is an important consequence of our construction that the particular space distribution of the impurity is completely arbitrary provided it is $L^2$ integrable. Hence, a specific form can be dictated by a physical system we would like to model. 
However, there exists a special class of impurities which are selected by a mathematical property of the model, i.e., the original kink (or antikink) profile remains unchanged. This means that besides the nice and useful BPS property of the antikink, which allows for a reduction of the second order field equation to the first order Bogomolny equation, we get an exact solution of the non-BPS kink, which, in general, would require to solve the full static EL equation.
Suppose that the scalar field in the model without impurity solves the equation
\begin{equation}
 \phi_{xx}=U_{\phi}\quad \Rightarrow\quad \phi_x=\pm\sqrt{2U}\,.
\end{equation} 
The solution remains unchanged in the BPS soliton-impurity model if the impurity satisfies the following equation 
\begin{equation}
 -\sigma U_\phi+\sqrt{2U}\sigma_x=0\,,
\end{equation} 
which, using the original no-impurity equation, can be rewritten as
\begin{equation}
 -\sigma\phi_{xx}\pm\phi_x\sigma_x=0\,.
\end{equation} 
In the first case with a minus sign it reduces to
\begin{equation}
 -(\sigma\phi_x)'=0\qquad\Rightarrow\qquad\sigma=\frac{\alpha}{\phi_x}\,.
\end{equation} 
In the second case we can divide the equation by $\sigma^2$ and obtain
\begin{equation}
 -\frac{\sigma\phi_{xx}-\phi_x\sigma_x}{\sigma^2}=-\left(\frac{\phi_x}{\sigma}\right)'=0
 \qquad\Rightarrow\qquad\sigma=\alpha\phi_x\,.
\end{equation} 
As we will see later, the parameter $\alpha$ controls the qualitative type of kink-impurity interaction (repulsive or attractive) in the non-BPS sector as well as its strength. 
 %%%%%%%%%%%%%%%%%%%%%%%%%%%%%%%%%%%%%%%%%
\subsection{An exact example}
%%%%%%%%%%%%%%%%%%%%%%%%%%%%%%%%%%%%%%%%%
Let us now consider the soliton-impurity model in the case of the $\phi^4$ potential 
\be
U=\frac{1}{2} (1-\phi^2)^2
\ee
where we suppressed the coupling constant. If the impurity is neglected we find the usual kink $K_0$ and antikink $\bar{K}_0$ solutions
\be
\phi=\pm \tanh x
\ee
which saturate the energy bound $E=4/3$ and have $Q=1$ and $Q=-1$ topological charge, respectively. There are also two vacuum solutions $\phi=\phi^v_{\pm}=\pm 1$ saturating the energy bound in the trivial sector. 

\vspace*{0.2cm}

Now we add the impurity in the way described above. In our example the prepotential is chosen to be $W(\phi)=(1-\phi^2)/\sqrt{2}$. Hence, it is not a positive definite function. Nonetheless, it reproduces the $\phi^4$ potential and therefore it is perfectly fine for our purposes. Observe that a different choice, for example $W(\phi)=|1-\phi^2|/\sqrt{2}$, can modify our findings. From now on, the energy reads 

\be
 E=\int_{-\infty}^{\infty} dx \left[\frac{1}{2} \phi_t^2+ \frac{1}{2} \phi_x^2 + \frac{1}{2} (1-\phi^2)^2+ \sqrt{2} \sigma (1-\phi^2)+ \sqrt{2} \sigma \phi_x\right] +\int_{-\infty}^{\infty} dx \sigma^2. \label{phi-en}
\ee
Then, the EL equation is
\be
-\phi_{tt}+\phi_{xx}+2\phi (1-\phi^2) +2\sqrt{2} \sigma \phi + \sqrt{2} \sigma_x=0. \label{EulerLagr}
\ee
The corresponding Bogomolny equation is
\be
  \frac{1}{\sqrt{2}} \phi_x + \sigma + \frac{1}{\sqrt{2}} (1-\phi^2)=0.
\ee
Again, this equation implies the full static Euler-Lagrange equation. Thus, its solutions are solutions of the full variational problem. The energy bound is 
\be
E \geq  -\frac{4}{3} Q
\ee
and is saturated for the BPS solution. 
 
To specify a concrete example, we choose the impurity in the kink-preserving form which guarantees that the non-BPS kink solution is given in an exact way 
\be
\sigma=\frac{\alpha}{\cosh^2 x}.
\ee
Observe that this impurity is localized exponentially and centered at $x=0$. Further, $\alpha$ is a real parameter.
Finally, the Euler-Lagrange equation is
\be
-\phi_{tt}+\phi_{xx}+2\phi (1-\phi^2) \left( 1+ \frac{\alpha \sqrt{2}}{\cosh^2 x (1-\phi^2)}\right) -2\sqrt{2}\alpha \frac{\tanh x}{\cosh^2 x}=0.
\ee

%%%%%%%%%%%%%%%%%%%%%%%%%%%%%%%%%%%%%%%%%
\section{Static solutions}
%%%%%%%%%%%%%%%%%%%%%%%%%%%%%%%%%%%%%%%%%
We shall find that we will be able to identify certain simple structures within general static solutions, so here we want to introduce some notation for these substructures. A kink and antikink are denoted by $K$ and $\bar K$, respectively, and a kink bound to the impurity is denoted by $K_0$. Further, topologically trivial lumps with asymptotic values $\pm 1$ are denoted by $\Sigma_\pm$.
%%%%%%%%%%%%%%%%%%%%%%%%%%%%%%%%%%%%%%%%%
\subsection{The BPS sector - the antikink-impurity state} \label{sect-3-1}
%%%%%%%%%%%%%%%%%%%%%%%%%%%%%%%%%%%%%%%%%
We start with the BPS sector defined by the Bogomolny equation
\be
  \frac{1}{\sqrt{2}} \phi_x = -  \frac{1}{\sqrt{2}} (1-\phi^2) - \frac{\alpha}{\cosh^2 x}.
\ee 

\begin{figure}
\centering
\includegraphics[height=6.cm]{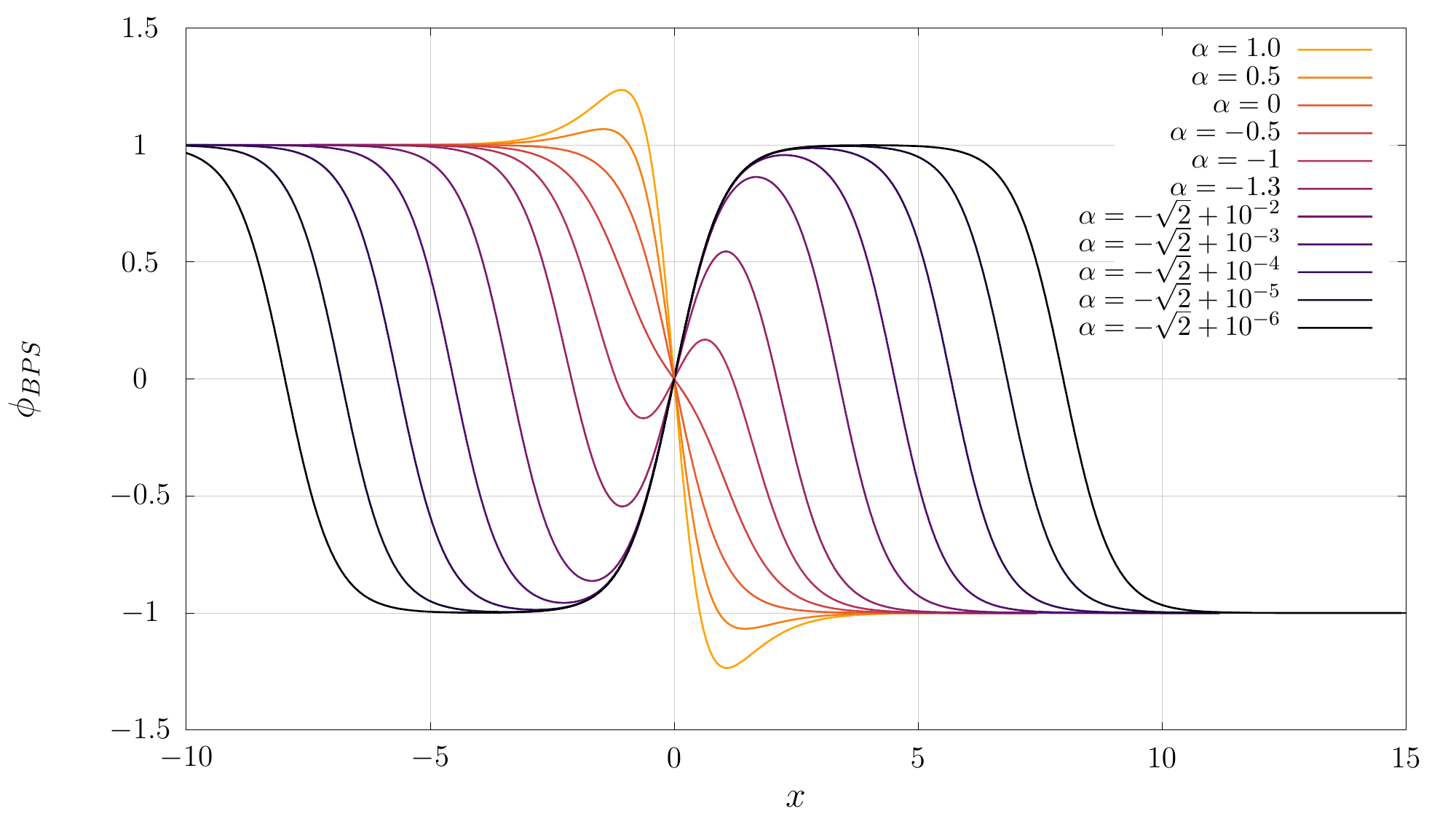}
\caption{Solutions of the Bogomolny equation for different values of $\alpha$ centered at $a=0$.}
\label{BPS}

\includegraphics[width=1\columnwidth]{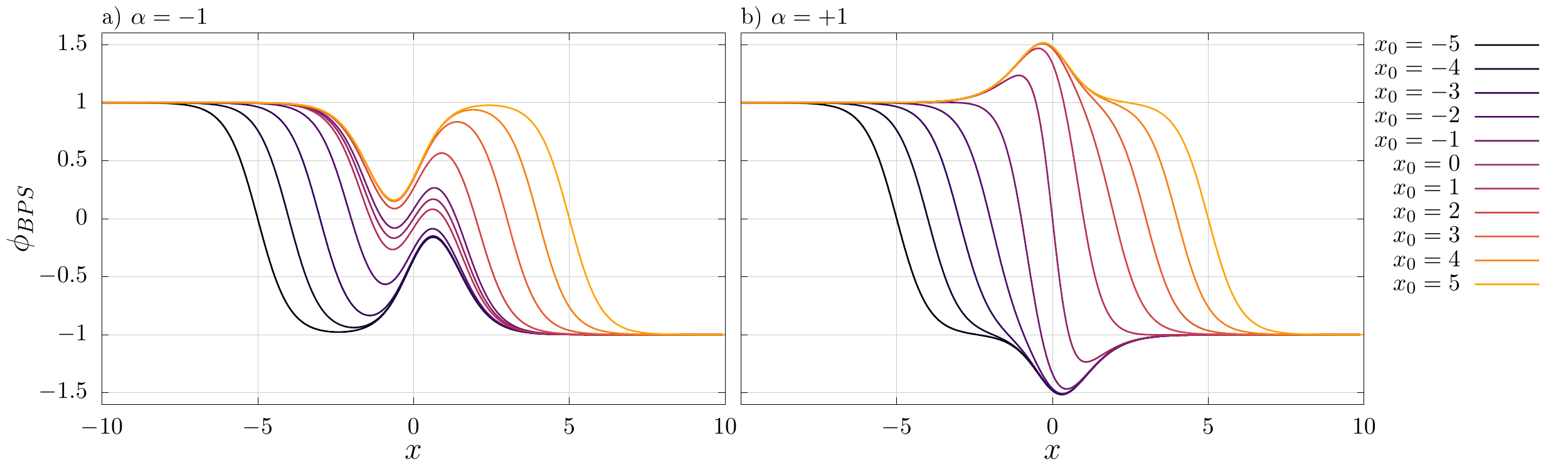}
\caption{Solutions of the Bogomolny equation for different positions of the topological zeros for $\alpha=\pm 1$.}
\label{BPSPositions}
\hspace*{-1.0cm}
\includegraphics[width=1\columnwidth]{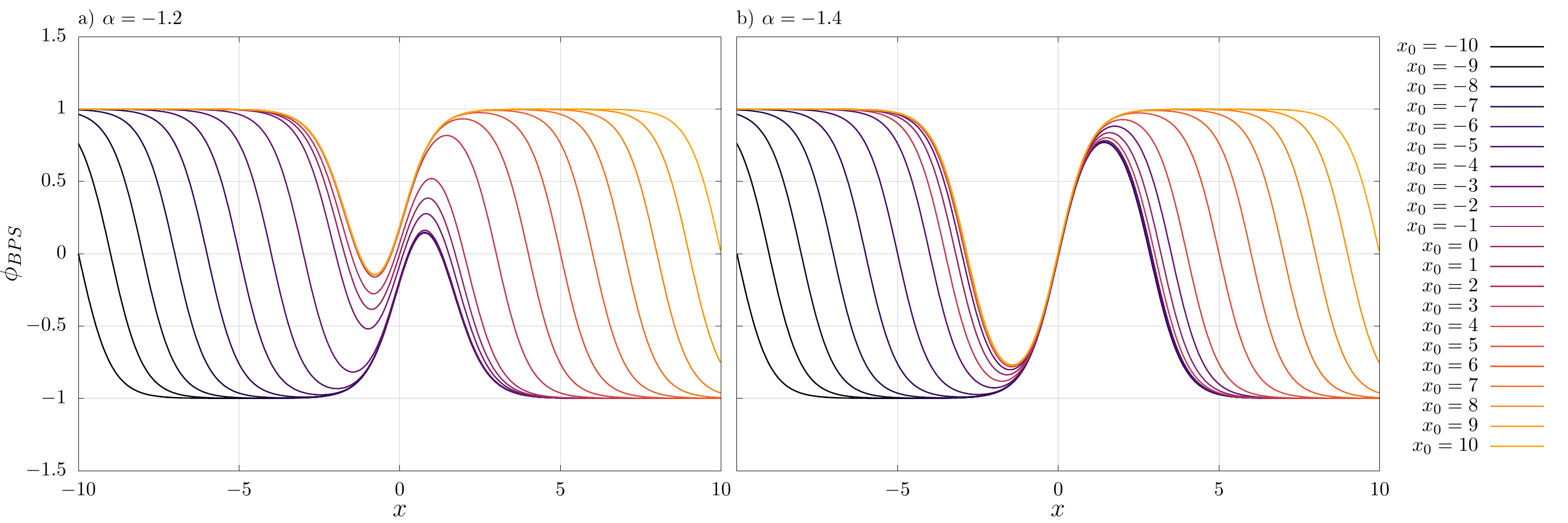}
\caption{The bottle-neck effect of the BPS solutions for $\alpha\to-\sqrt2$.}
\label{BPSPositions-1}
\end{figure}

The equation is invariant under the antisymmetric transformation $\phi(x) \rightarrow -\phi(-x)$. Therefore, in the simplest case, it supports odd-symmetric solutions. The results are plotted in Fig. \ref{BPS}. For $\alpha=0$ we obviously re-obtain the usual antikink solution of the pure $\phi^4$ theory, $\phi=-\tanh x$. As $\alpha$ increases, the antikink-impurity solution initially goes beyond the $\pm 1$ values and then, in the vicinity of the location of the impurity, i.e., at $x=0$, it exhibits a steeper and  steeper descent. Solutions exist for arbitrarily large values of $\alpha$. 

More interesting features happen for $\alpha <0$. Then, the BPS antikink is confined within the $\pm 1$ strip. For a sufficiently small $\alpha$, the originally decreasing profile starts to grow close to the location of the impurity, which in the odd-symmetric case corresponds to $\phi=0$. The critical value of the parameter $\alpha$ below which $\phi_x(x=0)$ takes a positive value is  $\alpha_{cr}=-1/\sqrt{2}$. As $\alpha$ further decreases, this inner structure begins to look like a {\it hidden} pair of kink and antikink. Hence, the full BPS charge -1 solutions look like a composite three particle object, consisting of a left antikink, a central kink located at the impurity and a right antikink. For $\alpha \rightarrow -\sqrt{2}$, the central kink approaches the $\tanh x $ profile, while the two antikinks get repelled to plus/minus infinity. Of course, as the impurity is exponentially localized, they take the $-\tanh (x\pm x_0) $ form, where $x_0\rightarrow \infty$. Hence, the Bogomolny equation supports $Q=-1$ solutions for all $\alpha > -\sqrt{2}$, with the energy $E=4/3$. However, as we approach $\alpha=-\sqrt{2}$, the just described inner structure emerges. The impurity hosts a hidden antikink-kink pair which together with the original antikink leads to a sort of three body bound state $\bar{K}K\bar{K}$, which is further bound to the impurity. In the limit $\alpha=-\sqrt{2}$ both antikinks are sent to infinity and the kink $\tanh x$ remains located at the impurity. Indeed, in this limit the Bogomolny equation is
\be
  \frac{1}{\sqrt{2}} \phi_x = -  \frac{1}{\sqrt{2}} (1-\phi^2) + \frac{\sqrt{2}}{\cosh^2 x}
\ee
and has $\phi(x)=\tanh x$ as its solution. Its energy is $E=-4/3$. However, the two antikinks contribute with $2\times4/3$, so the total energy agrees with the BPS energy in the $Q=-1$ sector. 

To summarize, the Bogomolny equation supports $Q=-1$ BPS solitons for $\alpha>-\sqrt{2}$, which smoothly (as described above), tends to an isolated BPS $Q=1$ kink for $\alpha=-\sqrt{2}$. Probably, we can even say that if $\alpha=-\sqrt{2}$, then both $Q=1$ and $Q=-1$ are BPS solutions. Thus, the model for this particular impurity is {\it fully BPS}, not {\it half-BPS} as occurs for all other impurities. However, remember that the $Q=-1$ solution contains two pure $\phi^4$ antikinks located at plus/minus infinity. Note also that the usual energy degeneracy of the BPS solitons is lifted. In the case $\alpha < -\sqrt{2}$ there are no finite energy BPS solutions. 

\hspace*{0.2cm}

It turns out that the Bogomolny equation supports also non odd-symmetric solutions. 
Since they are BPS solutions, their energy is exactly the same as the energy of the odd-symmetric BPS antikink (and of a free $\phi^4$ antikink). In the limiting case, they represent a configuration consisting of a well separated antikink (of the pure $\phi^4$ theory) and an impurity-induced topologically trivial lump, while all other intermediate configurations are also possible. In other words, we can say that in the BPS sector the antikink and the topologically trivial lump admit a {\it nonlinear superposition} which changes the shape of the BPS solution while the energy is kept constant and all the time saturates the bound. 
In Fig. \ref{BPSPositions} we show the solutions of the BPS equations with the condition $\phi(x_0)=0$ for $\alpha=\pm 1$, where $x_0$ can be identified with the position of the antikink. For large values of $|x_0|$ the solutions look like an antikink centered at $x_0$ and a lump solution attached to the impurity. A similar feature is presented in Fig. \ref{BPSPositions-1} where we consider $\alpha$ closer to $-\sqrt{2}$.
As all solutions have exactly the same energy, it costs no energy to transform one into the other, which physically means to move the antikink away from the impurity. In that sense, the binding energy between impurity and antikink is exactly zero. This is probably expected, because we are in the BPS sector where all solutions are zero-pressure (non-interacting) solutions. Hence, the anti-kink can be put in any location with respect to the topologically trivial lump. Below we rigorously prove that there is in fact a symmetry transformation for solutions of the BPS equation. This transformation represents the nonlinear superposition antikink-impurity law. Furthermore, as we demonstrate later, there is a corresponding zero mode in the spectrum of linear perturbations.

%%%%%%%%%%%%%%
\subsection{The generalised translational symmetry} \label{sect-3.2} 
%%%%%%%%%%%%%%

Here we demonstrate the existence of a symmetry transformation for solutions of the BPS equation
\be
BPS \equiv \phi_x + 1 - \phi^2 + \bar \sigma =0
\ee
(where we introduced $\bar \sigma = \sqrt{2} \sigma$ to get rid of the factor $\sqrt{2}$). We will use the Lie theory of symmetry, for details we refer to \cite{olver}. The idea is to introduce infinitesimal symmetry transformations via vector fields which act on the independent and dependent variables (in our case, $x$ and $\phi$). We shall, however, use the vector field in the simpler "evolutionary form", where derivatives w.r.t. the independent variable $\xi (x,\phi) \partial_x$ are replaced by a particular type of transformation $-\xi (x,\phi) \phi_x \partial_\phi$ acting on the dependent variable.  We consider a vector field $v = \psi \partial_\phi$ where, for the moment, $\psi = \psi (x,\phi,\phi_x)$. We also need the first prolongation of $v$ (because $BPS$ is a first-order equation),
\be
{\rm pr}^{(1)} v = \psi \partial_\phi + (D_x \psi) \partial_{\phi_x}
\ee
where the total derivative $D_x \psi$ is
\be
D_x \psi = \psi_x + \psi_\phi \phi_x + \psi_{\phi_x} \phi_{xx}.
\ee
Further, we need the first prolongation (the total derivative) of $BPS$,
\be
{\rm pr}^{(1)}(BPS) \equiv \phi_{xx} - 2\phi\phi_x + \bar\sigma_x =0.
\ee
Now we calculate the action of ${\rm pr}^{(1)}v$ on $BPS$,
\bea
{\rm pr}^{(1)}v(BPS) = D_x \psi - 2\phi\psi &=&0 \nonumber \\
\psi_x + \psi_\phi \phi_x + \psi_{\phi_x}\phi_{xx} -2 \phi \psi &=&0.
\eea
Next, we  make the simplifying assumption that $\psi = A(x) \phi_x + B(x,\phi)$. Here, $A(x)$ just generates standard (geometric) coordinate transformations, and $B$ generates generalised target space transformations.  Then we
eliminate $\phi_x$ and $\phi_{xx}$ with the help of $BPS$ and ${\rm pr}^{(1)}(BPS)$. This leads to some cancellations, resulting in
\be
(\phi^2 - 1 - \bar \sigma)A_x + B_x + (\phi^2 - 1 - \bar \sigma )B_\phi - \bar \sigma_x A - 2\phi B =0.
\ee
Now we assume the power series expansion $B = B^{(0)} (x) + B^{(1)}(x) \phi$ (up to first order is sufficient), expand the above expression in powers of $\phi$ (concretely, $\phi^2$, $\phi^1$ and $\phi^0$) and compare coefficients to find
\be
\phi^2: \quad A_x = B^{(1)} , \qquad \phi^1: \quad B^{(1)}_x = 2B^{(0)}
\ee
and
\be \label{A-eq}
\phi^0: \quad -2(1 + \bar \sigma) A_x + \frac{1}{2}A_{xxx} - \bar \sigma_x A =0.
\ee
But not all solutions of this equation are acceptable. The easiest way to see it is to go back to the case without impurity, $\bar \sigma =0$. Then the general solution of Eq. (\ref{A-eq}) is
\be\label{A-sol}
A = A_0 + A_c \cosh(2x) + A_s \sinh(2x)
\ee
where $A_0$, $A_c$ and $A_s$ are integration constants.  Here the only acceptable solution is the constant solution $A=A_0$ (corresponding to the translational symmetry), because solutions with nonzero $A_c $ or $A_s$ produce nontrivial transformations on $\phi$ (a nonzero $B$) which change the asymptotic values $\phi_{\pm} =\lim_{x \to \pm \infty} \phi $. The corresponding formal  "BPS solutions" have, therefore, infinite energy. In other words, the solutions generated from a kink solution by the action of the symmetry transformation with a nonzero $A_c$ or $A_s$ exist only locally but not globally.

\begin{figure}
\centering
\includegraphics[height=6.0cm]{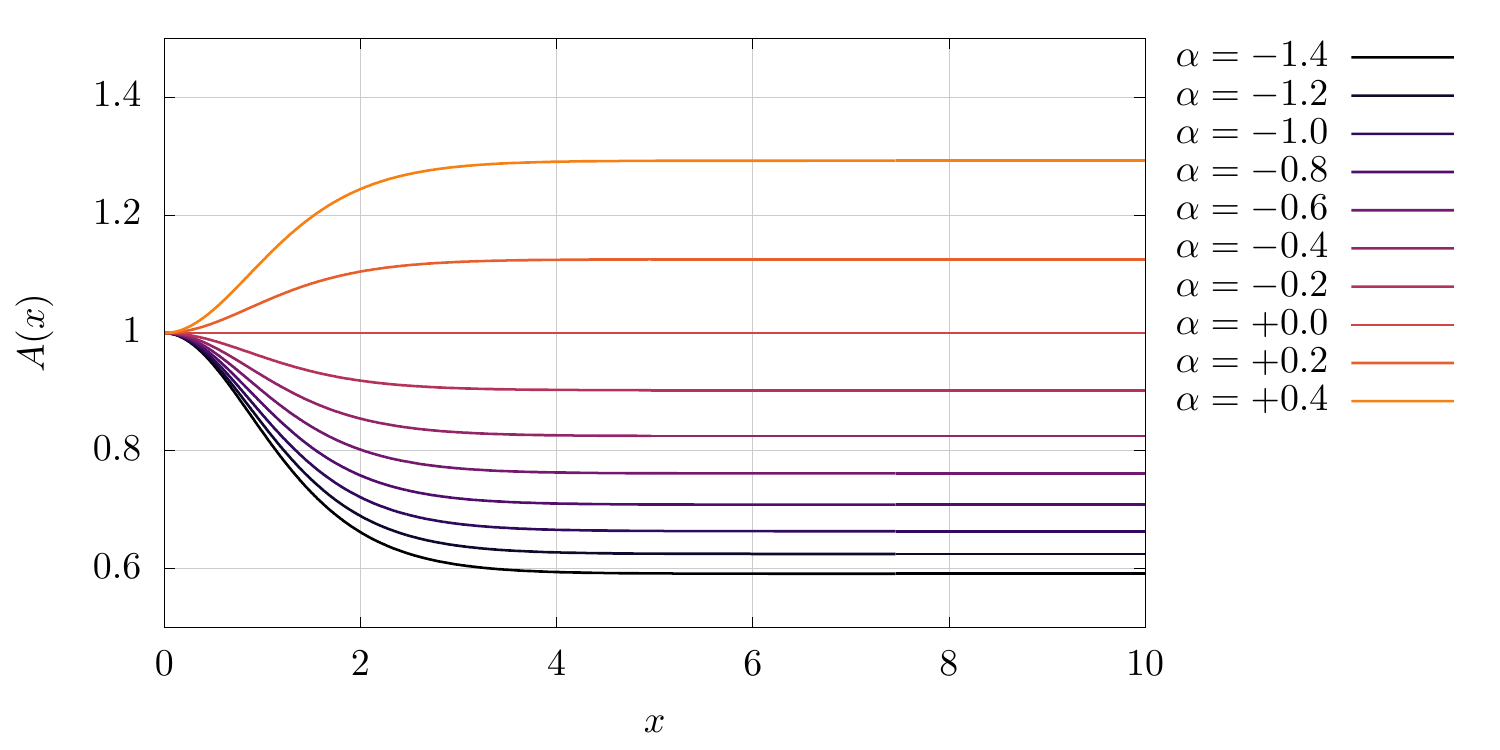} 
\caption{The function $A$ for several values of the impurity strength coupling $\alpha$.}
\label{sym-func}
\end{figure}

Even for a nonzero but localised impurity (i.e., an impurity which vanishes sufficiently fast in the limit of large $|x|$), an acceptable solution of Eq. (\ref{A-eq}) must approach the constant solution in the limit $x\to \pm \infty$, for the same reason. Unfortunately, it is not possible to integrate Eq. (\ref{A-eq}) analytically. 
For the impurity used in this paper,
\be
\bar \sigma = \frac{\sqrt{2}\alpha}{\cosh^2 x},
\ee
numerical solutions for a few values of $\alpha$ are presented in the Figure \ref{sym-func}. It turns out that for symmetric impurities $\sigma(x) = \sigma(-x)$, it is sufficient to consider symmetric functions $A(x)=A(-x)$, so it is enough to perform the numerical integration from $x=0$ to $x=\infty$.

Finally, we want to remark that there is no one-parameter family of lump solutions generated by these generalised translations, despite the fact that the lumps, too, are solutions of the BPS equation. Correspondingly, there is no zero mode in the linear spectrum of the lump. This implies that the lump (which represents the vacuum solution in the BPS sector) is {\em invariant} under the generalised translations, exactly like the vacuum solution of the model without impurity (the constant field $\phi = \phi^v_\pm = \pm 1$) is invariant under normal translations. The kink, on the other hand, does not belong to the BPS sector and, therefore, does not share this symmetry, which is a symmetry of the BPS equation but {\em not} of the energy functional.

%%%%%%%%%%%%%%%%%%%%%%%%%%%%%%%%%%%%%%%%%
\subsection{Analytical description of the antikink-impurity state}
%%%%%%%%%%%%%%%%%%%%%%%%%%%%%%%%%%%%%%%%%
%%%%%%%%%%%%%%%%%%%%%%%%%%%%%%%%%%%%%%%%%
\subsubsection{Weak/strong impurity expansion}
%%%%%%%%%%%%%%%%%%%%%%%%%%%%%%%%%%%%%%%%%
As we already know, for our choice of the impurity, the model possesses two distinguished values of the impurity strength. For $\alpha=0$ it reduces to the pure $\phi^4$ theory while for $\alpha=-\sqrt{2}$ the BPS $Q=-1$ solution is just a collection of infinitely separated solitons of the pure $\phi^4$ model. This leads to two possible expansions. 

In the limit for  $\alpha\to 0$, in the zeroth approximation the BPS solution is simply an antikink of the pure $\phi^4$ model. It is possible, however, to find the first $\alpha$ dependent term using the linearization technique. We expand the BPS equation around the antikink solution $\phi(x)=\phi_{\bar K}(x)+\sqrt{2}\alpha \eta(x)$ keeping only linear terms in $\alpha$. In $\mathcal{O}(\alpha)$ the equation takes the form
\begin{equation}
 \eta_x-2\phi_{\bar K}\eta+\frac{1}{\cosh^2x}=0\,,
\end{equation} 
which is solved by
\begin{equation}
 \eta(x) = -\frac{x}{\cosh^2x}+\frac{C}{\cosh^2x}\,,
\end{equation} 
where $C$ is an arbitrary constant. The part proportional to $C$ has the form of a translational mode of the antikink. Therefore, for the antikink fixed at the origin, we have the approximate solution
\begin{equation}
 \phi(x)\approx-\tanh x -\frac{\sqrt2\alpha\, x}{\cosh^2x}+\mathcal{O}(\alpha^2)\,.
\end{equation} 
The above approximation is in perfect agreement with the full numerical solution of the nonlinear BPS equation for $|\alpha|<0.2$.
Note that from (\ref{3-soliton}) and (\ref{alphaBPS}) we obtain in the limit $\alpha\to0$ 
\begin{equation}
  \phi(x)\approx-\tanh x -\frac{3\sqrt2\alpha\, \tanh x}{2\cosh^2x}+\mathcal{O}(\alpha^2)\,
\end{equation} 
which gives even slightly different asymptotic. This is not too surprising as the composite soliton picture is less applicable for $\alpha \rightarrow 0$.

For $\alpha\to-\sqrt2$ we can make yet another expansion
\begin{equation}
 \alpha=-\sqrt2 + \epsilon\,,\qquad\phi=\tanh x+\epsilon \eta+\mathcal{O}(\epsilon^2).
\end{equation} 
This expansion gives the following linearized BPS equation
\begin{equation}
 \eta_x-2\eta\tanh x +\frac{\sqrt{2}}{\cosh^2x}=0\,.
\end{equation} 
The solution to the above equation can be found analytically and the approximated solution up to first order in $\epsilon$ can be written as
\begin{equation}
 \phi(x)=\tanh x -\frac{\sqrt 2\epsilon}{3}\left[(2\cosh^2x+\tanh x)\tanh x+2C\cosh^2 x\right]\,,
\end{equation} 
where $C$ is an integration constant. 
The above solution is always divergent at least in one of the cases $x\to \pm \infty$, but is a valid solution as long as $|\eta|\ll 1$ or, in other words, as long as the solutions is \textit{close} to $\tanh x$.  
When $C=0$, the solution is odd and in the vicinity of $x=0$ it gives just some addition to the hyperbolic tangent. 
However, as $|x|$ becomes large, the solution diverges exponentially from $\tanh x$ and the above approximation is no longer valid. 
However, the solution gives quite a good approximation of the zeros of the solution, that is, the positions of the two antikinks. 
When $C=\pm 1$, the homogeneous part cancels one of the $\cosh^2 x$ from the inhomogeneous part of the solution, and the solution approaches the vacuum on one side. 
Such a solution approximates a $Q=0$ lump (see below). 

%%%%%%%%%%%%%%%%%%%%%%%%%%%%%%%%%%%%%%%%%
\subsubsection{The composite impurity picture} \label{sec-composite}
%%%%%%%%%%%%%%%%%%%%%%%%%%%%%%%%%%%%%%%%%

One can ask what happens with the observed hidden antikink-kink-antikink structure of the $Q=-1$ BPS solutions,  so well visible at $\alpha \rightarrow -\sqrt{2}$, when we consider bigger values of $\alpha$. Here we present some arguments that, although the constituent solitons are close to each other and finally lose their identity, the BPS solutions can always be described, with some approximation, as three-soliton states. To test this hypothesis, we consider the following exact expression representing a symmetric antikink-kink-antikink system, where the constituents are the solitons of the pure $\phi^4$ theory 
\be
\phi_{Q=-1}^{\bar{K}_0K_0\bar{K}_0}(x) = -\tanh (x+s) + \tanh (x) - \tanh (x-s). \label{3-soliton}
\ee
Here, $s$ is a distance between the constituent solitons, which should be computed applying a minimization scheme. We insert this trial function into the energy expression, which up to the boundary term is just the BPS equation squared. Then, we minimize it with respect to the separation variable $s$. This gives the following relation between the distance and $\alpha$
\be
\alpha = \sqrt{2} \frac{3s \left(4\cosh (2s)+\frac{1}{\cosh^2 s} \right) -7 \sinh (2s) -\sinh (4s) +3\tanh s}{-4s-8s\cosh(2s) +4\sinh(2s) +\sinh(4s)}.
\ee
One may verify that infinite separation $s\rightarrow \infty$ occurs for $\alpha \rightarrow -\sqrt{2}$. On the other hand, for $\alpha=0$ we recover the usual antikink solution $\phi=-\tanh x$ , which corresponds to the $s=0$ limit.  

\begin{figure}
\centering
\includegraphics[width=0.75\columnwidth]{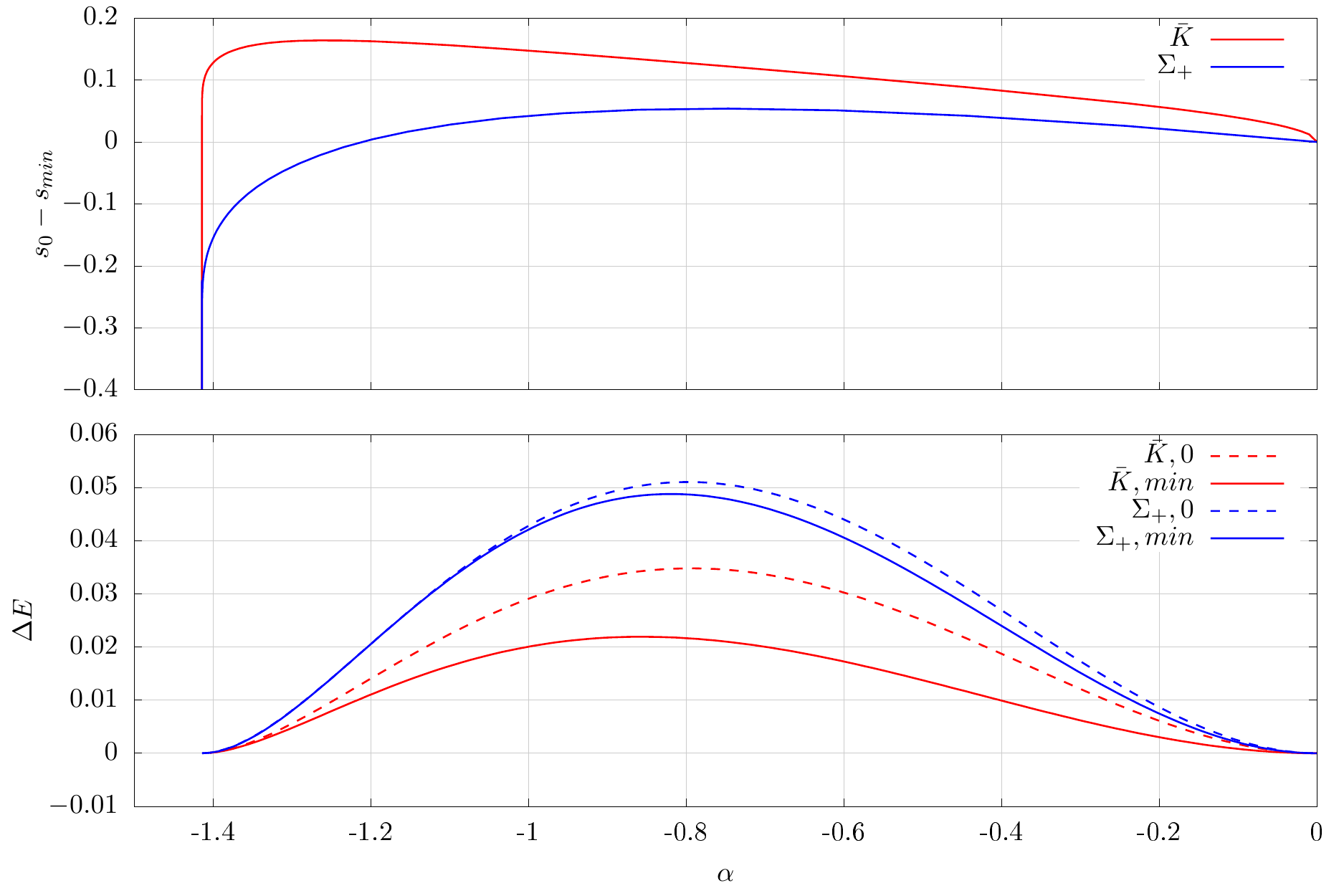}
\caption{Composite interpretation of the BPS solutions (antikink $\bar{K}$, lump $\Sigma_+$). Upper plot: discrepancy between separations obtained using the minimization of the energy ($BPS^2$) functional and the zero of the $BPS$ integral. Lower plot: energy excess of the solutions obtained from the approximations assuming trial functions, as compared to the real values ($E_\Sigma=0,\,E_{\bar K}=4/3$).}
\label{K*Kapprox}
\end{figure}

In Fig. \ref{K*Kapprox} (lower panel, solid red curve) we plot the difference between the energy of the $Q=-1$ BPS solution and its  antikink-kink-antikink approximation as a function of $\alpha$. The discrepancy is never bigger than $1.5\%$ (remember that the mass of the $Q=-1$ solution is always 4/3). 
It should be underlined that the three-soliton structure as well as the mutual distance between the constituents in the odd-symmetric solution is {\it an interpretation} of the numerical data for $\alpha$ bigger that a region close to $-\sqrt{2}$. However, the three-soliton configuration (\ref{3-soliton}) reproduces the true numerical solution quite well even for $\alpha \rightarrow 0$ which makes the three-soliton picture plausible.

\hspace*{0.2cm} 

\begin{figure}
\centering
\includegraphics[width=0.5\columnwidth]{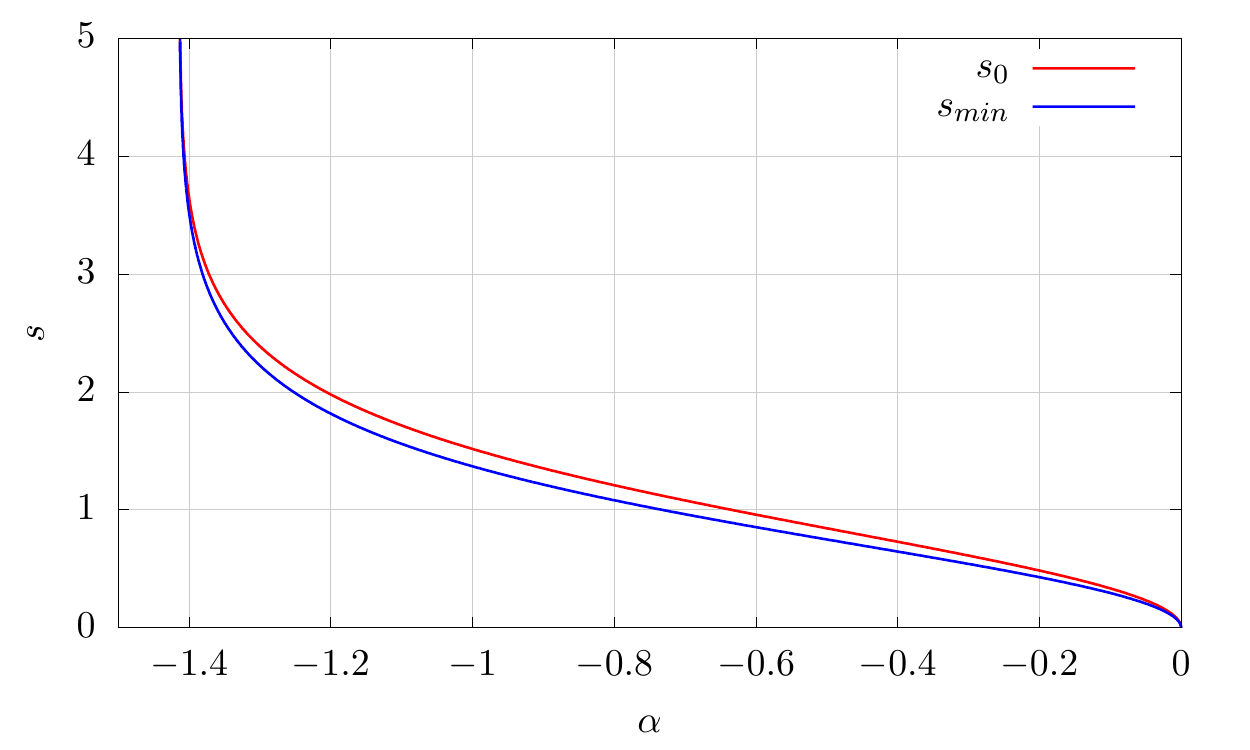}
\caption{The distance $s$ between the antikinks and the central kink in the $Q=-1$ BPS solution, as a function of the impurity parameter $\alpha$. Blue line - the energy minimization. Red line - the integration of the BPS equation.}
\label{kk}
\end{figure}

Surprisingly, a quite good approximation to the optimal separation distance (and therefore to the true solution and its energy) may be obtained if, instead of minimizing the energy functional, we use the BPS equation itself. This means that we insert the trial function into the BPS equation and integrate it over the whole space
\be
\int_{-\infty}^\infty \phi_x dx = \int_{-\infty}^\infty  (\phi^2-1) dx -\sqrt{2}\alpha \int_{-\infty}^\infty \frac{dx}{\cosh^2 x}
\ee
Obviously, the left hand side is just twice the topological charge of the assumed configuration. Also the integral over the impurity may be easily computed. Hence,
\be
-2+2\alpha \sqrt{2} = \int_{-\infty}^\infty  (\phi^2-1) dx 
\ee
Plugging in the approximate solution, we find the extremely simply formula
\be\label{alphaBPS}
\alpha=-\sqrt{2}\left(1-\frac{2s}{\sinh 2s} \right)
\ee
This reproduces the asymptotic behavior correctly. In the Fig. \ref{kk} we plot the position of the constituent antikink computed in both ways - by the energy minimisation $s_{min}$ (blue curve) and by the integration of the BPS equation $s_0$ (red curve). We see quite a good agreement. To better visualise it, we plot $s_{\min}-s_0$, see Fig. \ref{K*Kapprox}, upper panel, red curve. The observed singularity for $\alpha \rightarrow- \sqrt{2}$ is related to a discrepancy in subleading terms for $s_{\min}$ and $s_0$, which still tend to infinity in this limit. In the same plot we also show the discrepancy between the energy of the solution and the energy arising form the integration of the BPS equation (dashed red curve). Now, this approximation is slightly worse than the energy minimisation but is always smaller than $2.7\%$.

Note that the integration of the BPS equation bears some similarity with the derivation of the Bradlow law for the vortices in the Abelian Higgs model. 

Now, let us discuss why the minimisation of the integral of the BPS equation can at all give a reasonable result. It is a combination of two arguments. First of all, we have the inequality $\int (BPS)^2 \ge (\int BPS)^2$ where $BPS$ stands for the BPS equation, $BPS = \phi_x + \sqrt{2}\sigma + 1-\phi^2$.
Secondly, we have to assume that the trial function is sufficiently "good", that is, it gives an energy close to the Bogomolny bound for the optimal parameter values. This is equivalent to saying that $\int (BPS)^2$ is a small number, which implies that $(\int BPS)^2$ is even smaller and may be approximated by zero. The better the trial function (i.e., the smaller the energy for the optimal values of the parameters), the smaller will be $(\int BPS)^2$  and the better it is approximated by $(\int BPS)^2=0$.

\hspace*{0.2cm}

Interestingly, the well pronounced three soliton hidden structure of the $Q=-1$ BPS solution occurs in the strong coupling regime, i.e., where $\alpha$ takes the smallest possible value. 
In other words, when the impurity is strong then the $Q=-1$ solution reveals a structure of three weakly coupled constituents. 
On the other hand, when the impurity is weak, i.e., for $\alpha \rightarrow 0$ (which for example allows for a perturbative treatment of the impurity), then the hidden solitons are in some sense strongly coupled and lose their identity. 
All this suggests a sort of weak-strong correspondence between the strength of the impurity and the strength of the interaction between the hidden solitons.

%%%%%%%%%%%%%%%%%%%%%%%%%%%%%%%%%%%%%%%%%
\subsection{Topologically trivial lumps}
%%%%%%%%%%%%%%%%%%%%%%%%%%%%%%%%%%%%%%%%%
\begin{figure}
\centering
\includegraphics[width=0.75\columnwidth]{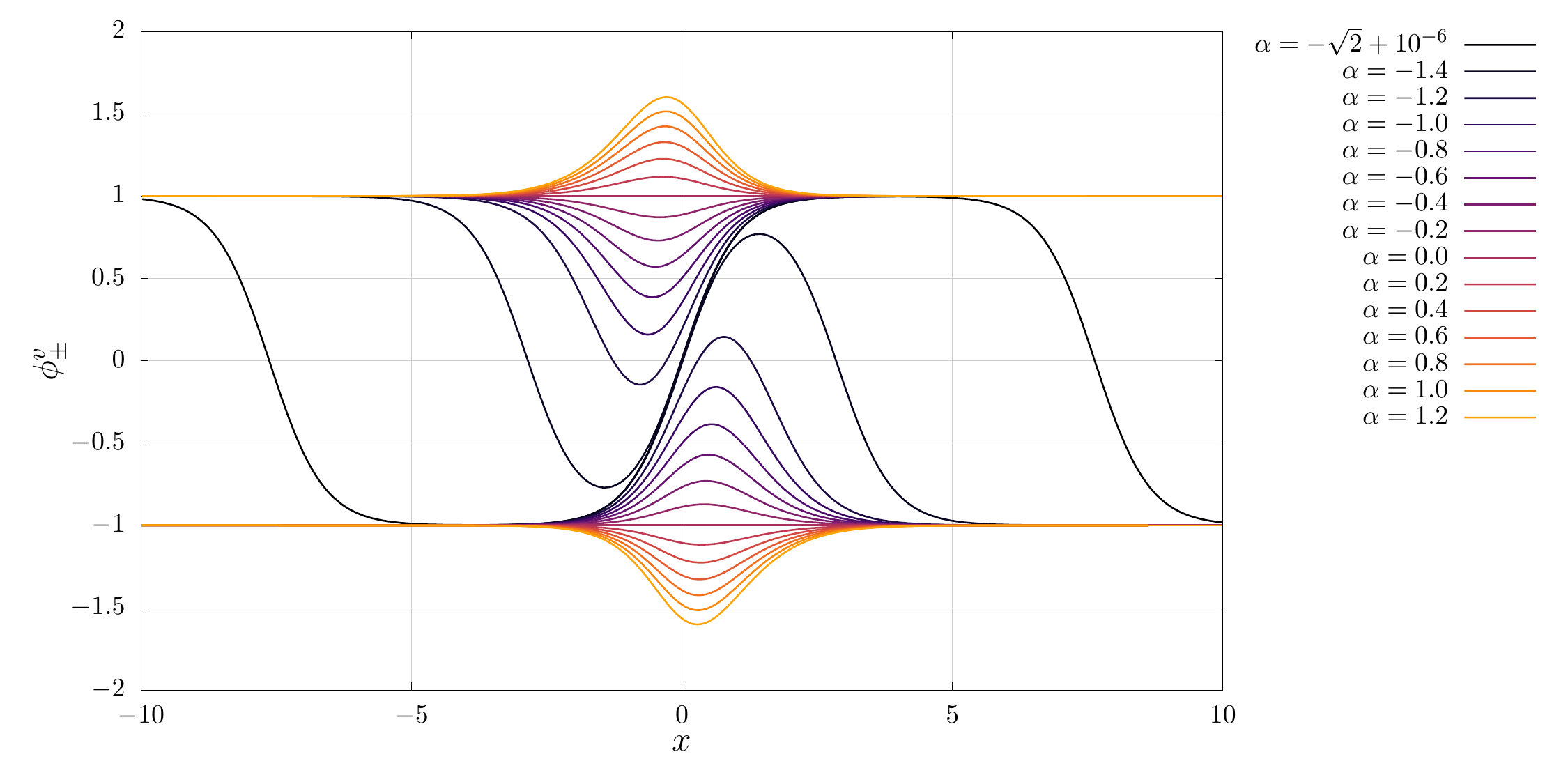}
\caption{Topologically trivial lumps $\Sigma_+$ and $\Sigma_-$ for different values of $\alpha$.}
\label{vacuum}
\end{figure}

It is a straightforward consequence of the nonlinear superposition of antikink and topologically trivial lump in the BPS sector that, after sending the antikink to infinity, we remain with the pure lump solution located on the impurity. As the infinitely separated antikink has the form $-\tanh(x-x_0)$ (with $x_0\rightarrow \infty$), its energy is $E=4/3$. Thus, the energy of the lump localised on the impurity is exactly 0. This means that the topologically trivial lump also saturates the energy bound and, in the above sense, is a BPS solution.  In Fig. \ref{vacuum} we plot the topologically trivial lumps $\Sigma_+$ and $\Sigma_-$ for different values of $\alpha$.  Note that the lump is not symmetric under the $x\rightarrow-x$ transformation. The symmetry center of the lump is shifted from $x=0$ where the impurity is located.

\begin{figure}
\centering
\includegraphics[width=0.495\columnwidth]{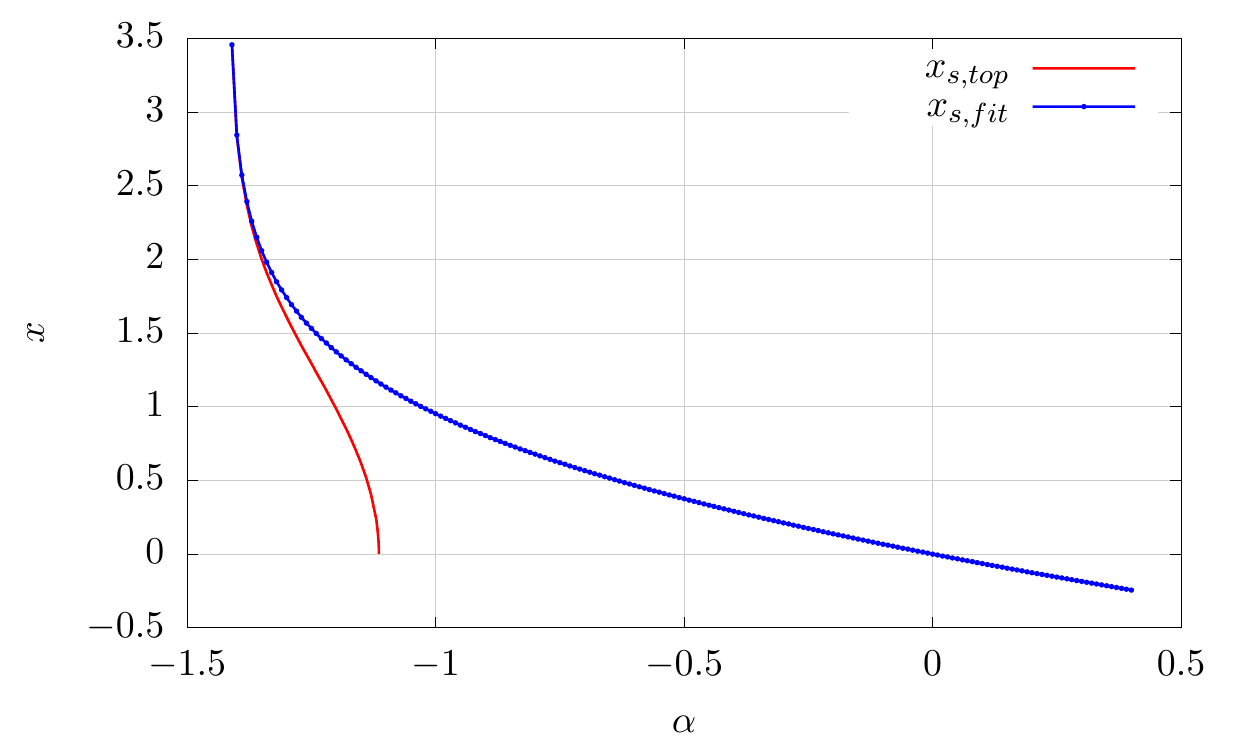}
\includegraphics[width=0.495\columnwidth]{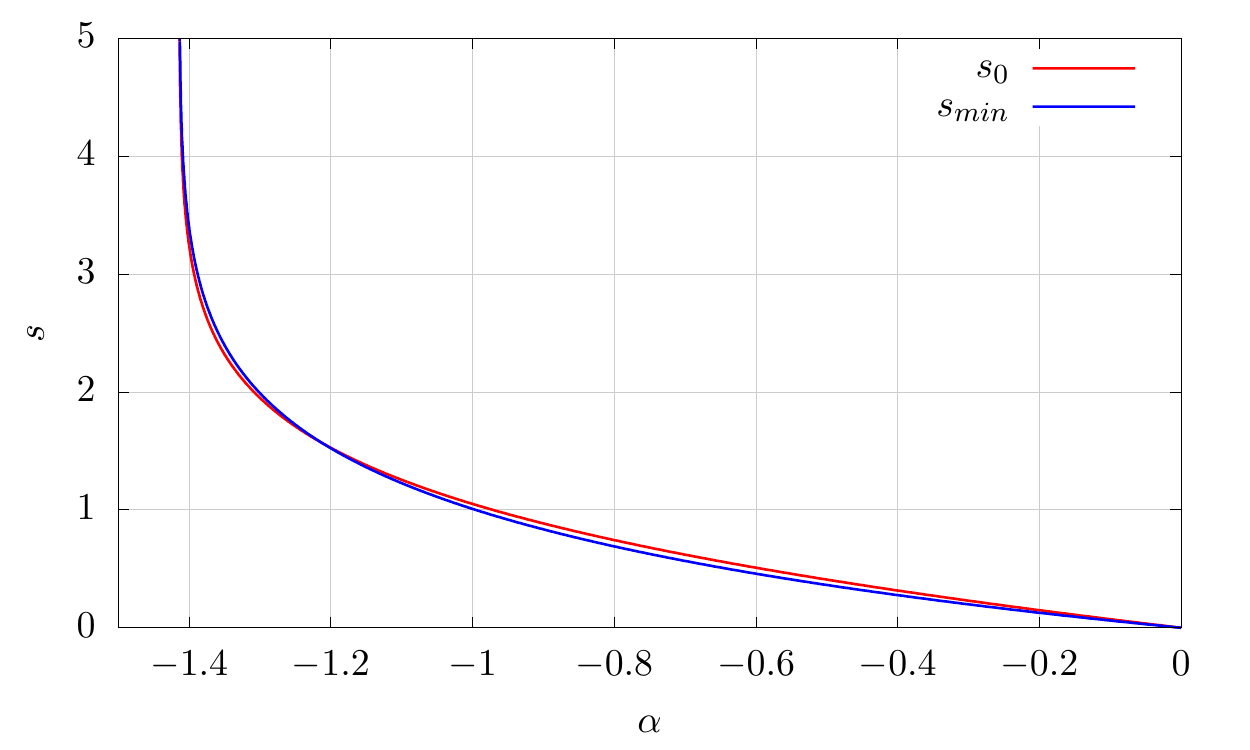}
\caption{Left: the distance between the hidden kink and antikink in the lump (blue line), together with the positions of the topological zeros (red  line). Right: the distance $s$ between the antikinks and the central kink in the $Q=0$ BPS solution in the fixed kink approximation, as a function of the impurity parameter $\alpha$. Blue line - the energy minimization. Red line - the integration of the BPS equation.}
\label{position}
\end{figure}

First of all, let us observe that in the limit when $\alpha \rightarrow -\sqrt{2}$, the topologically trivial lump, similarly as it happens for the solution in the $Q=-1$ BPS sector, clearly exhibits a hidden structure. In this limiting case, it consists of a kink located at the origin and an antikink sent to spatial infinity. 
Again, this picture, to some extent, is visible also for bigger values of $\alpha$. To verify this conjecture, we assume the following decomposition of the lump 
\be
\phi_{lump}^{K_0\bar{K}_0} = -1+\tanh(x-x_k) - \tanh (x-x_a) 
\ee
where $x_k$ and $x_a$ are the positions of the pure $\phi^4$ kink and the antikink, respectively. 
This trial function clearly takes into account the non-symmetric shape of the lump. We determine the optimal values of the fit parameters $x_k$ and $x_a$ by minimizing $\int (\phi_{num} - \phi_{lump}^{K_0\bar{K}_0})^2$, i.e., by a "least square" fitting procedure.
In Fig. \ref{position} (left) we plot $s_{min}=x_a-x_s$. In Fig. \ref{K*Kapprox} (lower panel, solid blue curve) we also plot the energy of the two soliton approximation. As the energy of the BPS lump is always 0, the energy scale to which our approximation should be compare is the mass of the antikink. Hence, we get an accuracy better than $3.75\%$. It is clearly seen that the kink-antikink interpretation of the lump works extremely well for $\alpha \rightarrow -\sqrt{2}$. When $\alpha$ grows, then the kink and antikink quickly approach each other and lose their identity. However, our simple two particle approximation reproduces the qualitative features of the true solution reasonably well, even for bigger $\alpha$, which supports our identification of the lump as a mixed state of kink and antikink. Of course, for higher $\alpha$, the discrepancy is always visible. This means that the true lump solution is not a simple superposition of these solitons but that a mutual, nonlinear interaction modifies the state. Therefore, for $\alpha$ outside a certain region close to $-\sqrt{2}$, the lump as a superposition of a kink and an antikink as well as their position inside the lump is again an interpretation which, however, works quite well. In particular, we may appreciate in Fig. \ref{position} that the distance between kink and antikink is zero for $\alpha =0$ and approaches infinity for $\alpha \to - \sqrt{2}$, as it must be.

In order to get an analytic insight into the separation distance between the hidden kink and antikink in the lump solution we apply a very simplified trial function where the position of the kink is {\it always} fixed at the origin
\be
\phi=-1+\tanh x -\tanh(x-s)
\ee
From the numerics we know that is it a good approximation {\it only} when $\alpha \rightarrow -\sqrt{2}$. Once $\alpha$ grows to 0 the kink also moves. But we may consider the above trial function as a {\it crude} approximation to the distance - not to the actual positions of the constituent solitons or the shapes of the solutions, especially as the position of the hidden kink $x_k$ is bounded from above and varies much less than its antikink counterpart. Then, again we may apply the standard energy minimization which gives
\be
\alpha=-2\sqrt{2} e^{s}\frac{-12 s \cosh s +9\sinh s +\sinh (3s) }{-5-4s -4(-1+2s)e^{2s}+e^{4s}}
\ee
while the integration of the BPS equation leads to 
\be
\alpha=-\sqrt{2}(s\coth s -s-1).
\ee
Again, both expressions reproduce the asymptotic behavior correctly. Furthermore, they coincide quite well - see Fig. \ref{position} (right panel). 
As the one parameter trial function does not reproduce the proper shapes for higher $\alpha$, the derived formulas could be less precise than in the $Q=-1$ sector, especially when $\alpha\rightarrow 0$. 

In the limit $\alpha\to 0$ we can linearize the BPS equation and obtain for $\phi=-1+\sqrt{2}\alpha\eta$
\begin{equation}
 \eta_x+\frac{1}{\cosh^2(x)}-2\eta=0.
\end{equation} 
This equation has an exact solution:
\begin{equation}
 \eta(x) = 2e^{2x}\left[\frac{1}{1+e^{2x}}+2x-\log\left(1+e^{2x}\right)\right] + Ce^{-2x}\,.
\end{equation} 
This time, the part coming from the homogeneous part has to vanish ($C=0$), otherwise the solution would not be be normalized. This is due to the fact that the lump is bound to the impurity and there is no translational freedom, as it happened in the case of the BPS antikink.
This approximation works perfectly where it should, that is $|\alpha|<0.2$. Clearly, the profile of the lump is not symmetric, and is not centered around $x=0$. Obviously, these features are not captured by the symmetric $K\bar K$ approximation in the limit of $\alpha\to 0$
\begin{equation}
 \phi(x)\approx-1-\frac{\sqrt 2}{2}\frac{\alpha}{\cosh^2x}+\mathcal{O}(\alpha^2)\,.
\end{equation}

%%%%%%%%%%%%%%%%%%%%%%%%%%%%%%%%%%%%%%%%%
\subsection{Non BPS kink-impurity state}
%%%%%%%%%%%%%%%%%%%%%%%%%%%%%%%%%%%%%%%%%

By construction, i.e., by the assumed form of the impurity, there is an exact non-BPS kink-impurity solution $K$ of the full second order static equation 
\be
\phi_{K}= \tanh x
\ee
for any value of $\alpha$.  Although the profile of the kink is $\alpha$ independent, its energy varies and reads
\be
E_{K}=\frac{4}{3} \left( 1+2\sqrt{2} \alpha + \alpha^2 \right) . \label{E kink}
\ee
 
Moreover, as we already remarked, for one isolated value of the parameter, $\alpha=-\sqrt{2}$, this kink-impurity configuration {\it is} also a proper BPS solution, i.e., it saturates the bound and solves the Bogomolny equation. The energy of this BPS kink is $E(\alpha=-\sqrt{2})=-4/3$ and obviously saturates the topological bound for $Q=+1$. 

The non-BPS kink-impurity solution $\phi_K$ appears to be stable for $\alpha \in [-\sqrt{2},0]$ while it develops an instability for $\alpha>0$. This means that there is a repulsive interaction between the kink and the impurity which expels the soliton to infinity. This can be easily proven if we compare the energy of the kink-impurity bound state (\ref{E kink}) and the energy of the infinitely separated topologically trivial lump with $E=0$ and free kink $E=4/3$
\be
E_K-\left( 0 +\frac{4}{3}\right) =\frac{4\alpha}{3} \left(2\sqrt{2}  + \alpha \right) 
\ee
which is positive for $\alpha >0$. Hence, for the bound state it is energetically favorable to decay into its infinitely separated constituents. 

\begin{figure}
\includegraphics[height=6.cm]{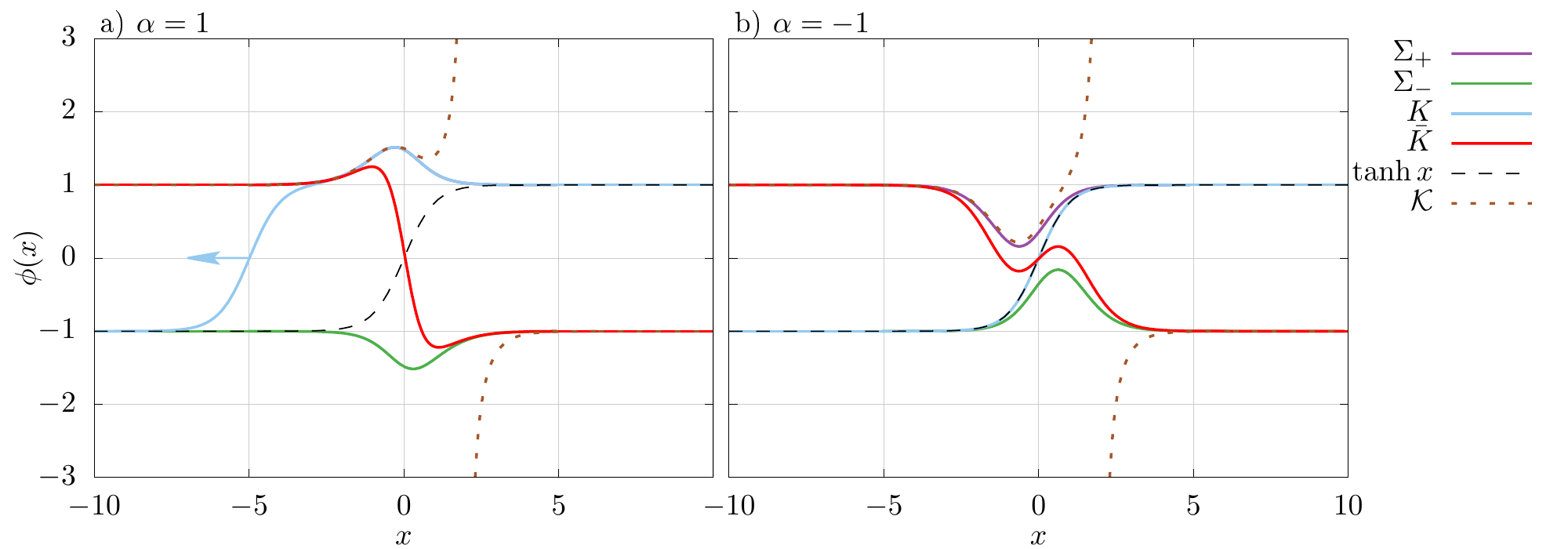}
\caption{Kink-impurity $K$, antikink-impurity $\bar{K}$ and topologically trivial lumps $\Sigma_{\pm}$ for the impurity model with $\alpha=1$ (left) and $\alpha=-1$ (right). The dotted purple line corresponds to a static singular solution.}
\label{sol}
\end{figure}

As an example, in Fig. \ref{sol} we show the kink, antikink and lump solutions for $\alpha=1$ and $\alpha=-1$. The antikink (red curve) is a stable, BPS solution in both cases. For $\alpha=-1$ a nontrivial inner structure (which grows to a kink-antikink pair as $\alpha\rightarrow -\sqrt{2}$) is already visible. The kink solution, $\phi_K=\tanh x$ is stable for $\alpha=-1$ and unstable for $\alpha=1$ where it decays into the vacuum impurity solution and an infinitely separated kink of the pure $\phi^4$ theory. To visualize this fact, we included a snapshot of the decay (kink running to the left) which was generated by introducing a small perturbation to the unstable kink solution (we remark that in Fig. \ref{sol} the lump $\Sigma_+$ which remains close to $x=0$ as a result of this decay cannot be distinguished from the pure lump solution $\Sigma_+$).
\vspace*{0.2cm}
 \begin{figure}
\centering
\includegraphics[height=6.cm]{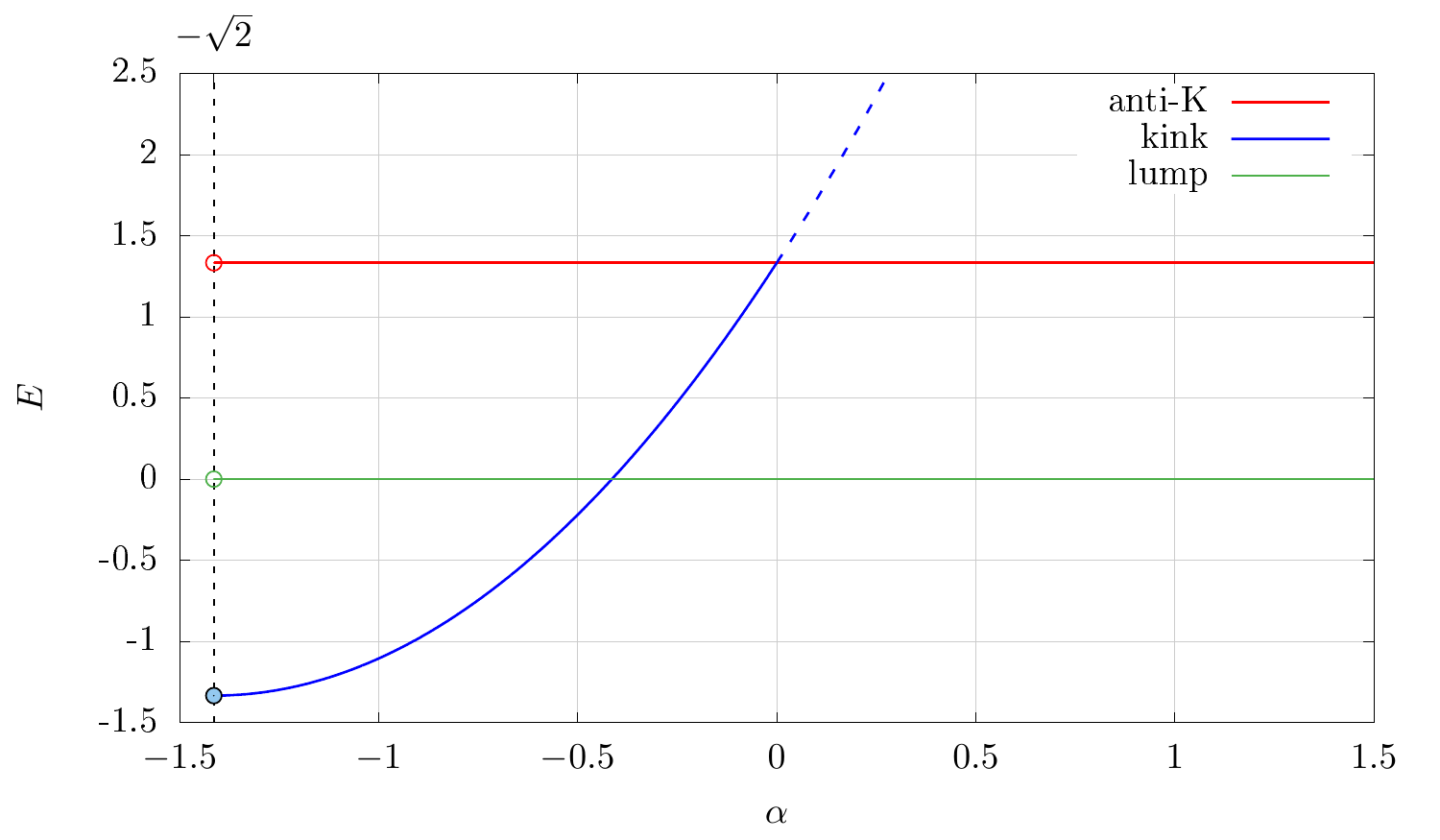}
\caption{Energies of the kink-impurity, antikink-impurity and topologically trivial lump solutions as functions of $\alpha$.}
\label{energy}
\end{figure}

Note that the second order equation has even more static solutions satisfying the appropriate boundary conditions (for $Q=\pm 1$) but developing a singularity at some point. 
For $\alpha=0$ these singular solutions $\mathcal{K}$ are given analytically in the form $\phi_\mathcal{K}(x)=\pm\coth(x-x_0)$.  
When considered on a full line, such solutions have infinite energy, but they can play an important role in models with restricted domains \cite{Dorey:2015sha}. In Figure \ref{sol}, the numerically found singular solutions are shown with dotted purple lines.

Let us summarize the spectrum of static solutions in the BPS $\phi^4$-impurity model as a function of the strength of the impurity (the parameter $\alpha$). In Fig. \ref{energy} we show the energies of the kink $\phi_K=\tanh x$ and the BPS antikink as a function of the parameter $\alpha$. 

The BPS antikink-impurity solution always saturates the energy bound $E=4/3$, which obviously enforces the topological stability of the solution. Strictly speaking, there are infinitely many solutions representing the antikink at an arbitrary distance from the the lump (always located on the impurity). Hence, we dynamically recover the translational invariance of the solitons in the pure $\phi^4$ model. Here, the change of the position of the antikink modifies the shape of the solution but leaves the energy unchanged. As we will see below, this corresponds to the existence of a zero mode. 

The stable kink-impurity solution exists for $\alpha \in [-\sqrt{2}, 0]$ and has energy smaller than the BPS antikink-impurity state. Its energy is, on the other hand, bigger than the bound, except for the very special case, $\alpha=-\sqrt{2}$, when it becomes a BPS solution with energy $E=-4/3$. Note that in this limit the energies of the kink and antikink exactly cancel. As a consequence, such a pair has the energy of the topologically trivial lump. In other words, the energy needed for the creation of a kink-antikink pair drops to 0 as we approach $\alpha=-\sqrt{2}$. This obviously coincides with our finding that for $\alpha=-\sqrt{2}$ the lump is a composition of the kink and antikink. Moreover, since the constituents are exponentially localized and their distance grows, the separation energy also quickly drops to 0.  

Surprisingly, for $\alpha = -\sqrt{2}+1$ the kink-impurity bound state and the topologically trivial lumps have exactly the same energies $E=0$ which may be interpreted as an enhancement of the degeneracy of the vacuum, which contains not only topologically trivial lumps but also the kink-impurity solution. For $\alpha< -\sqrt{2}+1$ the kink-impurity is the lowest energy state, and we have a unique and topologically nontrivial vacuum. Nonetheless, we will find that these facts do not have any impact on the dynamics of solitons in our model.

%%%%%%%%%%%%%%%%%%%%%%%%%%%%%%%%%%%%%%%%%
\section{Spectral structure}
%%%%%%%%%%%%%%%%%%%%%%%%%%%%%%%%%%%%%%%%%
In order to investigate the spectral structure of the model, we have to perturb the solutions in each topologically distinct sector. We introduce a small perturbation around the static kink-impurity, antikink-impurity and topologically trivial lump, $\phi=\phi_{static}+Ae^{i\omega t}\eta(x)+c.c.$, where the perturbation obeys 
\begin{equation}\label{linearized}
 -\eta''+(6\phi^2_{static}-2-2\sqrt2\sigma)\eta=\omega^2\eta\,
\end{equation} 
It is a well documented fact that the spectral structure of the static solutions plays an important and sometimes even a crucial role not only in the analysis of stability and relaxation, but also in processes of kink-antikink and kink-impurity scatterings \cite{Sugiyama:1979mi, Dorey:2011yw, Dorey:2017dsn}.
%%%%%%%%%%%%%%%%%%%%%%%%%%%%%%%%%%%%%%%%%
\subsection{ BPS antikink-impurity solution}
%%%%%%%%%%%%%%%%%%%%%%%%%%%%%%%%%%%%%%%%%

\begin{figure}
\includegraphics[width=\textwidth]{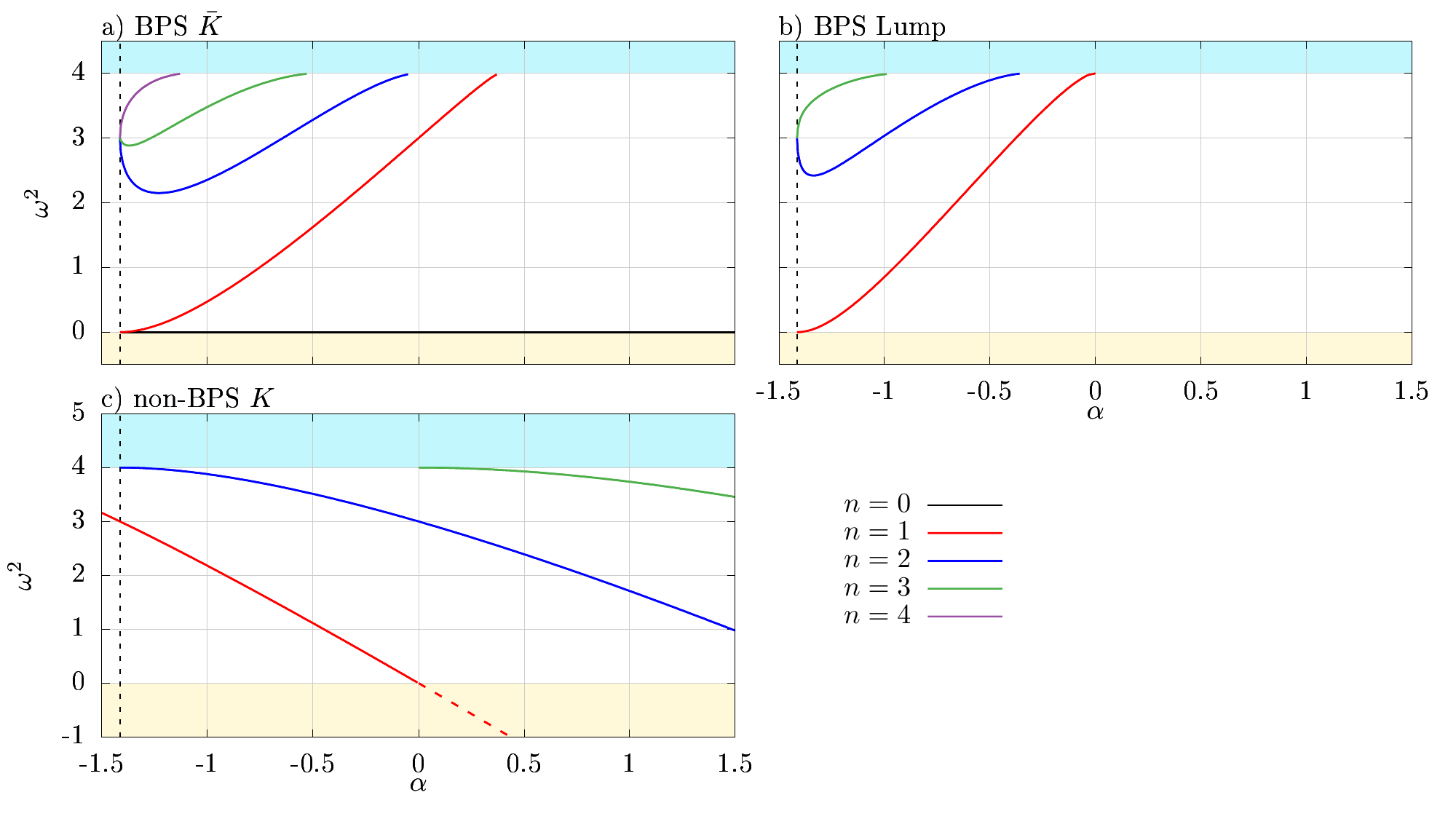}
\caption{Spectral structure of the antikink (left), the lump (right) and the kink (bottom).}
\label{spectral}
\end{figure}

We start with the BPS antikink-impurity solution. Unfortunately, we do not know the analytic form of the profiles of these solutions. However, because they are of the BPS type and obey first order equations, they are very easy to find numerically. We have found the eigen-frequencies of the bound vibrational modes of the BPS solution centered at $x=0$ using the linearized equation (\ref{linearized}) substituting $\phi$ with the numerically found antikink-impurity solution $\phi_{\bar K}$.
The results are gathered in Fig. \ref{spectral} (top left). 

Obviously, for $\alpha=0$ the defect has exactly the same spectral structure as the pure $\phi^4$ kink: a single mode with $\omega^2=3$ and a translational zero mode $\omega=0$. 
The first main result is that this zero mode exists for {\it any} acceptable value of the parameter $\alpha$. Hence, even if the impurity is coupled, the model recovers a sort of translational symmetry. This we have already understood as a nonlinear superposition of the antikink and the trivial lump which leaves the energy unchanged. 

Next, for $\alpha=-\sqrt{2}$ we know the exact BPS solution which now represents the kink. It has a single even mode with a frequency $\omega=\sqrt3$. Surprisingly, although the frequency of the mode coincides with the frequency of the pure $\phi^4$ kink, the potential generated by the kink trapped by the impurity is a P\"oschl-Teller potential with the same  depth as in the sine-Gordon model.

When $\alpha$ has a slightly larger value $\alpha=-\sqrt2+\epsilon$ the BPS antikink-impurity solution looks like a widely separated $\bar KK\bar K$ configuration. Both antikinks have the same structure as the free defects in the pure $\phi^4$ model. Namely, one translational mode $\omega=0$ and one oscillating bound mode $\omega=\sqrt3$. The whole configuration in the limit of $\alpha \to-\sqrt2$ has two modes - $\omega=0$ (twice degenerated) and $\omega=\sqrt3$ (three times degenerated). However, for nonzero $\epsilon$ the mutual interaction lifts the degeneracy and the modes split, which is clearly visible in Fig. \ref{spectral}. The modes with the frequency $\omega=\sqrt3$ split into three modes, one even (with the lowest frequency) and two odd modes. 
The mode with $\omega=0$ splits into two modes. The odd mode has a frequency which increases as $\alpha$ increases and becomes the single vibrational mode. It exists for $\alpha <0.39$ where it reaches the mass threshold. The second mode still has zero frequency and reflects the BPS-ness property of the solution. The excitation of this mode does not cost any energy, therefore the defect can be moved away from the impurity. The existence of this zero mode indicates the existence of a symmetry transformation generalising the translational symmetry of the model without impurity. We demonstrated the existence of this generalised translational symmetry in Section 3.2.

%%%%%%%%%%%%%%%%%%%%%%%%%%%%%%%%%%%%%%%%%
\subsection{The topologically trivial lump}
%%%%%%%%%%%%%%%%%%%%%%%%%%%%%%%%%%%%%%%%%

The spectral structure of the trivial lump possesses at least one oscillating mode for  $\alpha\in [-\sqrt{2},0)$. It reaches the mass threshold for $\alpha=0$, that is, in the limit of the pure $\phi^4$ theory, while it tends to a zero mode for $\alpha=-\sqrt{2}$. For the same value $\alpha = - \sqrt{2}$, two other oscillating modes merge to a single mode of $\phi^4$ theory with $\omega^3=3$. This is related to the fact that in this limit the topologically trivial lump can be viewed as a kink-antikink bound state where each of its constituents looks like a soliton of the pure $\phi^4$ model. Hence again, the number of degeneracies reflects the number of hidden constituents while their mutual interaction (for $\alpha >-\sqrt{2}$) leads to a splitting of the frequencies. 

It is interesting to note that the hidden structure of the lump, which first clearly appears at $\alpha=-1/\sqrt{2}$ and gets more visible as we approach $\alpha=-\sqrt{2}$, amounts to an appearance of a oscillating kink-antikink bound state in the spectrum of perturbations of the lump. As we clearly showed in a previous section, the lump can be described with very good accuracy as a pair of two solitons with a separation distance governed by the impurity. This picture results in the identification of $\omega_0$ as an oscillation frequency of the kink-ankikink pair. Precisely speaking, the kink is relatively strongly trapped by the impurity while the antikink performs oscillations whose period grows as we tend to the limiting $\alpha$. The mechanism of this behavior is simple. The negative impurity always attracts the kink while, at least in the BPS sector, it does not interact with the antikink. Hence, the kink is more and more frozen on the impurity, as we approach $\alpha=-\sqrt{2}$, while the antikink oscillates via an interaction with the kink. In the linear approximation, such a state exists for sufficiently negative $\alpha \rightarrow -\sqrt{2}$. However, since in this limit the separation energy of the pair also vanishes, in realistic processes there will always be a critical $\alpha>-\sqrt{2}$ for which the antikink will be released from the oscillating pair while the kink remains confined to the impurity. Such a behavior will be visible in many scattering processes. 

We remark that the vanishing of $\omega_0$ as $\alpha$ tends to $-\sqrt{2}$ leads to a very high stability of the oscillating topologically trivial lump (wobbling kink-antikink pair). The wobbling state can radiate but only by higher harmonics, which strongly suppresses its decay \cite{Dorey:2011yw, Romanczukiewicz:2017gxb, Adam:2017czk}.  

Note also that this oscillation mode approaches the mass threshold for $\alpha=0$. Moreover, for $\alpha \geq 0$ there are no oscillating modes at all. 

One should underline that we found a smooth transition between an oscillating excitation of the lump and an oscillating kink-antikink pair trapped on the impurity as we decrease $\alpha$. In other words, the $K\bar{K}$ oscillating state originates in the first oscillating mode. One may even say that {\it the exited lump is just the oscillating kink-antikink pair}, especially in the limit when $\alpha \rightarrow -\sqrt{2}$. Therefore, it can be interpreted as a linear effect. However, one should remember that the ground state above which the perturbation is performed, i.e., the lump, is a highly nonlinear solution. 

\begin{figure}
%\hspace*{-1.0cm}
\includegraphics[width=\columnwidth]{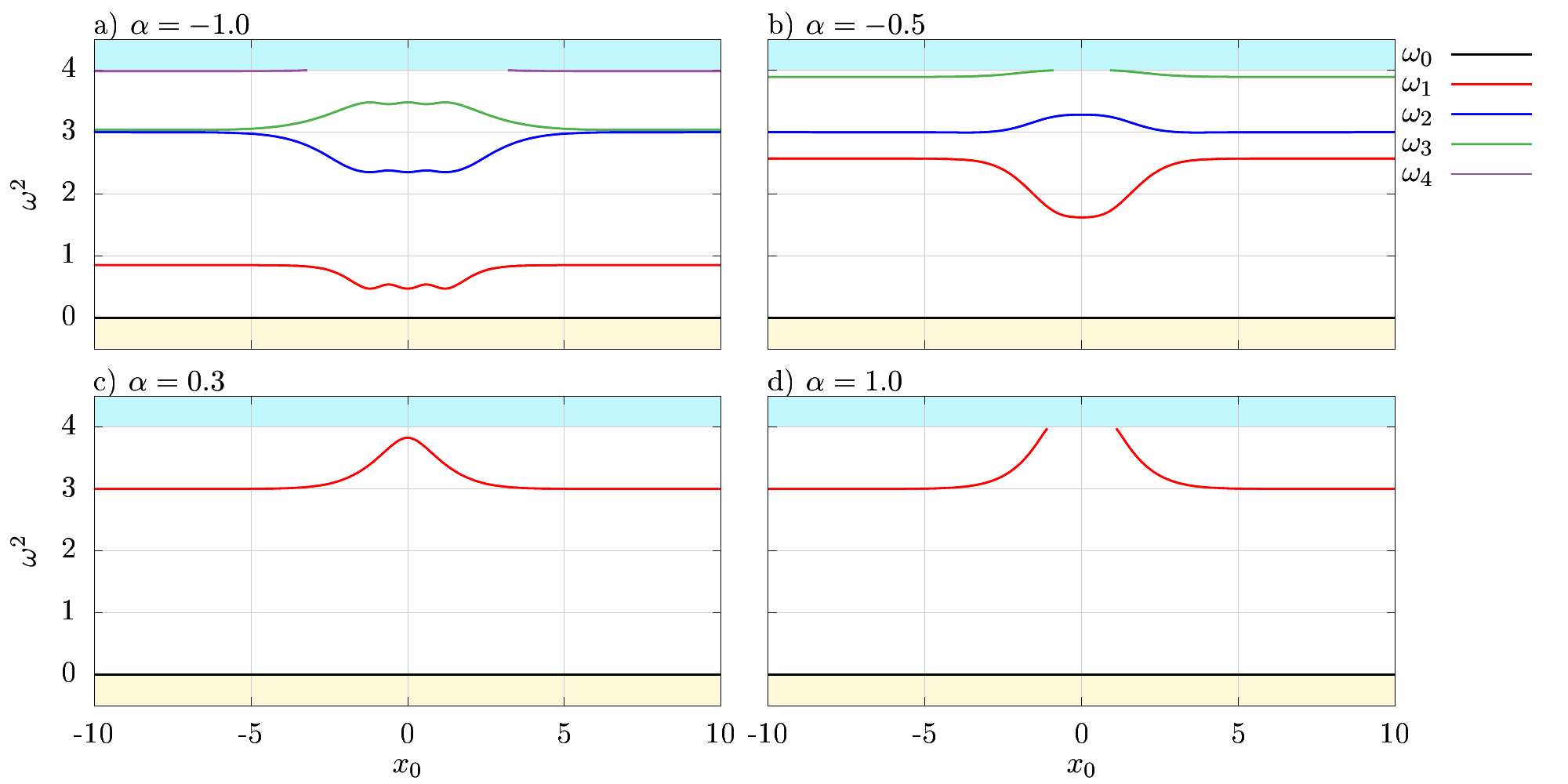}
\caption{Spectral structure of the BPS solution for different values of $\alpha$, as a function of the  topological zero of the antikink.}
\label{spectral2}
\end{figure}

\vspace*{0.2cm}

Since the interaction energy between the lump and the antikink is exactly zero, the position of the static antikink can be arbitrary. As the position changes so does the spectral structure. 
When the antikink is very far from the impurity, i.e., $|x_0|\gg1$, then the influence of the impurity is almost negligible. 
The spectrum consists of separated modes of the free $\phi^4$ soliton ($\omega=\sqrt 3$ and $\omega=0$) and the modes of the $\alpha$-dependent lump. 
As the defect approaches the lump, the modes starts to interact and for $x_0=0$ the structure becomes the one described in the previous subsection. The frequencies of the normalizable eiegnmodes are shown in Fig. \ref{spectral2} for four different values of $\alpha$. 
For large distances, the number of modes is the sum of all the modes of the antikink and the lump.
For example in Fig. \ref{spectral2} a) for $\alpha=-1$ there are five modes for $x_0=-10$, two of them are from the antikink and three from the lump (see Fig. \ref{spectral} b)). As the distance becomes smaller one of the modes enters the continuous spectrum, and for $x_0=0$ there are only four modes (see Fig. \ref{spectral} a)). For $\alpha=1$ Fig. \ref{spectral2} d) there are two modes: one oscillational mode of the antikink and one translational mode. For $x_0<2$ the oscillational mode becomes a narrow quasi-normal mode (QNM) and for $x_0=0$ its frequency becomes $\omega=2.478+0.096i$. Such a QNM can also play an important role in the dynamics of the system \cite{Dorey:2017dsn}. Note that for $\alpha<0.38$ the mode always stays below the threshold as in the Fig. \ref{spectral} c).
Note, that in \cite{Weigel:2018xng} it was conjectured that a similar disappearance of the shape modes of a colliding kink and antikink is responsible for the failure of the collective coordinate method applied to such a collision.
On the other hand, this distance-dependence of the spectral structure is crucial for the existence of the resonance bounce structure in $\phi^6$ model \cite{Dorey:2011yw}. 
%%%%%%%%%%%%%%%%%%%%%%%%%%%%%%%%%%%%%%%%%
\subsection{The kink-impurity solution}
%%%%%%%%%%%%%%%%%%%%%%%%%%%%%%%%%%%%%%%%%
Due to the very special form of the kink solution, we can analytically study its spectral structure
\begin{equation}
 -\eta''+\left(4-\frac{6+2\sqrt{2}\alpha}{\cosh^2 x}\right)\eta=\omega^2\eta\,.
\end{equation} 
This is a P\"oschl-Teller potential with a depth equal to $V_0=6+2\sqrt 2\alpha$. The bound modes of the potential are known
\begin{equation}
 \omega_n^2= 4-(\lambda-n)^2\,,\qquad 0\leq n\leq\lambda\,,\qquad\lambda(\lambda+1)=6+2\sqrt2\alpha\,.
\end{equation} 
Obviously, for $\alpha=0$ we recover the spectral structure of the kink in the usual $\phi^4$ theory. 
For example, the first bound mode has the frequency $\omega^2_0=0$. This mode is the translational zero mode of the kink, reflecting the translational symmetry of the model. 

The behavior of this mode in the impurity extended model strongly depends on the value of the parameter $\alpha$.  For negative $\alpha$, the mode transforms into an oscillating mode with frequency growing to $\omega^2=3$ at $\alpha=-\sqrt{2}$. This means that the pure $\phi^4$ kink is trapped by the impurity, forming a true kink-impurity bound state. Small perturbations  oscillate around the static solution with frequency $\omega_0$. For positive $\alpha$, the zero mode transforms into a negative frequency mode, $\omega^2<0$, which means that the static kink solution centered at $x=0$ is not a stable solution. Any small perturbation violating the symmetry results in the ejection of the kink by the impurity. As we see, the spectral structure or, more precisely, the sign of $\omega_0^2$, fully confirms the previously found structure of solitonic solutions. 

In the pure $\phi^4$ theory, the static kink solution has a second (shape) mode which is responsible for oscillations with $\omega_1=\sqrt 3$. This mode exists in the impurity model, as well. As the depth of the potential changes, so does the frequency of the mode and for the special value of $\alpha=-\sqrt{2}$ (or $\lambda=1$)  this mode disappears at the mass threshold. Interestingly enough, for this value of $\alpha$ the spectral structure is {\it identical} to the spectral structure of the kink in the sine-Gordon (sG) model, but because of the mass threshold differences ($m_{\phi^4}=2$ and $m_{sG}=1$) the mode corresponding to the translational mode of the sG kink ($\omega=0$) in our model oscillates with the frequency of the vibrational mode of the $\phi^4$ kink $\omega=\sqrt{3}$. We remind that in this limit the kink becomes a BPS solution and the model is fully BPS. 

 %%%%%%%%%%%%%%%%%%%%%%%%%%%%%%%%%%%%%%%%%
\section{Scattering processes} 
%%%%%%%%%%%%%%%%%%%%%%%%%%%%%%%%%%%%%%%%%

Having identified the structure of static solutions (asymptotic states structure) as well as their spectral properties, we can investigate scattering processes. As the first case, we consider the scattering of an incoming antikink (which at $t\rightarrow -\infty$ is just the antikink solution of the pure $\phi^4$ model) with the topologically trivial lump $\Sigma_-$ located on the impurity. As we know, such a antikink-impurity bound state lives in the BPS sector.  

%%%%%%%%%%%%%%%%%%%%%%%
\subsection{Moduli space approximation}
%%%%%%%%%%%%%%%%%%%%%%%
We start with a theoretical description of the scattering of a slow-moving antikink on the impurity within the framework of the moduli space approximation, using the generalised translation symmetry of Section \ref{sect-3.2} as a collective coordinate. The existence of this symmetry implies that there exists a one-parameter family of antikink BPS solutions.
Let us denote this family as $\phi(x;a)$ where $a$ is a parameter which uniquely defines the BPS solution. In our further examples we will assume that $a$ is the position of the topological zero $x_0$, which is a proper parameter for $\alpha>-1/\sqrt{2}$ (see Section \ref{sect-3-1}). However, in principle, $a$ can be any other parameter which uniquely identifies a solution, like, for example, a field value at a certain point.
Let us add a perturbation $\psi$ to the static BPS solution $\phi(x;a)$ in such a way that 
\be \label{norm-psi}
\phi(x;\tilde a)=\phi(x;a)+(\tilde a-a)\psi(x)
\ee
 is still a solution to the BPS equation for a different parameter $\tilde a$.
Assuming that, for a small change of the parameter $\tilde a=a+da$, the perturbation function $\psi da$ is also small, we can linearize the  BPS equation and obtain an equation for $\psi$
\begin{equation}\label{transformationEquation}
 \psi_x = 2\phi\psi.
\end{equation} 
This linearised BPS equation implies that $\psi$ is a zero mode, i.e., it obeys the mode equation (\ref{linearized}) for $\omega =0$, as may be checked easily.
The above equation can be formally solved
\begin{equation}
 \psi=Ce^{2\int\phi\,dx}\,,
\end{equation} 
where $C$ is an arbitrary constant. If we use the definition (\ref{norm-psi}) for $\psi$, however, then
this constant has a definite value for each particular choice of the parameter $a$. For example, when $a$ is the position of a topological zero $x_0$, the transformation $x_0\to \tilde x_0=x_0+dx_0$ is the translation of the topological zero by a small value $dx_0$. Note that the field value does not change (the transformation changes zero to zero, $\phi(\tilde x_0 ;\tilde x_0) = \phi (x_0 ;x_0)=0$) so the total derivative vanishes
\begin{equation}
 d\phi(x_0;x_0)=\phi_x(x_0;x_0)dx+\phi_{x_0}(x_0;x_0)dx_0=0,
\end{equation} 
from which it follows that $\psi(x_0)=\phi_{x_0}(x_0;x_0)=-\phi_x(x_0;x_0)$ since $dx_0=dx$.
Because $\phi(x_0;x_0)=0$ from the definition of $x_0$, we get from the BPS equation 
\begin{equation}\label{transInitialCondition}
 C=\psi(x_0) = \phi_{x_0} (x_0;x_0)=-\phi_x (x_0;x_0) =1+\sqrt{2}\sigma(x_0)
\end{equation} and finally 
\begin{equation}
 \psi(x)=\left[1+\sqrt2\sigma(x_0)\right]\exp\left(2\int_{x_0}^x\phi(x')\,dx'\right)
\end{equation} 
For $\sigma\equiv0$, $\phi(x)=-\tanh(x-x_0)$, $\int\phi=-\log\cosh(x-x_0)$ and eventually $\psi=\cosh^{-2}(x-x_0)$, which is exactly the translational mode of the antikink.

\begin{figure}
\includegraphics[width=0.495\textwidth]{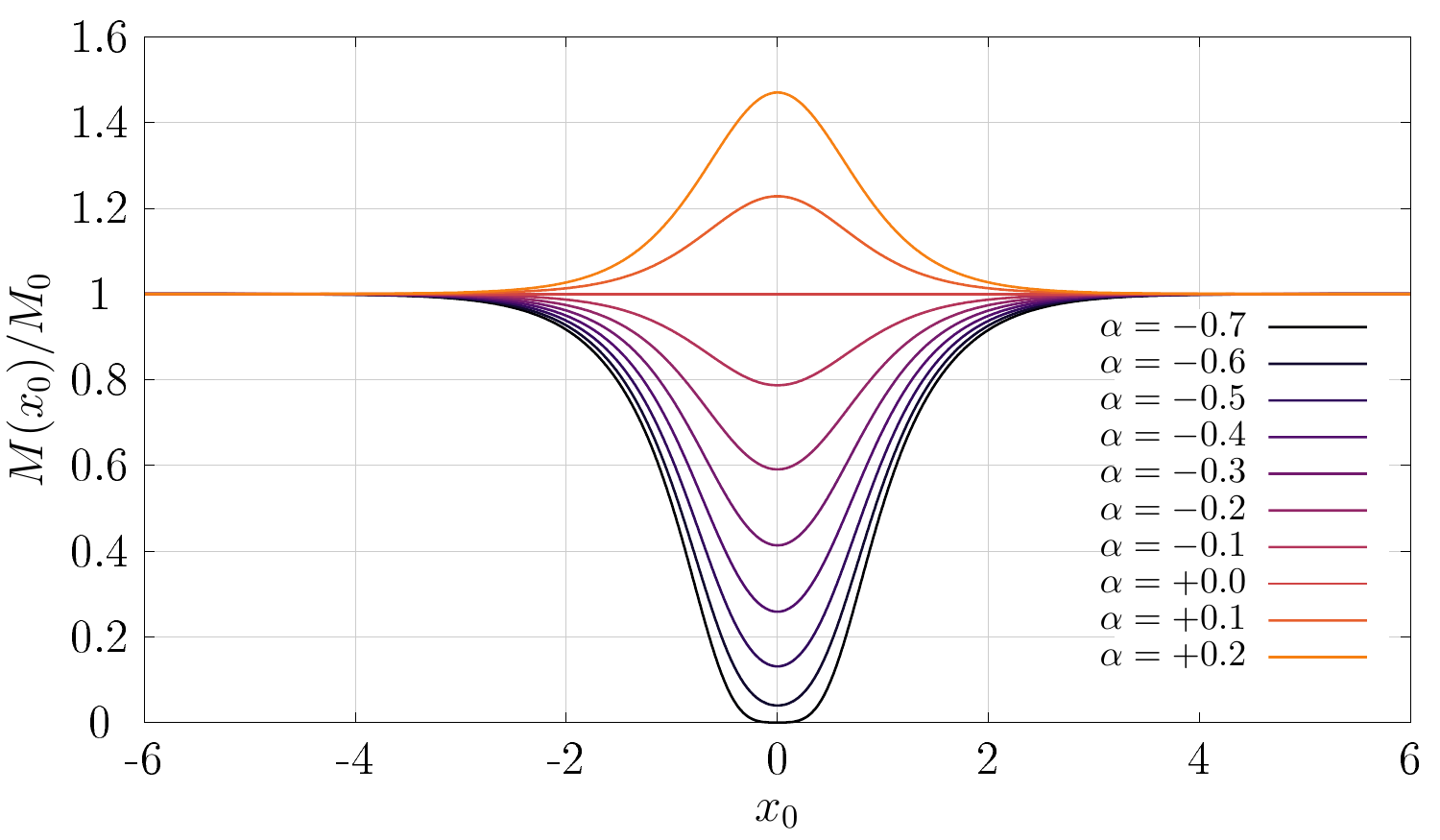}
\includegraphics[width=0.495\textwidth]{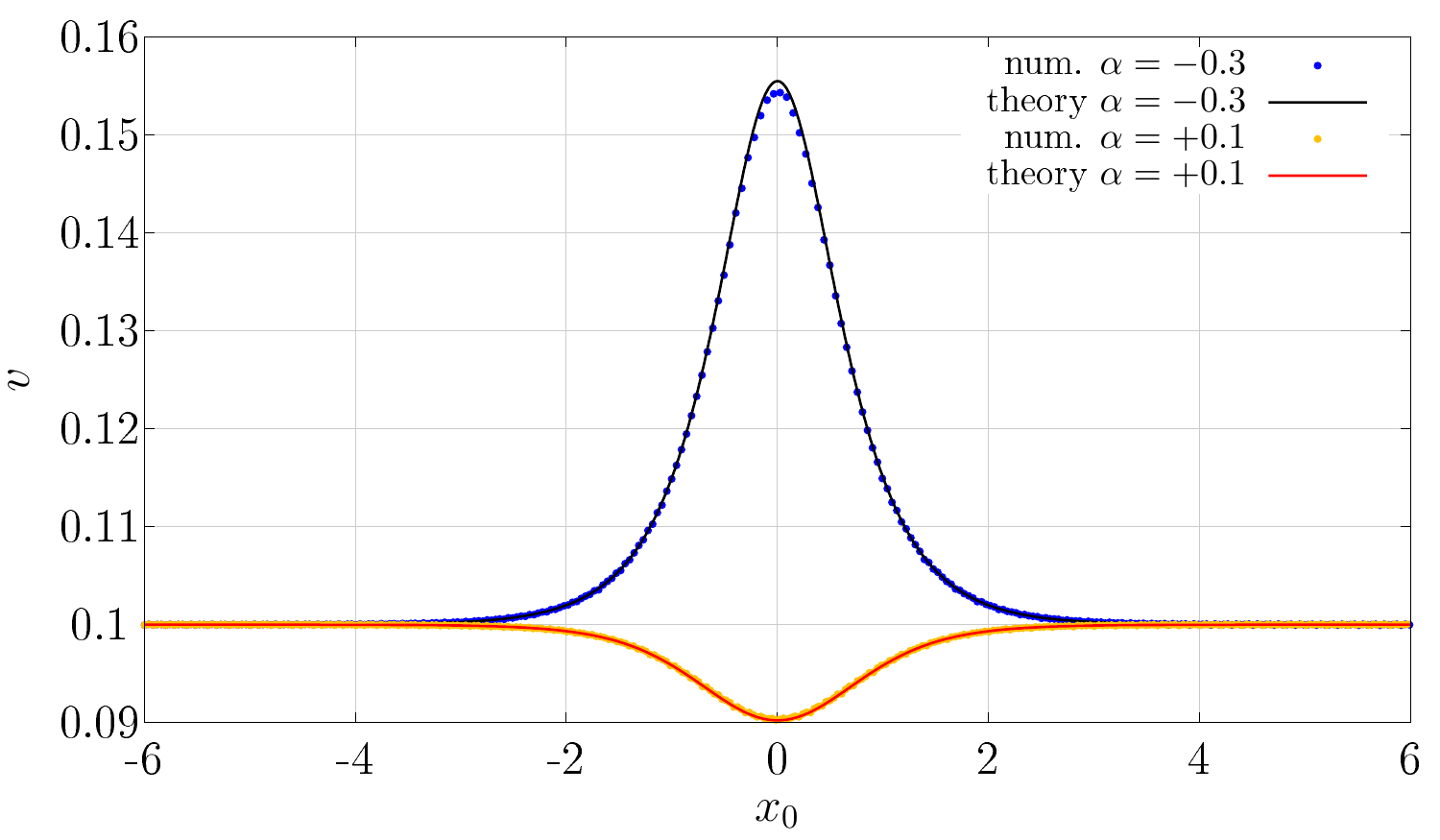}
\caption{Effective mass of the BPS solution (left) and velocity of the zero of the antikink, measured from the full PDE evolution (dots) compared with the velocity obtained from equation (\ref{modulivelocity}) (solid lines) for initial velocity $v=0.1$.}
\label{moduli}
\end{figure}

Next, we want to study the antikink-impurity scattering in the moduli space approximation, where the only modulus (collective coordinate) is given by the parameter $a$. That is to say, we limit ourselves to a slow evolution of the nearly BPS state,
\begin{equation}
 \phi\approx\phi_{\bar K}(x;a(t))
\end{equation} 
where $a(t)$ is a time-dependent parameter describing  the static BPS solution. Let us consider slow (adiabatic) evolution $0<|\dot a|\ll 1$. We assume that the change is slow enough so that  no Lorenz contraction nor deformation of the solution is visible. In other words, the configuration $\phi_{\bar K}(x,a(t))$ always maintains the shape of the corresponding static solution. The parameter $a$ can, for instance, be the position of the topological zero, which for $\alpha>-1/\sqrt2$ can be any real number uniquely defining the BPS solution.
The Lagrangian can then be written as 
\begin{equation}
 L= \int_{-\infty}^\infty\!dx\;\;\frac{1}{2} \phi_t^2 - \frac{1}{2} \phi_x^2 - U - 2\sigma \sqrt{U} - \sqrt{2} \sigma \phi_x - \sigma^2 
\end{equation} 
or, when the BPS property is used
\begin{equation}
 L=\int_{-\infty}^\infty\!dx\;\;\frac{1}{2} \phi_t^2=\frac{1}{2}M(a)\dot a^2\,,\qquad M(a)=\int_{-\infty}^\infty\psi^2\;dx\,.
\end{equation} 
If $a$ is indeed the position of topological zero, $M(a\to\pm\infty)=M_0=4/3$, which is the mass of the kink/antikink in the free $\phi^4$ model. In the above, $M(a)$ plays the role of an effective, kinetic  mass (or inertia) of the BPS solution. 
The equation of motion can be written as
\begin{equation}
 \ddot a M=-\frac{1}{2}\dot a^2M_a
\end{equation} 
which can be integrated to a first order equation
\begin{equation}\label{modulivelocity}
 \dot a = \dot a_0\sqrt{\frac{M_0}{M(a)}}\,,
\end{equation} 
where $\dot a_0$ is an arbitrary constant. For $a$ being the position of the antikink $x_0$, $\dot x_0$ is simply its initial and final velocity. 
Note that the velocity of the antikink after the collision will be the same as before the collision. 
For $\alpha>0$, the effective mass $M(x_0)>M_0$ (Figure \ref{moduli}) with a maximum at $x_0=0$ which means that the minimal velocity of the topological zero should be observed as it crosses  $x=0$.
On the other hand, for $-1/\sqrt{2}<\alpha<0$ the effective mass has a minimum at $x_0=0$ which reveals the maximal velocity of the topological zero at $x=0$. If we identify the position of the topological zero with the position of the antikink, this implies that the impurity repels the moving antikink for $\alpha >0$ and attracts it for $\alpha <0$.

In our numerical procedure of calculating $M(x_0)$ for fixed $x_0$, we have integrated twice (from  $x=x_0$ to $L=20$ and from $x=x_0$ to $-L$) the system of three ODEs: the BPS equation with an initial condition $\phi(x_0)=0$,  equation (\ref{transformationEquation}) with an initial condition (\ref{transInitialCondition}) and $\mu_x=\psi^2$, and calculated the difference $M(x_0)=\mu(L)-\mu(-L)$.

%%%%%%%%%%%%%%%%%%%%%%%%%%%%%%%%%%%%%%%%%
\subsection{Antikink-impurity scattering - numerical results}
%%%%%%%%%%%%%%%%%%%%%%%%%%%%%%%%%%%%%%%%%

\begin{figure}
\centering
\includegraphics[width=0.75\textwidth]{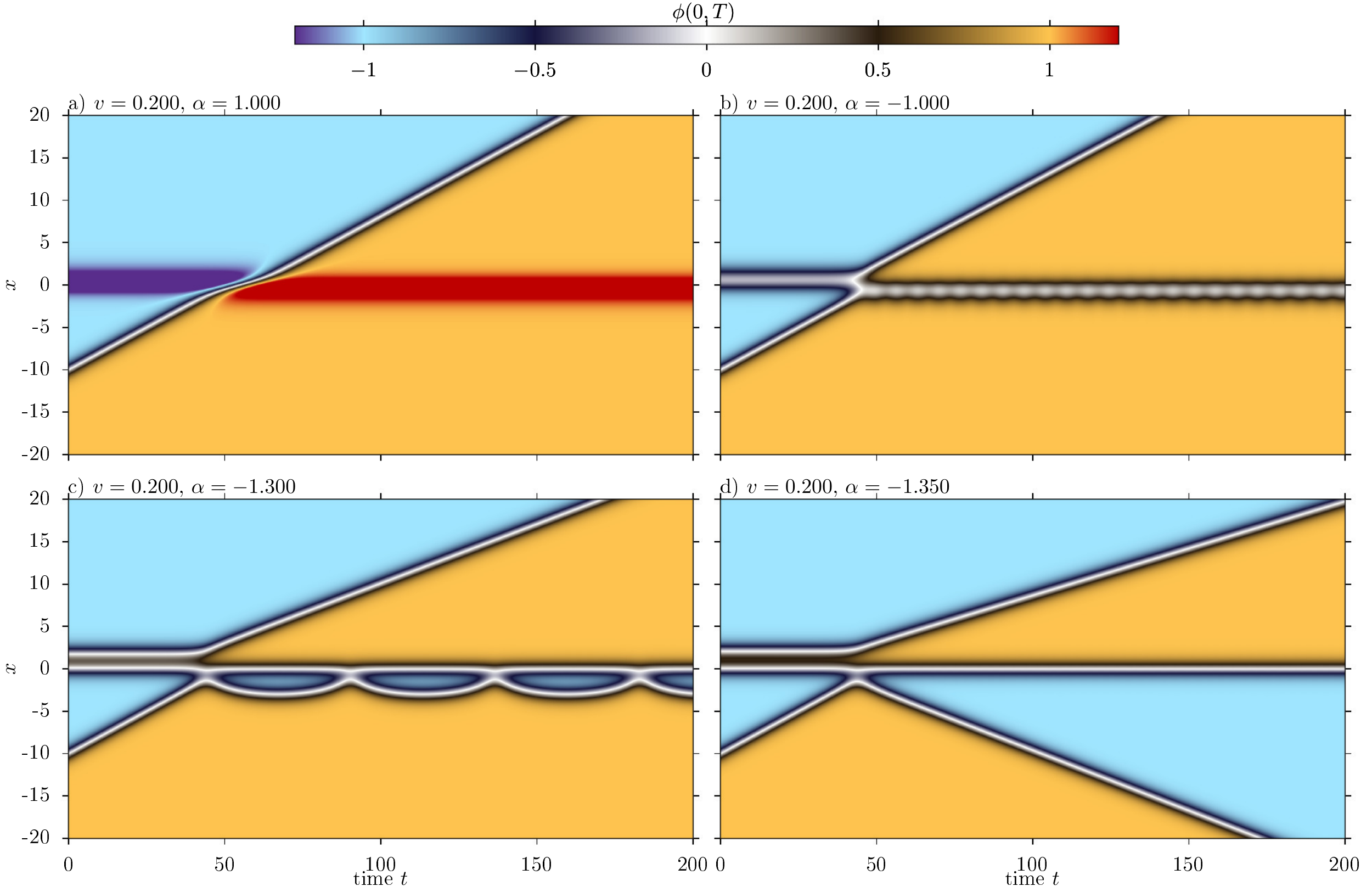}
\caption{Different scenarios of antikink-impurity ($\bar K+\Sigma_-$) collisions for $v=0.2$.}
\label{Collisions1}
\end{figure}
For various choices of the parameter $\alpha$ and the initial velocity of the antikink we have simulated the collision of the antikink and the impurity by numerically solving equation (\ref{EulerLagr}) with the following initial conditions:
\begin{equation}
 \phi(x,t\approx0) = 1-\tanh(\gamma(x-x_0+vt))+\phi_{\Sigma_-}(x)\,,\qquad\gamma=(1-v^2)^{-1/2}.
\end{equation} 
Note that due to the Lorenz contraction the antikink loses its BPS property, therefore we expect that the collisions will become less and less elastic as the initial velocity increases.

In Fig.  \ref{Collisions1} we present the four main qualitatively different types of scattering for not too high velocity. Here the initial velocity of the incoming antikink is $v=0.2$. 

For a positive $\alpha$, we found an almost elastic collision,  Fig.~\ref{Collisions1}a). The final state is formed by the antikink traveling to infinity and the lump $\Sigma_+$. The velocity of the outgoing free antikink is conserved, and the only effect of the scattering is the phase shift in the trajectory. As we know from the spectral analysis, there is no oscillating mode which could be excited during such a collision, which explains the observed elasticity.

For not too small negative $\alpha$, the outgoing state is delayed since the impurity can now host an oscillation, which easily may transfer some part of the kinetic energy of the incoming antikink. See Fig.~\ref{Collisions1}b) with $\alpha=-1$ where the oscillating frequency is $\omega\approx 0.9219$.  However, the velocity of the antikink is again almost unchanged as the amplitude of the oscillations is still quite small. 

For even smaller values of $\alpha$, the inner structure of the lump begins to play a crucial role in the scattering process, as the hidden kink-antikink pair is more and more visible. In Fig.~\ref{Collisions1}c) we see an example with $\alpha=-1.30$. The interaction still occurs via an excitation of the oscillating mode. However, now this mode has the previously found interpretation as a wobbling of the kink-antikink pair. The oscillating mode seems to be largely excited, but because the frequency is small $\omega=0.3103$, the energy stored in that mode is rather small, and the velocity of the outgoing antikink is only slightly reduced. Moreover, the collision looks more like the capture of the antikink by the impurity forming antikink-kink bound state and ejecting an antikink from the right hand side of the topologically trivial lump. 
In addition, the kink from the hidden $K\bar{K}$ pair seems to be relatively strongly confined by the impurity. The BPS antikink bounces around the almost motionless kink. This picture coincides with the previously announced mechanism of the formation of a wobbling kink-antikink pair. Here the oscillating antikink-kink pair is again trapped on the impurity due to the attractive force between the impurity and the constituent kink. Note also that the first oscillating mode corresponds with good accuracy to the frequency of the antikink oscillating around the kink. 

\begin{figure}
\centering
\includegraphics[height=6.0cm]{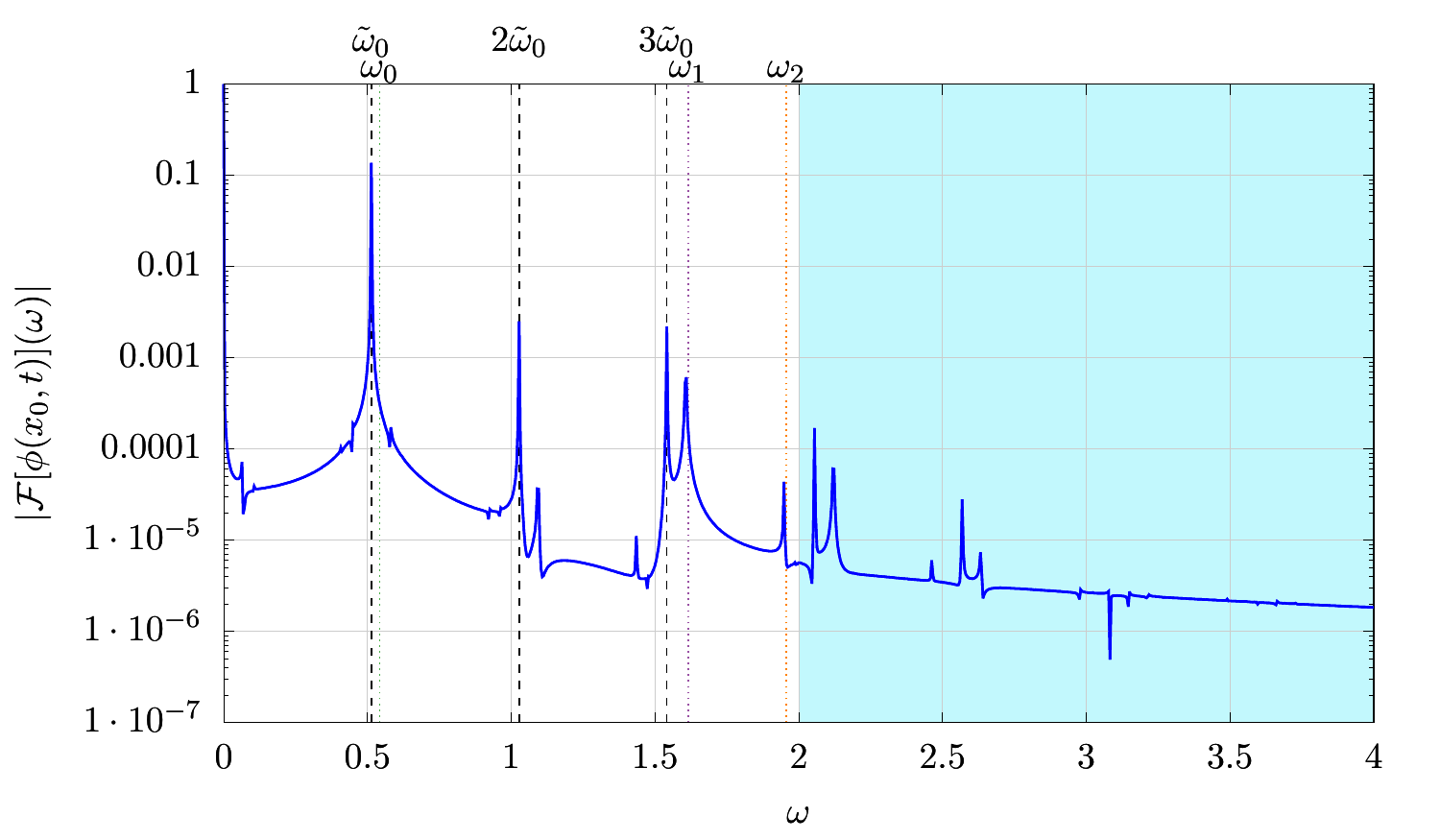}
\caption{Power spectrum for the exited lump (interpreted as an oscillating kink-antikink pair) for $\alpha=-1.2$. The shown values $\omega_0$, etc. correspond to the linear analysis and, therefore, do not exactly coincide with the peaks of the full (numerical) nonlinear power spectrum.}
\label{relax}
\end{figure}

It should be underlined that the wobbling kink-antikink pair forms a very stable, long living object. This can be explained by studying the power spectra for a long time evolution, see Fig. \ref{relax}. Here $\alpha=-1.2$ and the field is measured at $x=-2$ with $t\in [200,1600]$. The frequencies of the antikink oscillation (which is the frequency of the first mode) together with two higher harmonics as well as two higher modes lie below the mass threshold $\omega =2$. Hence, the first propagating harmonics is the fourth one, which amounts to a rather high stability of the oscillating pair \cite{Dorey:2015sha, Romanczukiewicz:2017gxb, Adam:2017czk}. In addition, we see that this pair can be fully understood as an excited lump state. 

Finally, for $\alpha < -1.31$  the BPS antikink is bounced back from the vacuum. The collision knocks out another antikink moving towards $+\infty$ leaving a bound kink  at $x=0$. This is plotted in Fig.~\ref{Collisions1}d) where $\alpha=-1.35$. This can also be explained easily. The separation energy of the kink and antikink in the oscillating lump tends to zero as the separation length between the solitons grows with decreasing $\alpha$. Hence, any surplus of energy is enough to release the antikink from the oscillating pair and send it back to minus infinity. 

To summarize, the antikink-lump scattering process can be very well understood taking into account the hidden kink-antikink structure of the topologically trivial lump, and with the help of the knowledge of the corresponding spectral structure (an equivalence between the wobbling of the  kink-antikink pair and the excited lump). The fact that the scattering looks to a large extent like an elastic process is probably related to the existence of the moduli space (zero mode) which, on the other hand, is a direct consequence of the BPS nature of the antikink-impurity solution. 
One may perhaps conclude that the antikink-impurity scattering occurs close to the BPS limit. 

\vspace*{0.2cm}

\begin{figure}
\centering
\includegraphics[height=9.0cm]{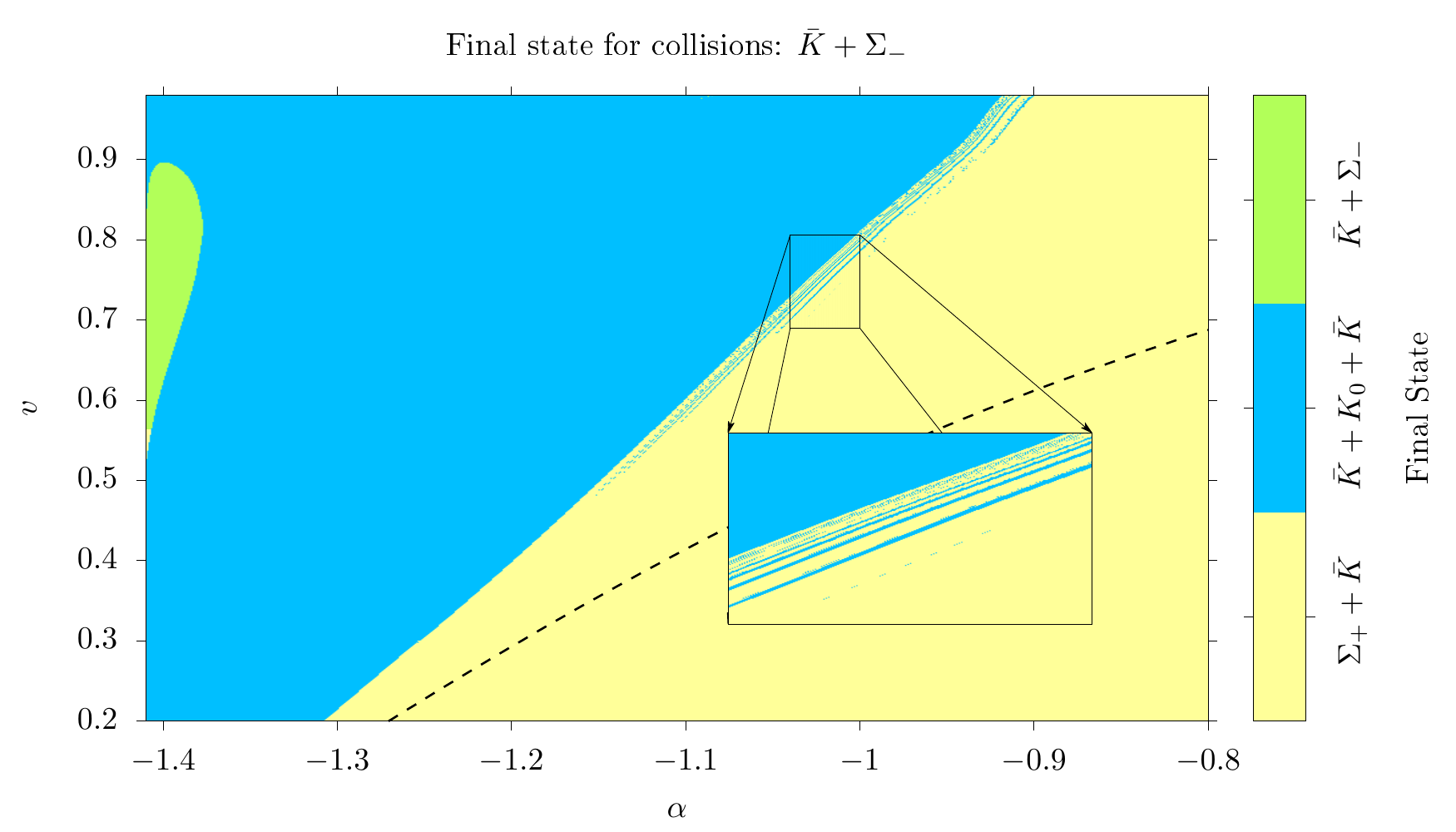}
\caption{Final states in the antikink-impurity collisions}
\label{K*Sigma-final}
\end{figure}

\begin{figure}
\centering
\includegraphics[width=0.75\textwidth]{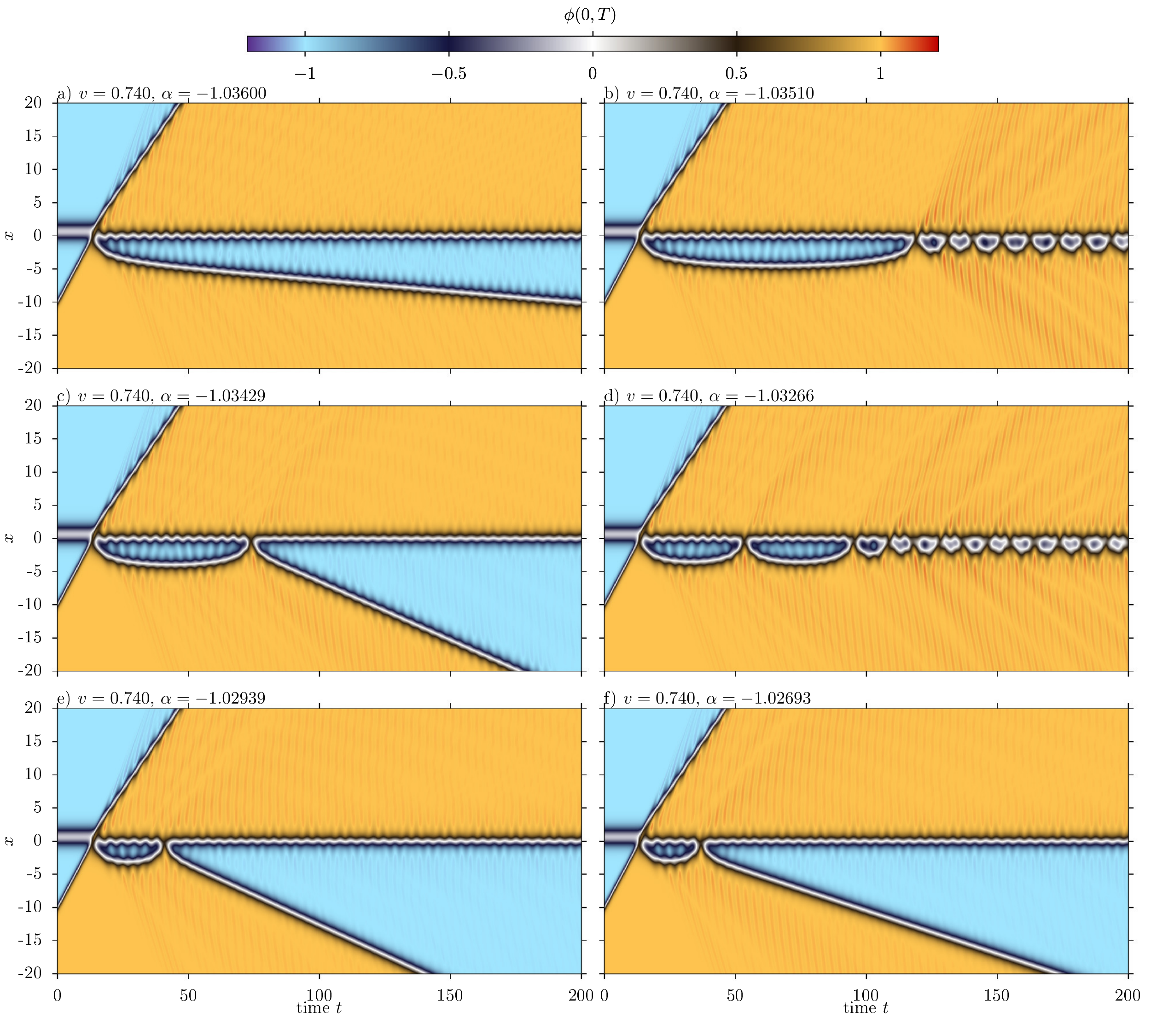}
\caption{Chaotic behaviour of antikink-impurity ($\bar K+\Sigma_-$) collisions with multibounce windows for $v=0.74$.}
\label{Collisions1B}
\end{figure}

This simple picture gets more complicated as the initial velocity grows and higher nonlinear effects begin to be more important - see Fig. \ref{K*Sigma-final}. Then, a fractal structure emerges on the border between the one-particle and the three-particle final states. This is a manifestation of the typical chaotic behaviour of the pure $\phi^4$ model. In Fig. \ref{Collisions1B} we demonstrate such a chaotic regime for the initial velocity of the incoming antikink $v=0.74$.

On the other hand, the qualitative description of the final state transition can be easily given. As $\alpha \rightarrow -\sqrt{2}$, the energy of the kink-antikink pair tends to 0 and therefore even a very slow initial antikink can create such a pair. Moreover, in this limit the pair is very well pronounced within the topologically trivial lump located on the impurity. All this leads to a three particle final state with an almost unchanged velocity of the incoming antikink. When $\alpha$ grows, it costs more energy to create the pair. Hence, the incoming particle must have higher kinetic energy or equivalently higher velocity. Furthermore, the hidden pair is less visible in the lump. As a consequence, the incoming antikink may simply go through the impurity without the creation of a kink-antikink pair. 

The transition line,  which separates the three body from the one body final state, can be found by assuming that the full kinetic energy of the incoming antikink will be used for the production of the (static) antikink and kink-on-impurity final state
\be
\frac{4}{3} \left( \frac{1}{\sqrt{1-v^2}}-1\right) = \frac{4}{3} \left( 2+2\sqrt{2}\alpha+\alpha^2\right)
\ee
We plot this line in Fig. \ref{K*Sigma-final}. The apparent discrepancy might find an explanation by the formation of some very long-living oscillating kink-antikink pairs (excited lumps), which is in fact relevant for the yellow regime above the border line.  

Observe that the in principle possible (strongly inelastic) case, where the antikink is captured by the impurity, never occurs. This is due to the BPS nature of the antikink-impurity state with the moduli space described by a zero mode, being just the distance between the antikink and the lump located on the impurity.  Instead, only for a very narrow strip near $\alpha\to-\sqrt2$ and high velocities, the antikink can bounce back from the impurity.

In fact, the antikink-impurity process, at least for small velocities, is very close to the static BPS regime and, as we have shown in the previous subsection, it gives a quite precise description within the collective coordinates approximation.
A similar description is often used in the case of vortices in the Abelian Higgs model in (2+1) dimensions at critical coupling. 

Last but not least, we observe very little radiation in the scattering process at low and medium velocities. This, together with the appearance of a breather, is typically understood as a manifestation of integrability in the system. Here the role of the breather may be played by the excited lump, which is just a very stable wobbling kink-antikink pair. Therefore, we may say that the antikink-impurity scattering reveals some properties of an integrable sector of the full theory.

\begin{figure}
\centering
\includegraphics[width=0.75\textwidth]{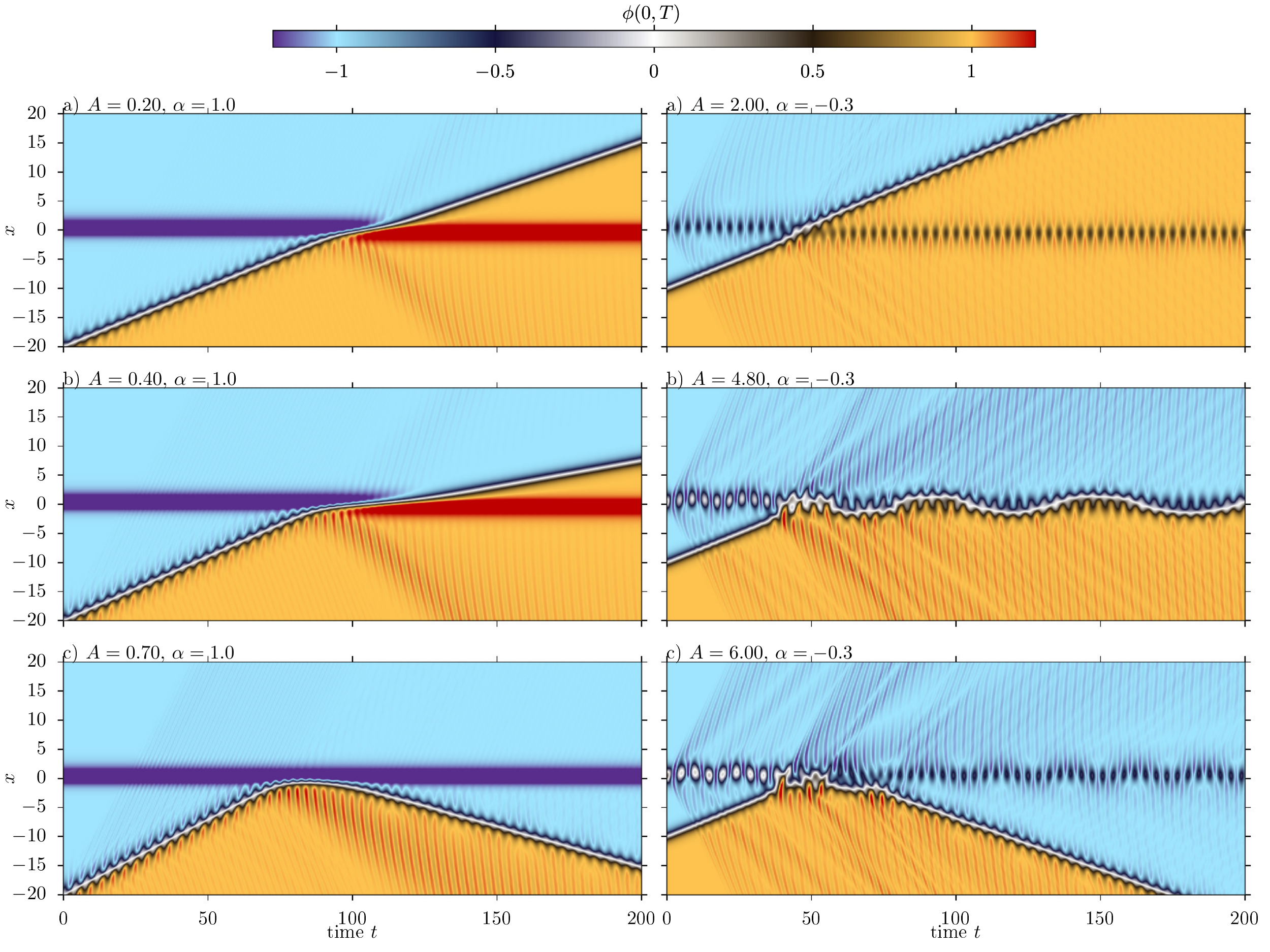}
\caption{Different scenarios of perturbed antikink-impurity ($\bar K^*+\Sigma$, left panel) and antikink-perturbed impurity ($\bar K+\Sigma^*$, right panel) collisions. }
\label{Collisions5}
\end{figure}

\vspace*{0.2cm}

It might be surprising that the space of possible outcomes of the collisions is rather poor in the scattering of the antikink with the lump. Having in mind the richness of phenomena in the pure $\phi^4$ model \cite{Sugiyama:1979mi, Campbell:1983xu, Anninos:1991un} and the even more complex spectral structure of antikinks and lumps, one could have anticipated a repetition of the fractal structure of multibounce collisions also in this model. In fact, we have seen some remnants of such a structure, but for a very narrow range of parameters at high velocity and near the critical value $\alpha=-\sqrt 2$. 
On the other hand, this is the type of behaviour we expect for BPS solutions with the translational symmetry. Even though there is a rich spectral structure, the antikink passes through the impurity in a relatively smooth (adiabatic) manner. The excitation of the internal and radiation modes is almost negligible. The main mechanism standing behind the resonance structure is that the modes get excited during the collision and store enough energy to bind defects for some time. In the scattering of BPS antikinks, this is not the case. However, when the modes are excited prior to the collision, the picture changes. We have performed a set of simulations of the antikink with excited internal modes colliding with a lump, given by initial conditions
\begin{equation}
\phi(x,t) =1 - \tanh\,x'\,+A\frac{\tanh x'}{\cosh x'}\cos(\omega t') + \phi_{\Sigma_-}(x)
\end{equation} 
where 
\begin{equation}
 x'=\gamma(x-a-vt)\,,\qquad t'=\gamma(t-v(x-a))\,,\quad \gamma=\frac{1}{\sqrt{1-v^2}}
\end{equation} 
are Lorenz transformed coordinates of the moving antikink.
We have found that for $\alpha>0.4$ the defect can bounce from the impurity. 
As the antikink approaches the impurity, the frequency of the  excited internal mode increases. For $\alpha>0.38$ the mode enters the continuous part of the spectrum, becoming  a quasinormal mode. The impurity acts like a barrier and the only direction the mode can radiate is backwards. 
However, this radiation is bounded with the defect and tries to pull the defect back.  This scenario is shown in the left panel of Figure \ref{Collisions5}.

For an attractive impurity $\alpha<0$ we could not find a similar process. However, the lump can have its own modes excited. We excited the lump increasing its height by a factor $A$. In such a case, the anktikink can also bounce back Figure (\ref{Collisions5}f)), but can also be captured (\ref{Collisions5}e)). However, the excitation of the lump has to be very large, and for $A=5$ it is more appropriate to consider it as some oscillon-type state bound to the impurity. The capturing-bouncing  process seems to be also chaotic. 

Looking from a wider perspective, a plausible explanation of the richer structure of the antikink-lump scattering with one object being excited is that the initial state is not a solution of the BPS sector. 

\begin{figure}
\hspace*{-1cm}\includegraphics[width=0.53\textwidth]{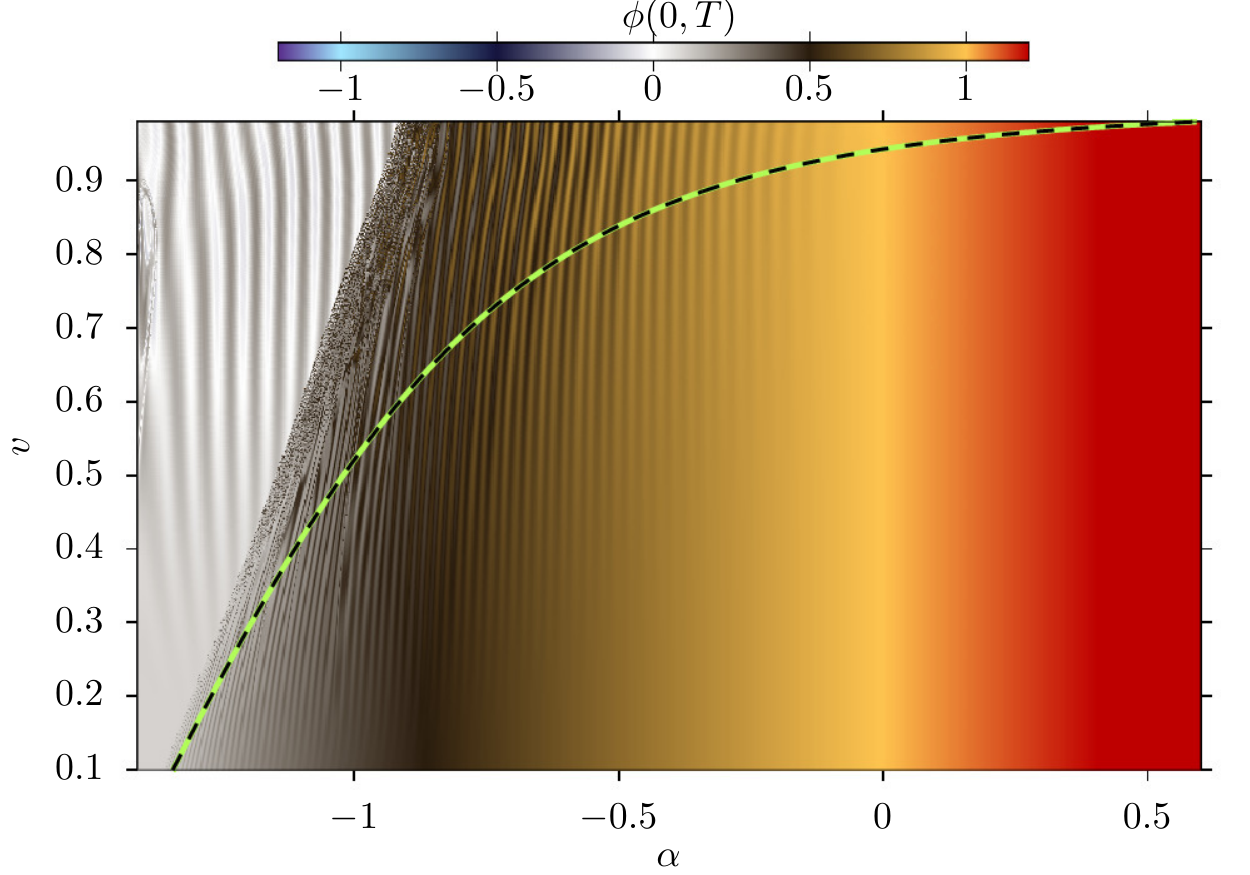}
\hspace*{-4mm}\includegraphics[width=0.53\textwidth]{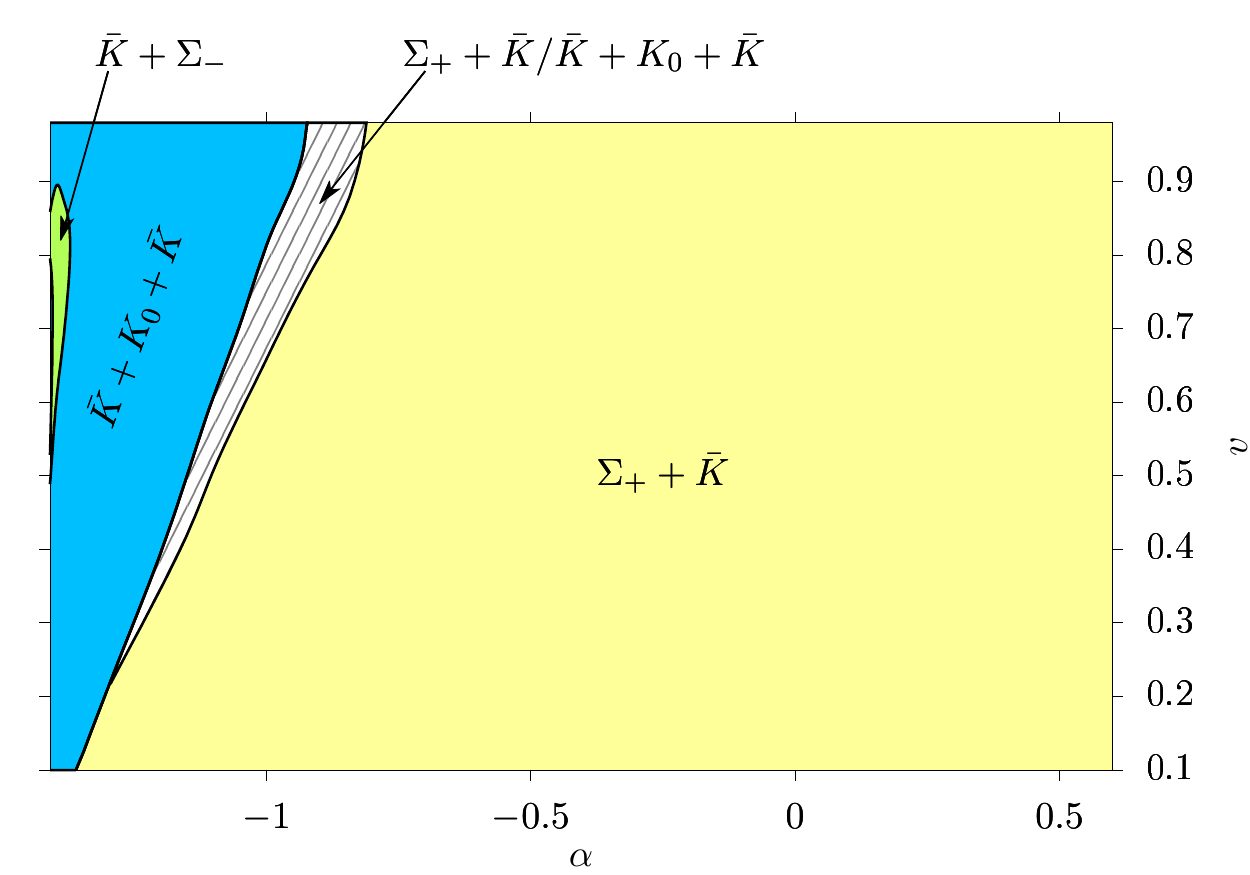}
\caption{Different scenarios of antikink-impurity ($\bar K+\Sigma_-$) collisions. The left plot shows the field value at center at time $T$ after the collision. The right plot shows the products of the collisions. Line filled regions correspond to chaotic behaviour. }
\label{Collisions0Diag}
\end{figure}

To summarize all the possible results of the collisions, it is helpful to study the final state at a specific point, say $x=0$ at certain time. We choose $T=100+v/|x_0|$. This gives some information about the final state of the whole system, Figure \ref{Collisions0Diag}. However, the information can also be misleading, especially near $\alpha\to 0$ when all final states have similar values 
$\phi_f\equiv\phi(0,T)\approx 0$ or when a few different results can have the same value at the center of the collision. One hint that can help in identifying those regions is looking at dislocations of constant phase lines. In many cases, the final state continuously depend on parameters until the structure of results suddenly changes. This can be seen for example in the Figure \ref{Collisions0Diag} near $\alpha\approx 1.4$ and $v\approx 0.8$.
Below we summarize the possible results of the antikink-impurity collisions. 
The last column is the color we use to mark  the final state in the phase diagram (for short we use R-red, O-orange, G-gray, B-blue, I-indigo):
\begin{equation}
 \bar K+\Sigma_-\longrightarrow
 \left\{\begin{array}{lcccc}
\Sigma_++\bar K&\quad \text{smooth passage}\quad&\phi_f\approx 1&\quad\text{RO}\quad\\
\bar K+K_0+\bar K&\text{lump decay}& \phi_f\approx 0&\text{G}\\
\bar K+\Sigma_-&\text{bounce}&\phi_f\approx 0&\text{G}&\;\;\alpha\to-\sqrt{2}\;\;\\
\end{array}
\right.
\end{equation} 
Note that the last possibility happens when $\alpha$ is very close to the critical value $-\sqrt 2$. For this value the field admits similar values $\phi(0,T)\approx 0$ for all possibilities. 
Only a more careful analysis of the whole scattering process gives the appropriate identification.
Those regions were marked on the right panel of Figure \ref{Collisions0Diag} with different colors. There is yet another region, between smooth passage and lump decay, where the system behaves chaotically (the boundary has some fractal properties). Such regions we mark with line patterns.

%%%%%%%%%%%%%%%%%%%%%%%%%%%%%%%%%%%%%%%%%
\subsection{Kink-impurity scattering}
\begin{figure}
\centering
\includegraphics[width=0.75\textwidth]{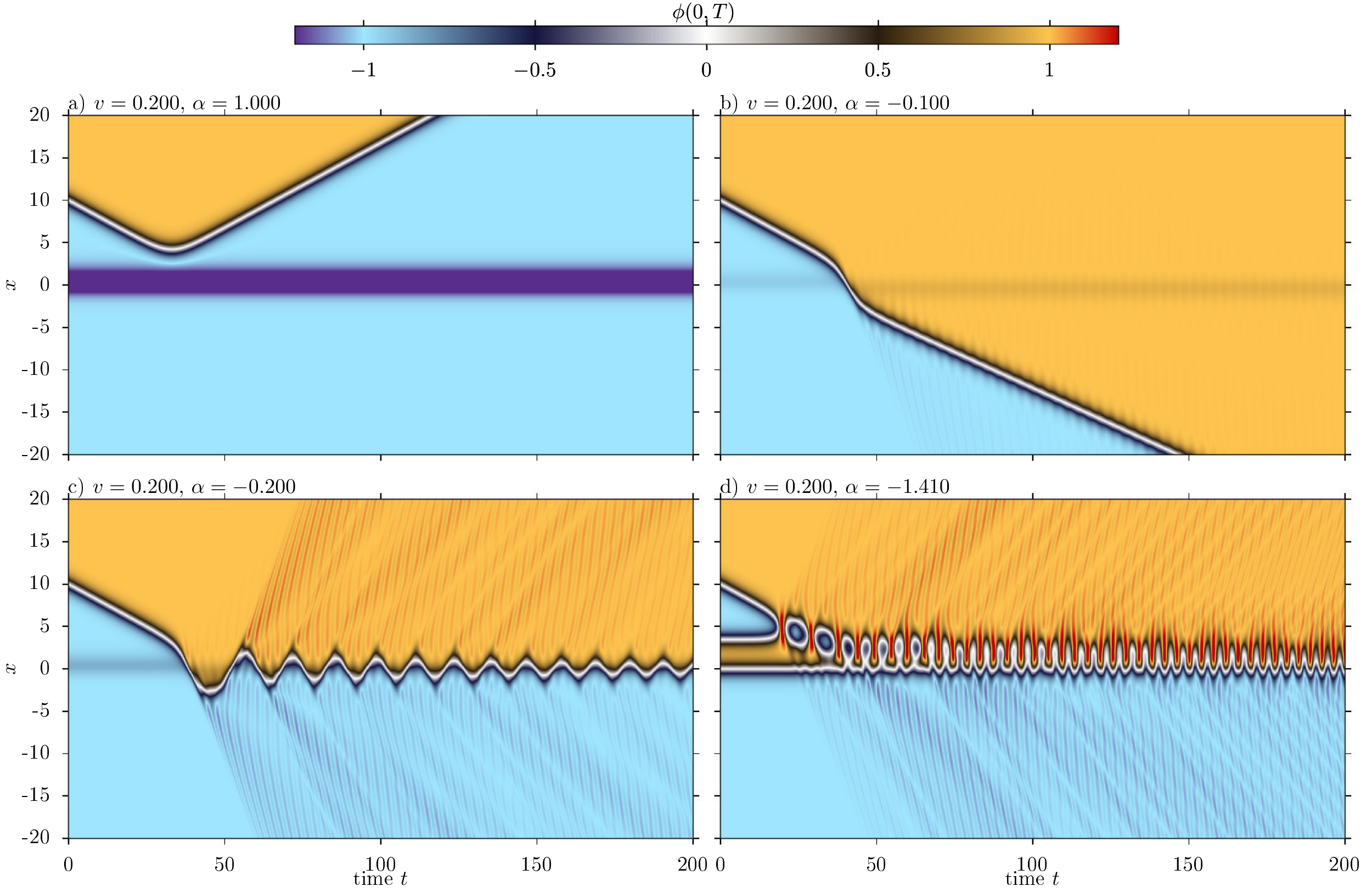}
\caption{Different scenarios of kink-impurity ($\Sigma_-+K$) collisions for $v=0.2$}
\label{Collisions2}
\end{figure}

%%%%%%%%%%%%%%%%%%%%%%%%%%%%%%%%%%%%%%%%%
As the next process, we consider the scattering of an incoming kink (from the right) on the topologically trivial lump $\Sigma_-$ located on the impurity. The main possibilities occurring at $v=0.2$ are presented in Fig. \ref{Collisions2}. 

For positive $\alpha$, the impurity always repels the kink as there is no kink-impurity bound state. Thus, the incoming kink is elastically reflected back - see Fig. \ref{Collisions2}a) with $\alpha=1$. 
 \begin{figure}
 \centering
%\hspace*{-1.0cm}
\includegraphics[width=0.75\textwidth]{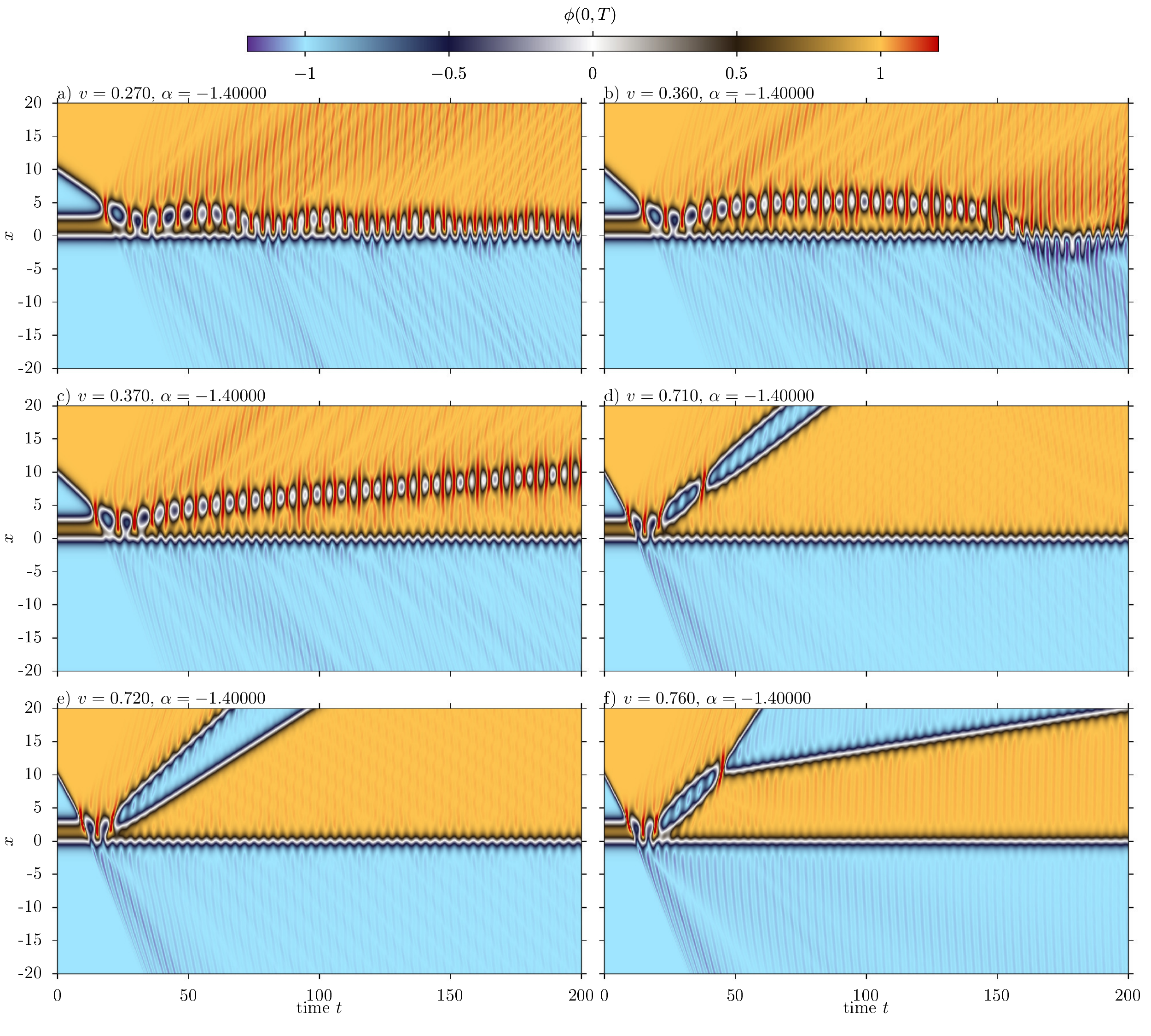}
\caption{Chaotic behaviour of the oscillon after the antikink-impurity ($\Sigma_-+K$) collisions for $\alpha=-1.4$.}
\label{Collisions2B}
\end{figure}

 Of course, for sufficiently large initial  velocities the kink should have enough energy to go through the impurity. In the limiting case, we expect that the field should go through a static configuration representing the kink at $x=0$. On the one hand, this is the lowest energy satisfying the condition that $\phi(0)=0$ in the $Q=1$ topological sector but, on the other hand, for positive values of $\alpha$, the lowest energetic state is when the kink is far away from the impurity. Therefore the static kink configuration is a saddle point. The kink with energy below the energy of the static unstable kink is unable to penetrate the potential barrier generated by the impurity. Therefore the limiting case gives the estimation for critical velocity when the kink can go through to the other side. Using the energetic argument we find that
\begin{equation}
 \frac{4}{3}\frac{1}{\sqrt{1-v_{cr}^2}}=\frac{4}{3}\left(1+2\sqrt2\alpha +\alpha^2\right)\,.
\end{equation} 

Once $\alpha$ becomes negative, the kink can go through it. For small negative $\alpha$, such a transition does not significantly change the velocity of the incoming kink as is visible in Fig. \ref{Collisions2}b) where $\alpha=-0.1$. If we make $\alpha$ smaller, then the attractive force of the impurity overcomes the kinetic energy of the kink, which is now trapped by the impurity forming an excited bound state, as presented in Fig. \ref{Collisions2}c) ($\alpha=-0.2$). It oscillates and, slowly radiating energy, decays into the static kink-impurity solution. When $\alpha \rightarrow -\sqrt{2}$, the inner structure of the lump starts to play a role. Then, the capture process is a bit more involved. The incoming kink annihilates with the antikink from the hidden kink-antikink pair of the lump. In the shown example they form an oscillon which goes towards the impurity. This subprocess is inherited from the pure $\phi^4$ model, as both solitons are quite far away from the impurity (remember it is exponentially localized). The remaining kink behaves as a spectator and is initially unaffected but later is perturbed (excitation of the oscillating mode) by the remains of the annihilation process. Note that in contrast to the previous case, we find a lot of radiation emitted during the kink-impurity scattering. This is, of course, related to the fact that the static kink-impurity solution is not a BPS configuration. 
\begin{figure}
\centering
\includegraphics[width=0.75\textwidth]{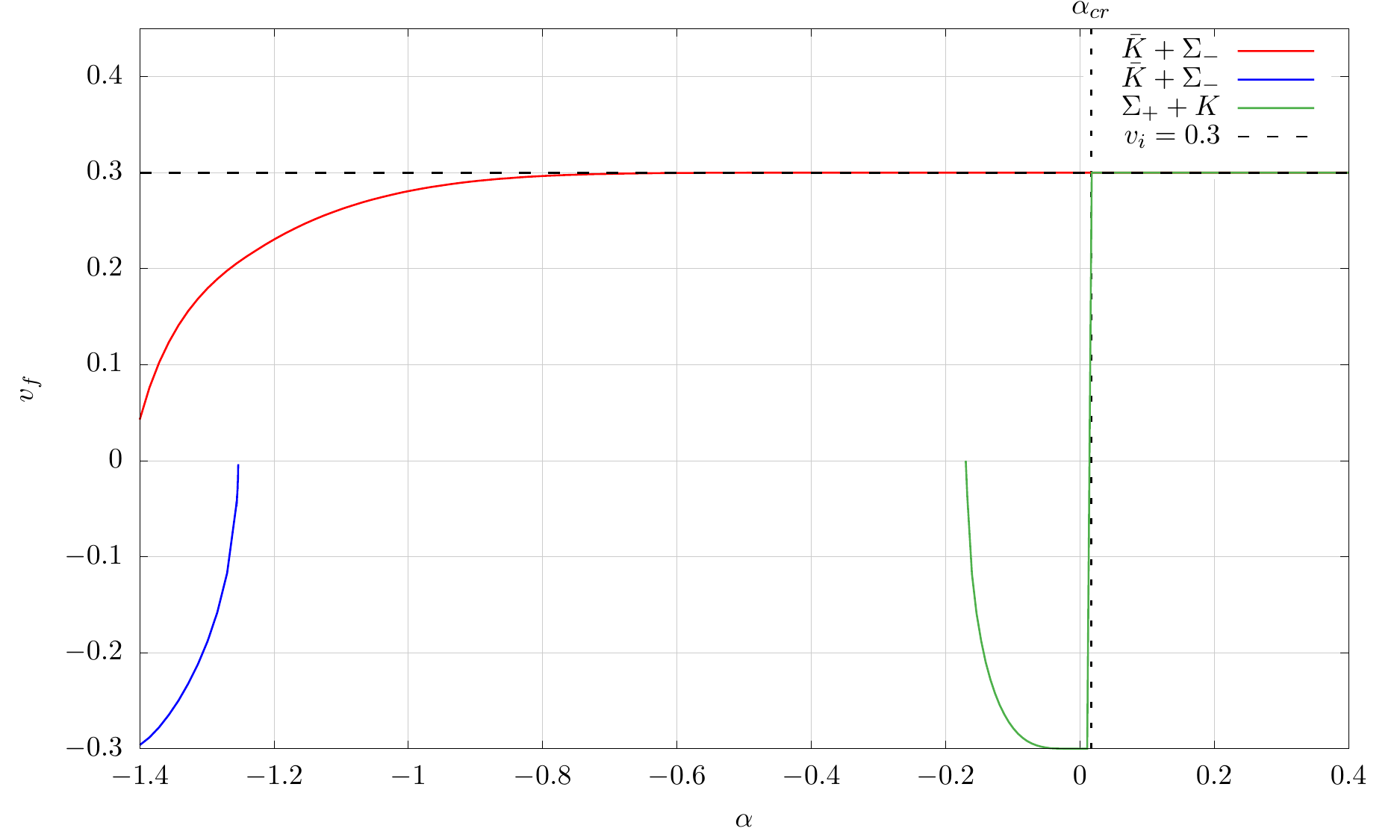}
\caption{Velocities of free defects propagating after impurity-defect type collisions.}\label{Velocities}
\end{figure}
\begin{figure}
\hspace*{-1cm}\includegraphics[width=0.53\textwidth]{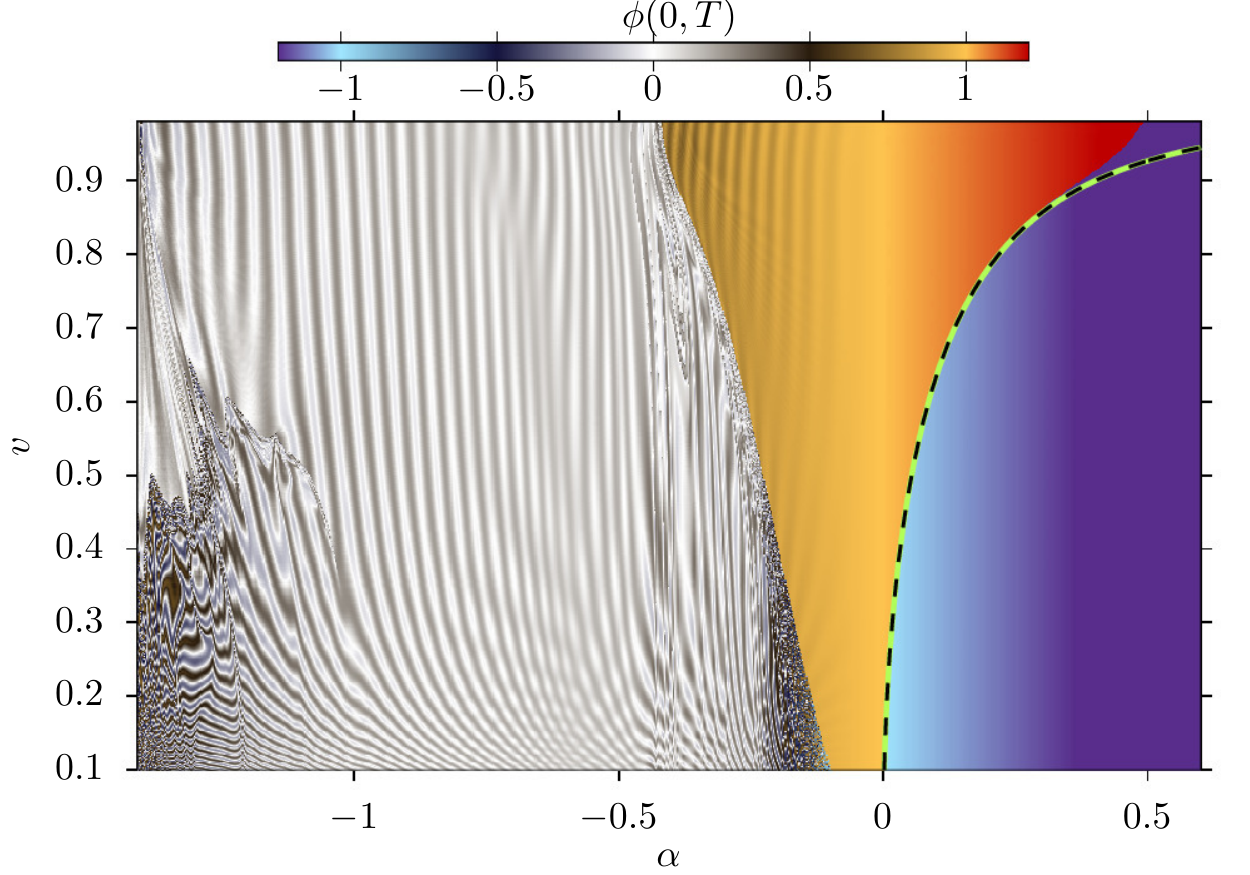}
\hspace*{-4mm}\includegraphics[width=0.53\textwidth]{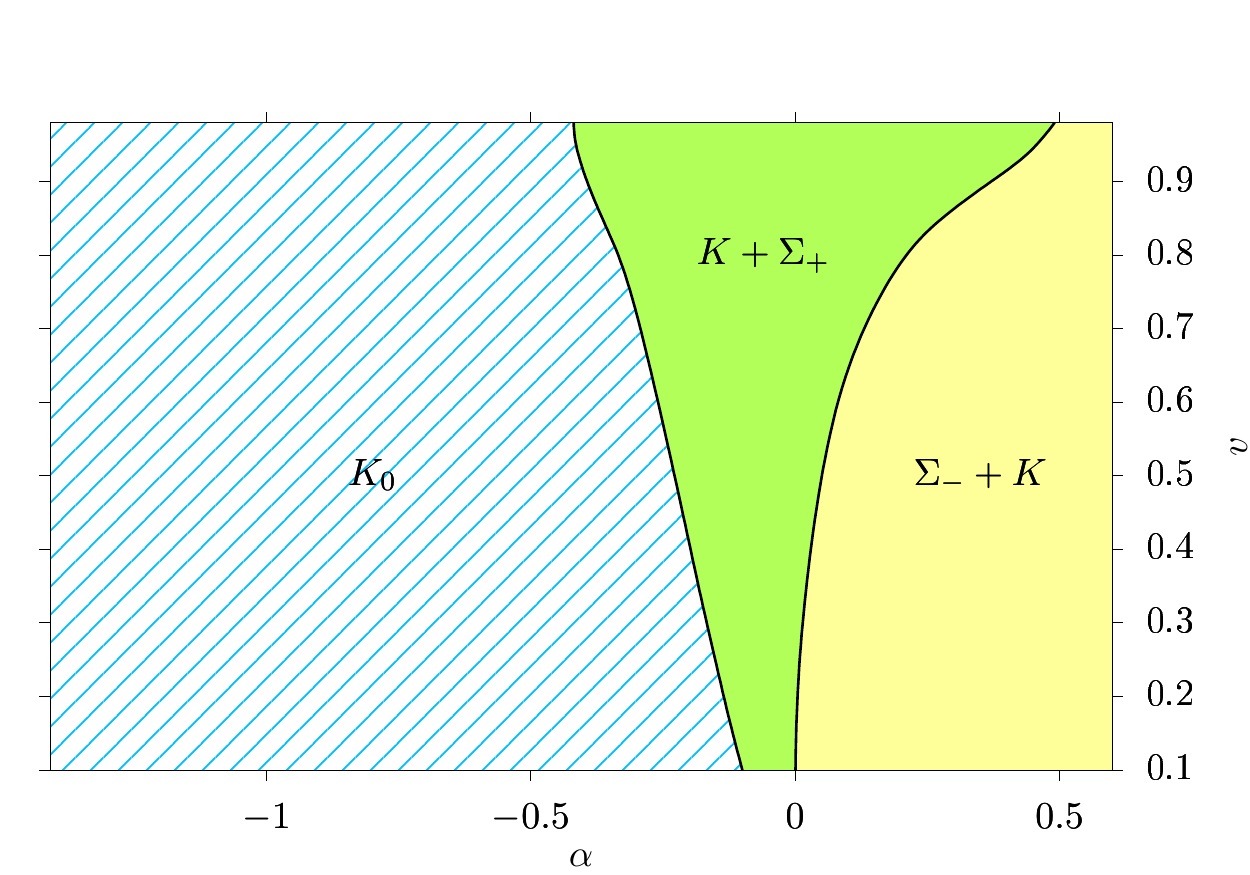}
\caption{Different scenarios of kink-impurity ($\Sigma_-+K$) collisions. The left plot shows the field value at center at time $T$ after the collision. The right plot shows the products of the collisions. Line filled regions correspond to chaotic behaviour. }
\label{Collisions1Diag}
\end{figure}

To summarize the defect-impurity collisions, we measured the final velocity of defects after the collision for initial velocity $v_i=0.3$ as a function of $\alpha$. The red curve in plot \ref{Velocities} shows the final velocity of the antikink in the antikink-lump collision. For a large range of values $\alpha>-0.7$, the final velocity is almost constant and almost equal to the initial velocity. This means that the energy exchanged between the defect and the impurity is negligible. This proves that the collisions of the BPS solutions with the generalised translational symmetry are almost elastic. In the range $-1.25<\alpha<-0.7$ the scattering is less  elastic and more  energy is transferred to the lump. Below $\alpha<-1.25$  the lump is destroyed in the collision and a second antikink is ejected from the impurity, leaving a kink tightly bound to the impurity. This also proves that in some sense, for values of $\alpha$ close to $-\sqrt 2$ the lump has a structure of a kink-antikink pair stabilized by the impurity.
The second observation is that in the case of kink-impurity scattering the kink which is reflected from the impurity also loses very little of its initial kinetic energy. However, just after the transition is possible (marked by dotted line and $\alpha_{cr}$ in the Figure \ref{Velocities}) the kink indeed can go through the impurity but loses more and more energy as the strength of the impurity increases. For $\alpha<-0.17$ the kink is captured by the impurity. The energy is then radiated out from the internal modes of the configuration described in the previous section.

In Figure \ref{Collisions1Diag} we have shown the final states of $\phi(0,T)$ (left) and the identification of the final state (right)
\begin{equation}
 \Sigma_-+K\longrightarrow
 \left\{\begin{array}{lcccc}
\Sigma_-+ K&\quad\text{bounce}\quad&\phi_f\approx -1&\text{I}\\
K+\Sigma_+&\text{passage}& \phi_f\approx 1&\quad\text{RO}\quad\\
K_0&\text{capture}&\phi_f\approx 0&\text{G}&\\
\end{array}
\right.
\end{equation} 
The last reaction, during which the kink in captured by the impurity, can produce a lot of radiation in order to get rid of the excess of energy. 
Sometimes this radiation takes the form of an oscillon. 
The oscillon can be recaptured by the impurity and adopted eventually as one of the internal modes of the kink, or it can pass through the impurity or can be torn apart into a kink-antikink pair. 
All of these processes lead to similar final values at the center of collisions, $\phi(0,T)\approx 0$. 
However, the excitation of the captured kink, or the kink with a recaptured oscillon, lead to different excitation amplitudes, which can be seen as discontinuities of the constant phase line in the lower left corner of the left plot in Figure \ref{Collisions1Diag} or near the boundary with the passage region.

%%%%%%%%%%%%%%%%%%%%%%%%%%%%%%%%%%%%%%%%%
\subsection{Antikink-(kink-on-impurity) scattering}
%%%%%%%%%%%%%%%%%%%%%%%%%%%%%%%%%%%%%%%%%
\begin{figure}
\hspace*{-1.0cm}
\includegraphics[height=12.0cm]{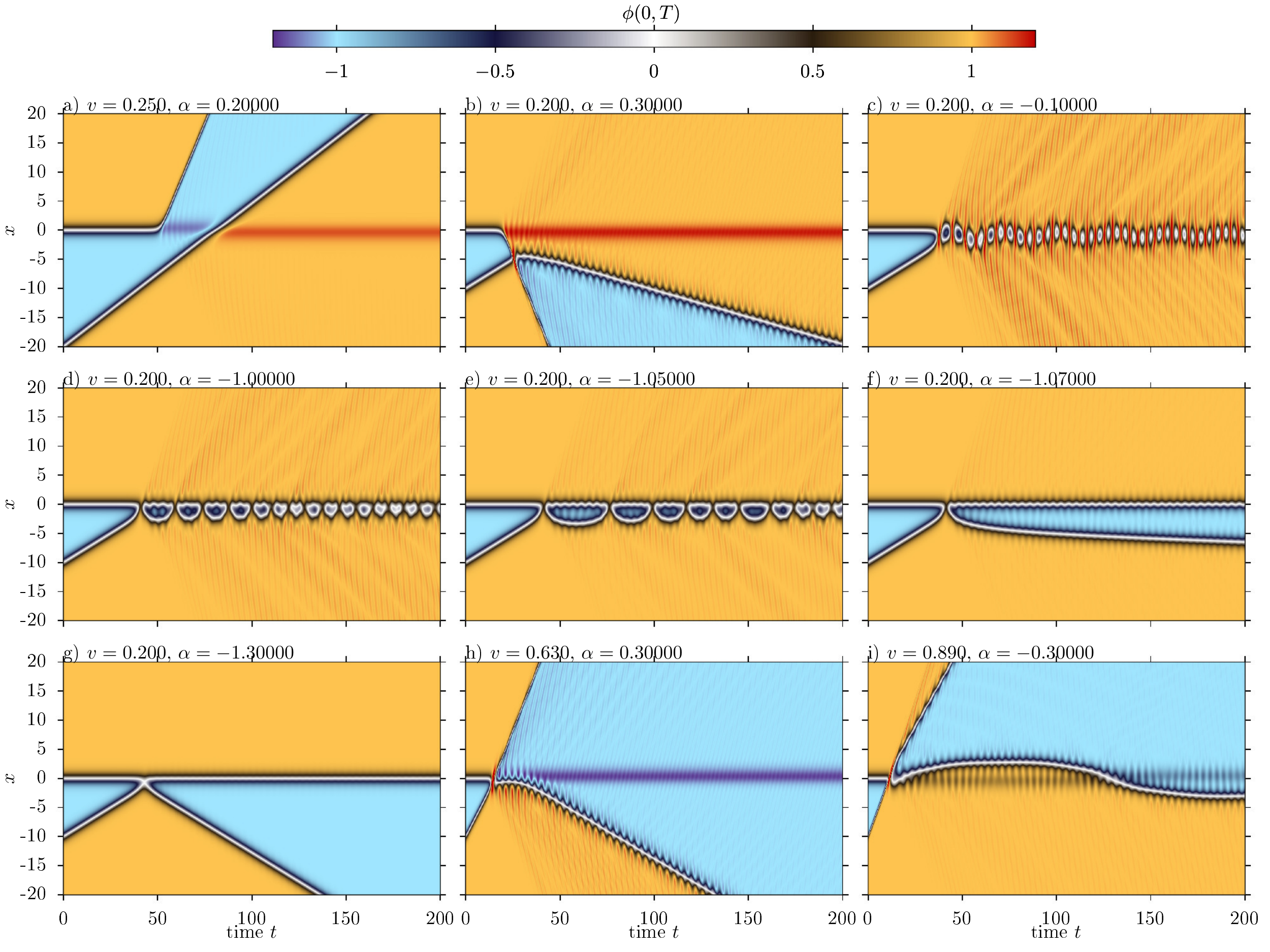}
 \caption{Different scenarios of antikink-(kink-on-impurity) ($\bar K+K_0$) collisions for $v=0.2$}
\label{Collisions3}
\end{figure}
\begin{figure}
%\hspace*{-1.5cm}
\includegraphics[width=\textwidth]{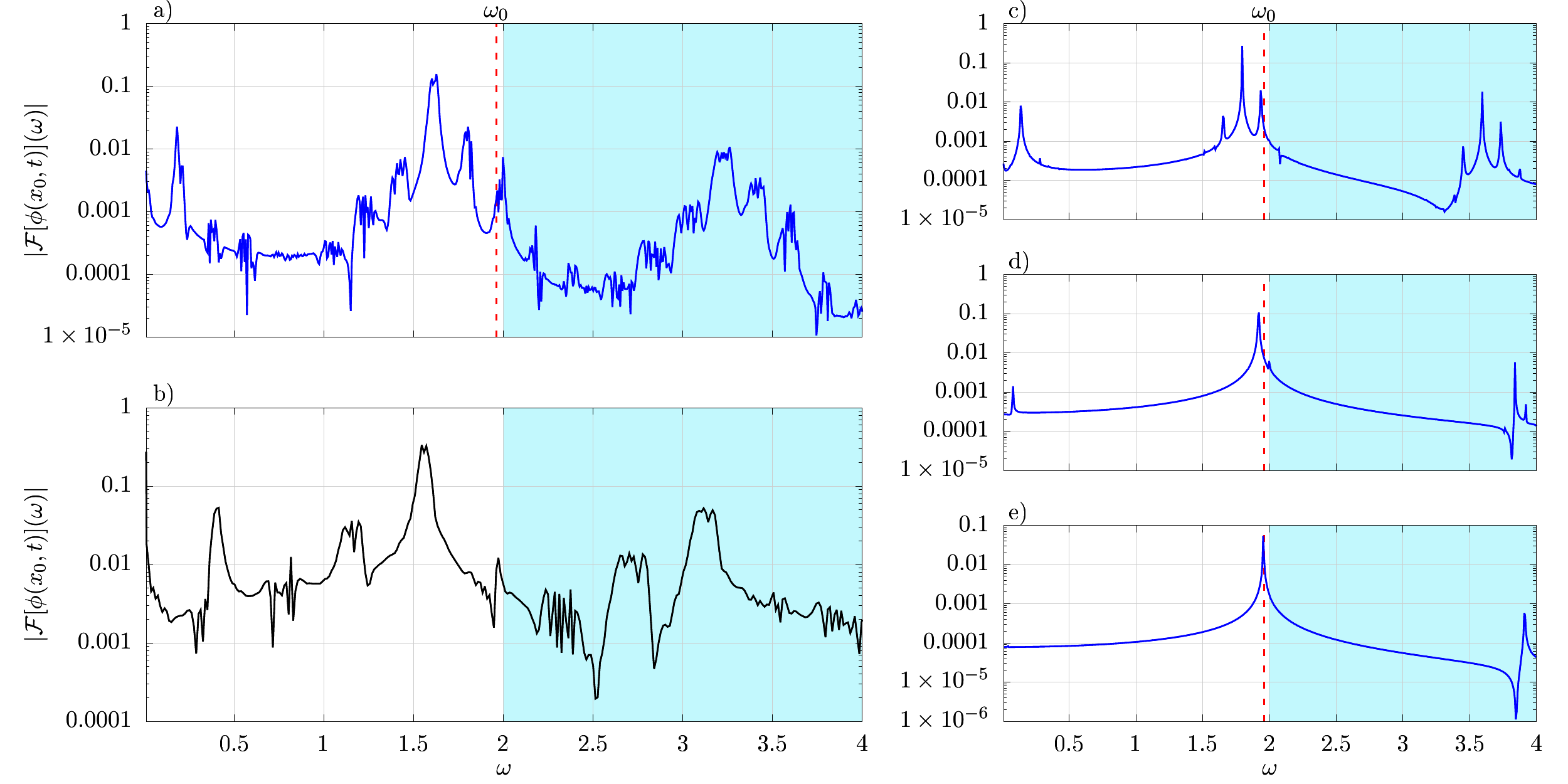}
\caption{(a) Power spectrum of the field after the collision $\bar K+K_0$ for $\alpha=-0.1$, $v=0.2$ measured at $x=x_0=-2$ compared with (b) the power spectrum of the oscillon $\alpha=0$ and (c-e) in the damped evolution with increasing damping term. }
\label{relax2b}
\end{figure}

Now we analyze the scattering of the incoming antikink with the kink-impurity bound state. We remind that such a bound state is stable only for negative values of $\alpha$. In Fig. \ref{Collisions3} we present the main qualitatively different scenarios of the scattering at the velocity of the incoming antikink $v=0.2$. 

For positive $\alpha$, the kink-impurity is an unstable configuration. The kink can be ejected both towards positive  \ref{Collisions3}a) and negative $x-$values \ref{Collisions3}b).  The incoming antikink can then either go through the impurity as in the previous case or can capture the kink from the impurity - see Fig. \ref{Collisions3}b) with $\alpha=0.3$. The mechanism is straightforward. The antikink attracts the kink which is simultaneously repelled by the impurity. However, the energy stored in the unstable state is high enough to avoid the annihilation. Both the incoming antikink as well as the released kink are scattered to minus infinity.

For negative $\alpha$, the incoming antikink is attracted by the kink and both get annihilated by forming an oscillon, as shown in Fig. \ref{Collisions3}c) with $\alpha=-0.1$. As the impurity is  weak, the behavior is similar to the pure $\phi^4$ theory. However, the oscillon is trapped by the impurity and performs a sort of coherent wobbling around the impurity. In Fig.  \ref{relax2b} we analyse this particular case in detail.
Fig. \ref{relax2b} a) shows the power spectrum of the field measured at $x=-2$ for the time range $t\in [1000,2000]$. 
There are certain peaks in the spectrum. 
The highest peak corresponding to the basic frequency $\omega\approx 1.6$ is very different from the frequency obtained in the linear theory $\omega_0=1.96$ corresponding to the only eigen-mode of the lump. 
The power spectrum resembles the typical power spectra of oscillons created via defect collisions in the pure $\phi^4$ model (Fig. \ref{relax2b} b). 
Therefore we can identify this object as a trapped oscillon. 
Oscillons are famous for their long and slow decay. 
To increase the rate of the decay we added an additional damping term $\gamma\phi_t$ to the equation for times in the range $200<t<1000$. 
After that time we calculated again the power spectrum. 
As the damping term increases, the spectrum becomes more and more similar to the spectrum predicted by the linear approximation. 
After a small damping one can see that a single frequency is even more dominant than in the undamped case. 
But the frequency is still much below the eigenfrequency due to the nonlinearities. As the solution is damped even more, the frequency becomes closer to the eigenfrequency of the lump. 
Therefore we can expect that the trapped oscillon would be absorbed by the lump and assimilated as the mode of the lump. A similar behaviour was analyzed in more detail in \cite{Romanczukiewicz:2017gxb}.
 Note that although the oscillon is a long living oscillating object, its decay is a more violent phenomenon than in the case of the wobbling kink-antikink pair in the antikink-lump scattering. For example, the amount of radiation emitted in such a process is bigger. In any case, the intermediate topologically trivial state observed here, i.e., the oscillon, behaves differently if compared to the excitation of the lump (which is just the wobbling kink-antkikink pair for $\alpha <0$). This may further support our interpretation of the antikink-impurity scattering as a quasi-integrable process. Here, as there is no static BPS antikink-kink-on-impurity solution, the observed scattering does not exhibit any integrable-like properties and is to a high extent governed by the properties of the scattering in the pure $\phi^4$ model. 
\begin{figure}
\centering
\includegraphics[width=0.75\textwidth]{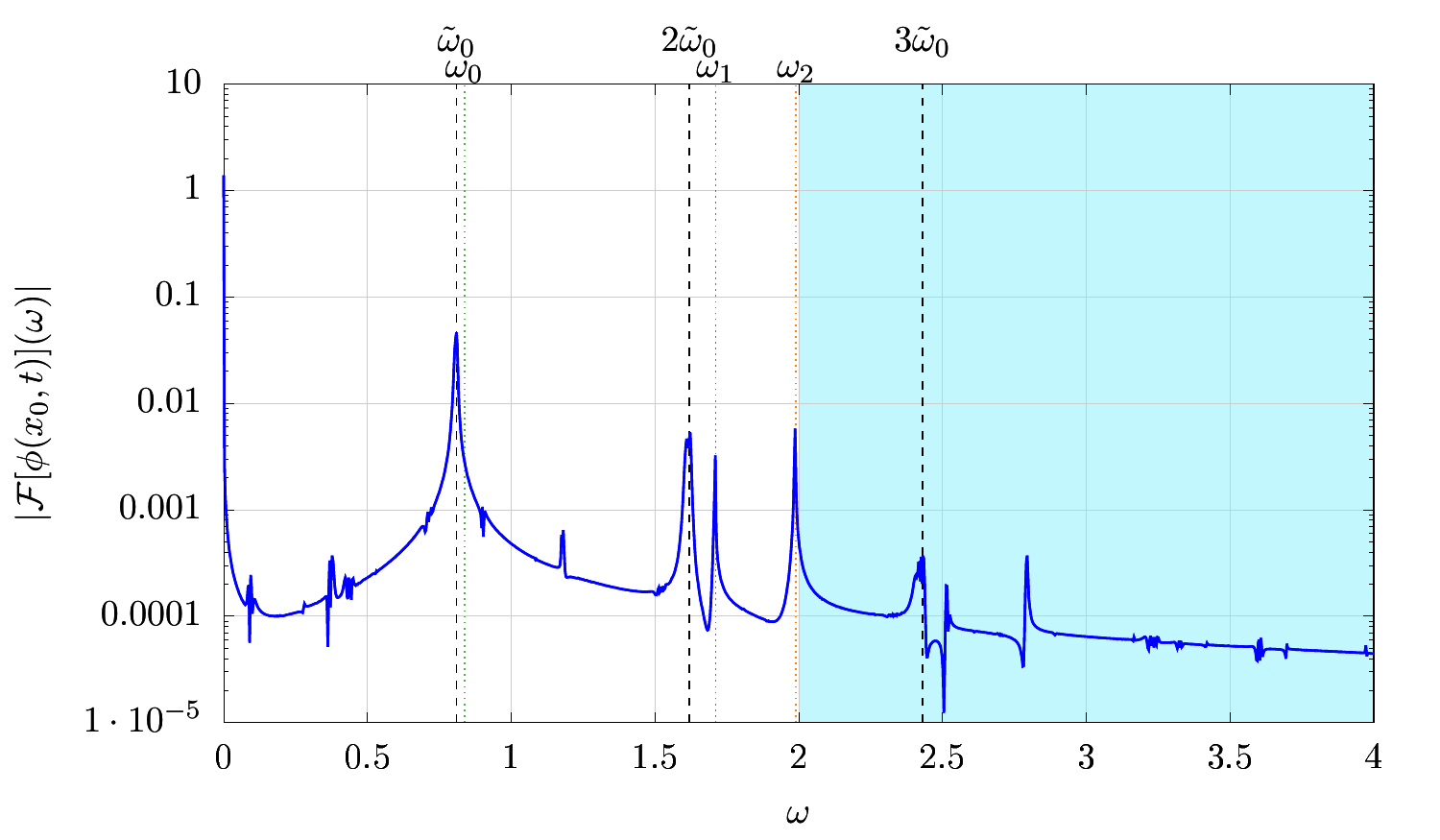}
\caption{Power spectrum of the field after the collision $\bar K+K_0$ for $\alpha=-1.05$, $v=0.2$ measured at $x=x_0=-2$. Eigenfrequencies of the kink $\omega_i$ and multiplicities of the basic frequency $n\tilde\omega_0$ are also marked.}
\label{relax2d}
\end{figure}

For smaller negative $\alpha$, we observe the appearance of the wobbling state, see Fig. \ref{Collisions3}d) and Fig. \ref{Collisions3}e) with  $\alpha=-1.0$ and $\alpha=-1.05$ respectively. However, it is much less stable than in the antikink-impurity collision. It radiates a significant amount of energy and decays quite quickly into the topologically trivial lump. 
Although the perturbation seems to be quite large, all the oscillations can be easily connected with the eigenfrequencies of the trapped kink and their multiplicities (see Fig. \ref{relax2d}).
%{\color{blue} Probably there is an additional mode excited on the constituents, especially on the kink, which allows for a fast emission of the radiation. One possibility is that the oscillon is captured by the kink - has to be checked.} 
Again, as $\alpha$ decreases, one of the constituents, that is, the kink, behaves as a spectator only, with almost no oscillations around its initial position. 

\begin{figure}
%\hspace*{-1.5cm}
\includegraphics[width=\textwidth]{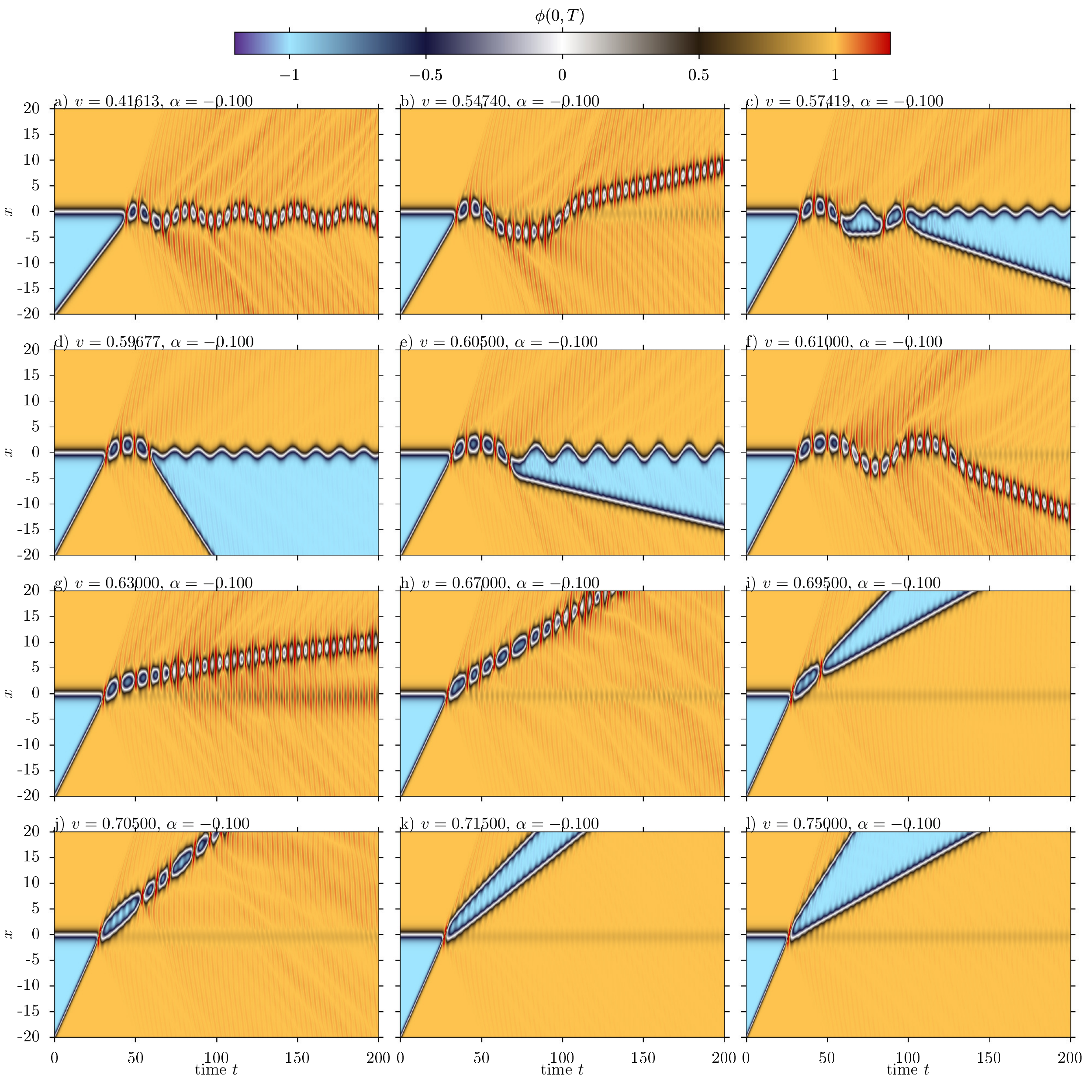}
\caption{Different scenarios of antikink-(kink-on-impurity) ($\bar K+K_0$) collisions with fixed impurity $\alpha=-0.1$ and different initial velocities of the incoming antikink.}
\label{Collisions3a}
\end{figure}

There is a critical $\alpha$ at which the scattering process reveals another final state. Namely, the initially created topologically trivial state breaks into a kink trapped to the impurity and a free antikink reflected to minus infinity  - see Fig. \ref{Collisions3}e). Here this happens at $\alpha=-1.07$. After that, the scattering tends to an elastic one once we approach $\alpha=-\sqrt{2}$. 

 A high velocity antikink can also push the kink from the impurity \ref{Collisions3}h) and \ref{Collisions3}i). The antikink can be ejected from the impurity leaving the lump at the center of the collision \ref{Collisions3}i) or replace the kink \ref{Collisions3}h) at the impurity. Note that we previously described similar processes for the excited antikink and lump collisions, which are shown in Figure \ref{Collisions5}.  

The picture gets much more complicated once we vary the initial velocity of the incoming antikink. Again, a chaotic structure emerges, which seems to be inherited from the pure $\phi^4$ theory. In Fig. \ref{Collisions3a} we plot some particular cases for a fixed impurity $\alpha=-0.1$ while we change the velocity of the incoming antikink. 
It is clearly seen that there are annihilation, bouncing and simple scattering windows both at smaller and higher velocities. However, the velocity (energy of the incoming antikink) determines whether the remainders of the scattering (as, for example, an oscillon) are bounded to the impurity, reflected or transmitted. 

It is also visible that the oscillon may be trapped by the impurity Fig. \ref{Collisions3a}a) or behave as a free object Fig. \ref{Collisions3a}g), h) and j). Furthermore, it can also be confined to the impurity for a short time, making several oscillation as a whole object around the impurity, and then escape to plus or minus infinity - see Fig. \ref{Collisions3a}b) and Fig. \ref{Collisions3a}f),  respectively. 

Because this process is quite complex, and in some cases can be regarded as a three-body collision, also the space of final states is quite complex.
\begin{equation}
 \bar K+K_0\longrightarrow
 \left\{\begin{array}{lcccc}
\Sigma_+&\quad\text{annihilation}\quad&\phi_f\approx 1&\text{RO}\\
 \bar K+K_0&\text{bounce}& \phi_f\approx 0&\quad\text{G}\quad\\
\bar K +\Sigma_- +K&\text{bounce/knock}&\phi_f\approx -1&\text{I}&\\
\Sigma_-+\bar K +K&\text{passage/knock}&\phi_f\approx 1&\text{RO}&\\
\{\bar K,\Sigma_\pm, K\}&\text{ejection}&\phi_f\approx 1&\text{RO}&\text{two stage}\\
\bar K_0+K&\text{BPS capture}& \phi_f\approx 0&\quad\text{G}\quad\\
\end{array}
\right.
\end{equation} 

In the right ($\alpha>0$) lower (small $v$) corner of the phase diagram for this process the dominant color is red, indicating that $\phi(T,0)\approx 1$. However, in this region the initial state, a kink on top of the impurity, is unstable. 
The incoming antikink provides a small disturbance resulting in the early ejection of the kink. 
The defects then collide outside the impurity, which is similar to the collisions known from the pure $\phi^4$ model. 
The products (defects or oscillons) can hit the impurity again after the first stage of collision. This two stage process is very chaotic. 

For high velocities and a weakly attractive impurity, the kink can be kicked from the impurity and be replaced by the antikink. 
Capturing of the antikink is an unusual situation and must be connected with the excitation of the internal mode of the antikink. Such excitations also lead to a large sensitivity to the change of parameters of collisions.

Another chaotic behaviour can be seen between annihiliation $\Sigma_+$ and bounce  processes ($\bar K+K_0$).

\begin{figure}
\hspace*{-0.5cm}\includegraphics[width=0.53\textwidth]{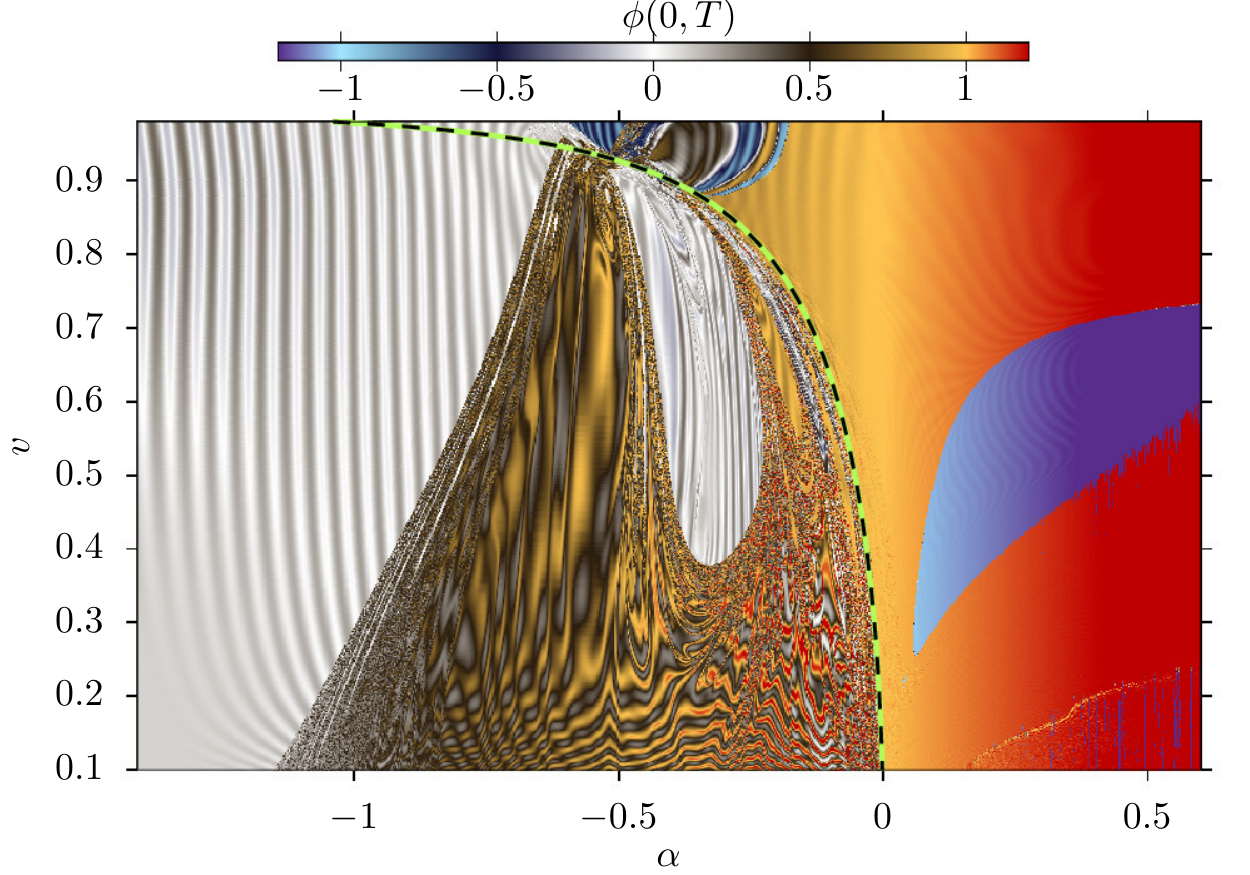}
\hspace*{-4mm}\includegraphics[width=0.53\textwidth]{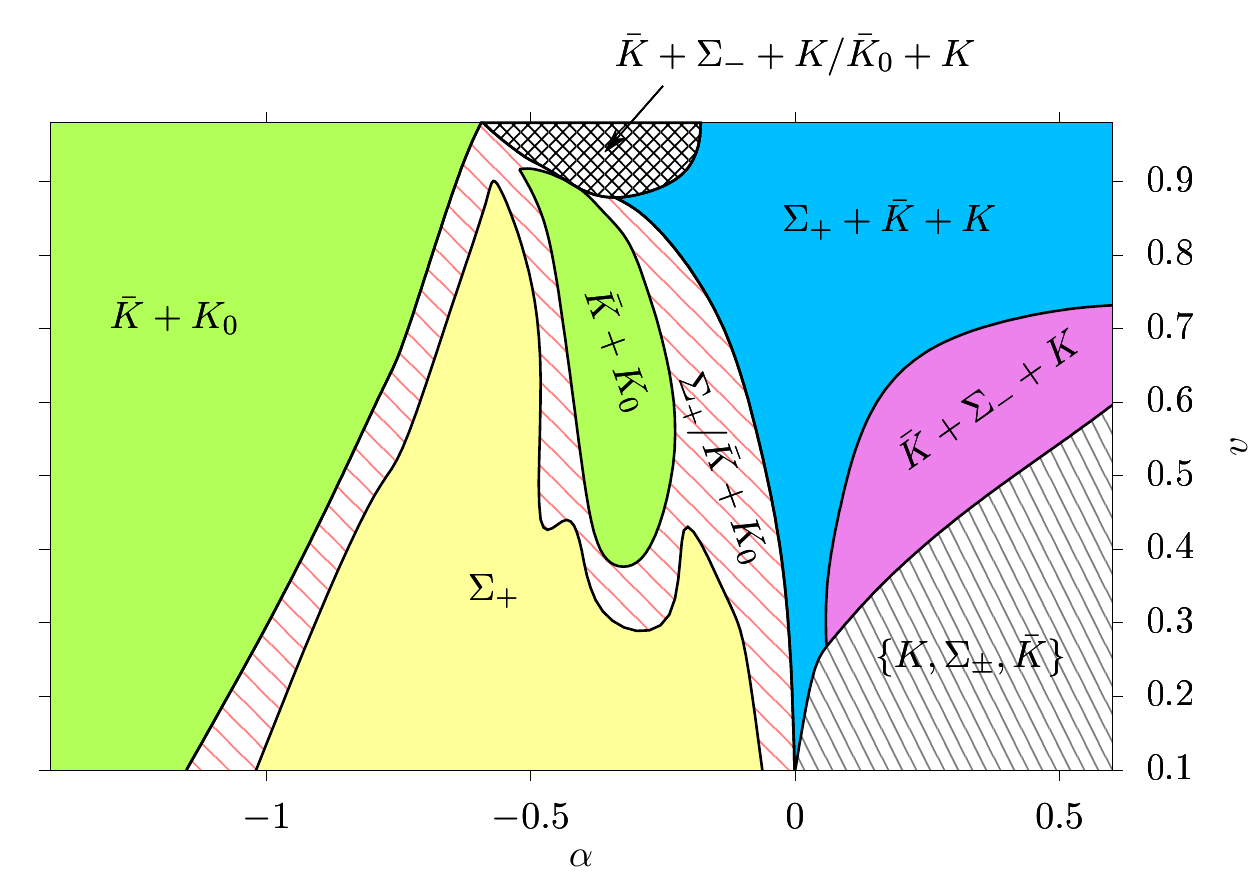}
\caption{Different scenarios of antikink-(kink-on-impurity) ($\bar K+K_0$) collisions. The left plot shows the field value at center at time $T$ after the collision. The right plot shows the products of the collisions. Line filled regions correspond to chaotic behaviour. }
\label{Collisions2Diag}
\end{figure}

The dashed green-black line on the left plot in the Figure \ref{Collisions2Diag} corresponds to the critical velocity below which the antikink has not enough kinetic energy to free the kink bound to the impurity. 
Below that velocity, the antikink can only bounce back or annihilate with the kink.

%%%%%%%%%%%%%%%%%%%%%%%%%%%%%%%%%%%%%%%%%
\subsection{Kink-(antikink-on-impurity) scattering}
\begin{figure}
\includegraphics[width=\textwidth]{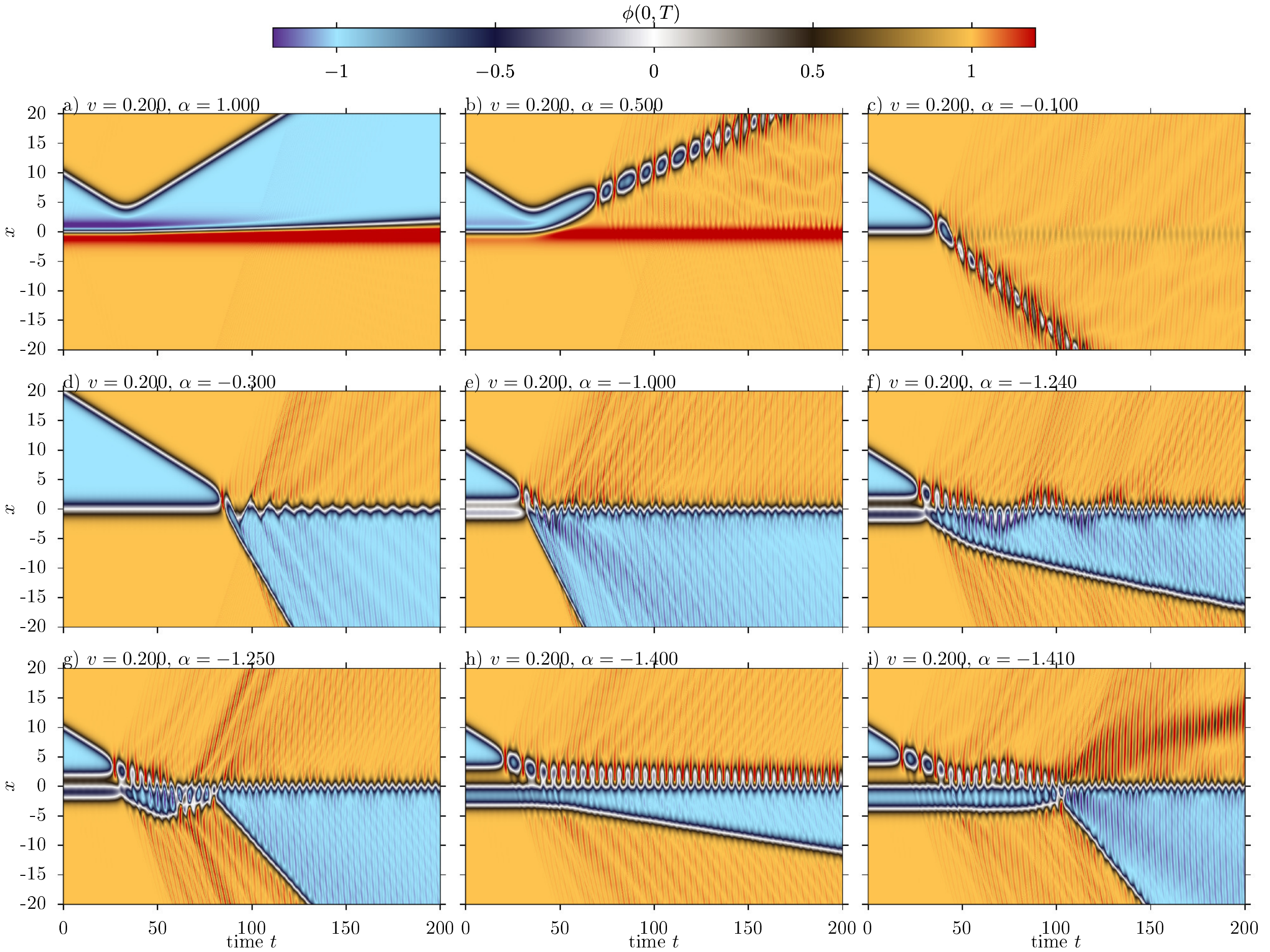}
\caption{Different scenarios of kink-(antikink-on-impurity) ($K+\bar K_0$) collisions.}
\label{Collisions4}
\end{figure}

%%%%%%%%%%%%%%%%%%%%%%%%%%%%%%%%%%%%%%%%%
Finally, we consider the incoming kink scattering on the antikink-impurity state. As there are infinitely many energetically equivalent antikink-impurity configurations, the main ingredients of the process are the kink-impurity interaction (repulsion for $\alpha>0$ and attraction for $\alpha<0$) as well as the kink-antikink attraction. In Fig. \ref{Collisions4} we plot several scenarios which can occur during this scattering. Here $v=0.2$. 

First of all, for positive $\alpha$ the incoming kink gets reflected by the impurity. However, it interacts with the antikink and attracts it. As a consequence, the antikink follows the kink. If $\alpha$ is big ($\alpha=1$ then the impurity strongly repels the kink, which leads to a weak kink-antikink interaction. As a result we get an almost elastic reflection of the kink and a slowly moving antikink. - see Fig. \ref{Collisions4}a). For smaller but positive $\alpha$, the kink-antikink interaction grows (the repulsive force decreases) and the antikink can eventually hit the kink which leads to an annihilation process where an oscillon is formed - Fig. \ref{Collisions4}b). Once $\alpha$ becomes negative, the impurity attracts the kink, and the solitons annihilate on the other side of the impurity - Fig. \ref{Collisions4}c). However, if we further decrease $\alpha$, i.e., increase the kink-impurity attraction, then the kink gets trapped by the impurity, while the antikink is repelled to minus infinity, as demonstrated in Fig. \ref{Collisions4}d) with $\alpha=-0.3$. For even smaller $\alpha$, the inner structure of the antikink located on the impurity starts to look like a three-particle state i.e., antikink-kink-antikink. Then the scattering process in a first approximation can be understood as a sequence of two-body interactions. In Fig. \ref{Collisions4}e) ($\alpha=-1.0$), the incoming kink annihilates with the first antikink again, by forming a sort of oscillon. The remainder of the annihilation excites the kink located on the impurity, which basically does not move from that point. The last antikink, on the other hand, gets expelled to minus infinity. This pictures is repeated in Fig. \ref{Collisions4}f)- Fig.  \ref{Collisions4}i) where $\alpha=-1.24, -1.25, -1.40$ and $-1.41$. Notably, the details of the annihilation process as well as the properties of the expelled antikink reveal some chaotic features. 

In this scenario the possible results of the collisions are

\begin{equation}
 \bar K_0+K\longrightarrow
 \left\{\begin{array}{lcccc}
\bar K+K_0&\quad\text{replacement}\quad&\phi_f\approx 0&\text{G}\\
\bar K+\Sigma_-+K&\text{bounce}& \phi_f\approx -1&\quad\text{B}\quad\\
\Sigma+&\text{annihilation}& \phi_f\approx 1&\quad\text{OR}\quad\\
\{\{\bar K,K\},\Sigma_+\}&\text{passage}& \phi_f\approx 1&\quad\text{OR}\quad\\
\end{array}
\right.
\end{equation}
\begin{figure}
\hspace*{-1cm}\includegraphics[width=0.53\textwidth]{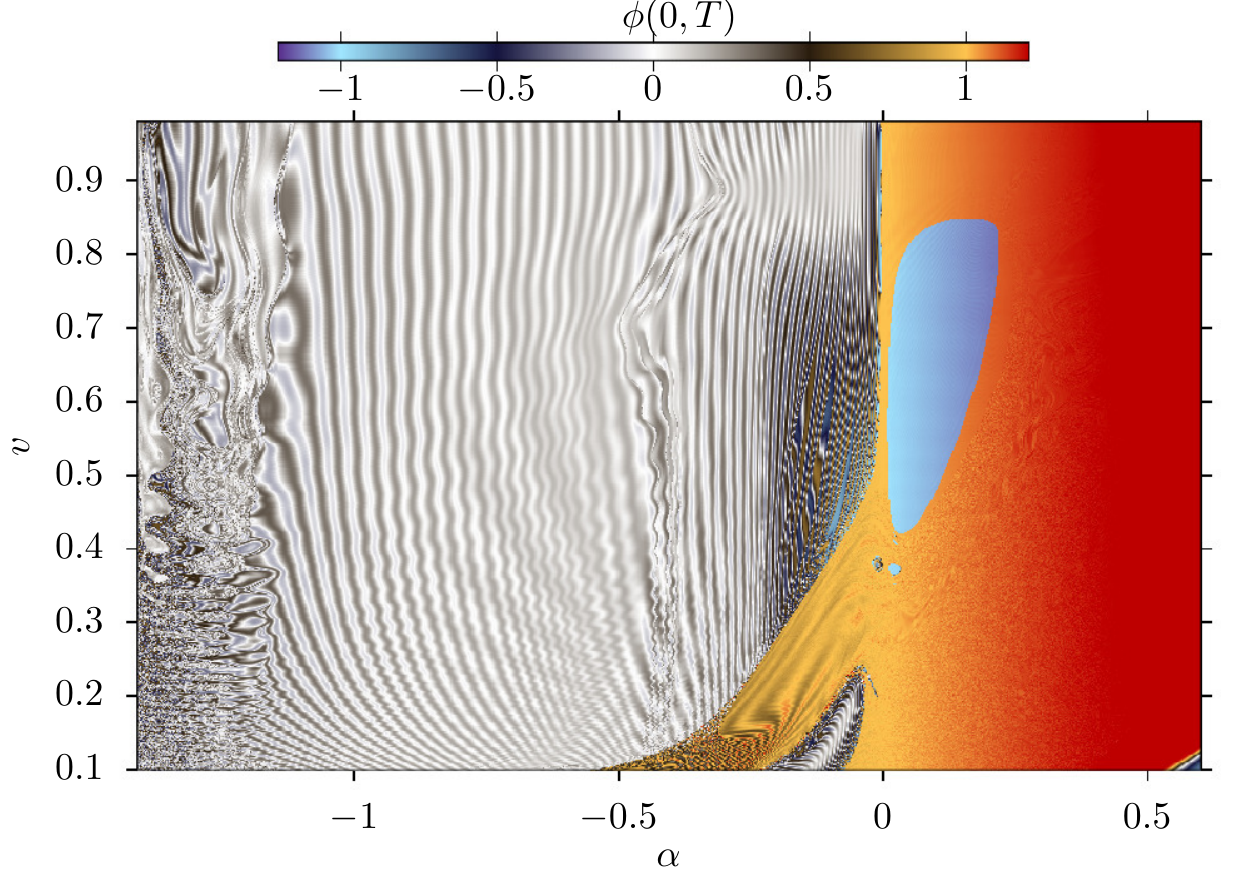}
\hspace*{-4mm}\includegraphics[width=0.53\textwidth]{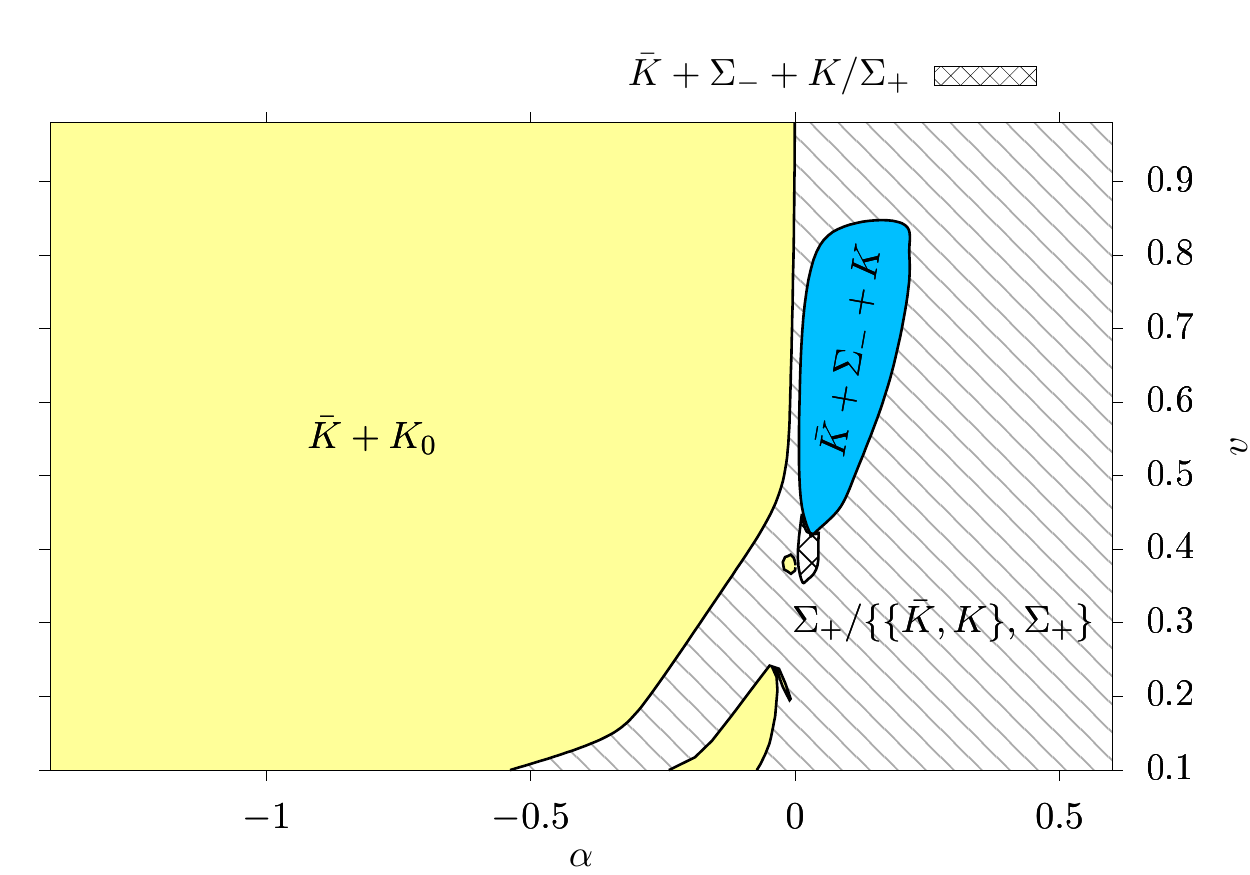}
\caption{Different scenarios of kink-(antikink-on-impurity) ($\bar K_0+K$) collisions. The left plot shows the field value at center at time $T$ after the collision. The right plot shows the products of the collisions. Line filled regions correspond to chaotic behaviour. }
\label{Collisions3Diag}
\end{figure}
Here in the last reaction as a result a lump around $\phi=1$ remains and a pair of defects is ejected in one of the directions. The pair could also annihilate forming an oscillon. It is difficult to analyze all these chaotic processes from just the values of $\phi(0,T)$. 

%%%%%%%%%%%%%%%%%%%%%%%%%%%%%%%%%%%%%%%%%
\section{Summary}
%%%%%%%%%%%%%%%%%%%%%%%%%%%%%%%%%%%%%%%%%
In the present work, we carefully investigated the properties of a $\phi^4$ model coupled with a BPS preserving defect, where its particular spatial distribution was fixed by a purely computational convenience. Namely, it preserved the form of the non-BPS kink solution, which for any value of the impurity coupling constant, was given by the same, simple analytical formula. Moreover, in our example, the BPS-ness was preserved in the $Q=-1$ and $Q=0$ sectors. The main findings are like follows. 

First of all, we identified in the BPS sector a generalized translational symmetry which transforms a charge $Q=-1$ solution into an another one without changing its energy. This corresponds to the observation that there exists an infinite family of energetically equivalent BPS solutions which represent $Q=-1$ solitons with an arbitrary separation from the impurity. This is also reflected in the appearance of a zero mode in the linear perturbation spectrum. Such a transformation exists for any value of the impurity strength $\alpha$ and, obviously, for $\alpha=0$ it reduces to the usual translational symmetry of the antikink. The generalized translation acts also on the topologically trivial solutions called lumps, which solve the Bogomolny equation in the charge zero sector. However, in this case it maps the lump into itself and no zero mode shows up. These properties are not related to a specific potential and form of the impurity. Instead, we expect to find them in any model with a BPS preserving defect. 

Secondly, the BPS solutions, that is, the antikink and the lump, possess a hidden multi-soliton structure where its constituents are, in a first approximation, just the pure $\phi^4$ theory kink and antikink. This structure is clearly visible in the limit when $\alpha\rightarrow -\sqrt{2}$ (which represents the maximally negative impurity) while, after some interpretation, can also be identified for other values of the impurity strength. In addition, the distance between the constituent components can be found not only by the standard minimisation of the energy functional with respect to the trial function parameters, but also via the integration of the Bogomolny equation over the full space with the trial function inserted. 
This observation goes beyond the model considered here and may find some applications to theories with a non trivial (multi-particle) BPS sector. 

Thirdly, as expected, we observed a profound kink-antikink asymmetry.  In contrast to the antikink, the kink, i.e., a soliton with topological charge $Q=1$, is not a BPS solution. That is, it obeys the second order EL equation instead of the first order Bogomolny equation. Furthermore, it does not saturate the pertinent topological energy bound. Since the kink is not a BPS solution, there is no generalised translation symmetry, and no zero mode exists. Hence, the kink has a fixed position with respect to the impurity. For positive $\alpha$, the impurity repells the kink and no bound state exist, while for negative $\alpha$ the kink is attracted by the defect and forms a stable bound configuration. Its energy is always smaller than the energy of the antikink. The kink becomes even the lowest energy state, that is lighter than the topologically trivial lump, for sufficiently negative strength of the impurity. Finally, when $\alpha$ tends to the smallest negative value $-\sqrt{2}$, the energy goes to the topological bound and even the kink is a BPS configuration. At this value of the impurity coupling constant, the model, in some sense, becomes a fully BPS theory.  

The kink-antikink asymmetry is clearly visible in scattering processes. Interestingly, there is a scattering process which occurs in a sort of close-to-BPS regime, where we scatter the incoming antikink with the lump located on the impurity. In the static case, as it is a BPS solution,  the antikink can be located at an arbitrary distance from the lump without changing its energy. Here, the initial state is, of course, a non-BPS state due to the non-static nature of the incoming antikink. As a result, we observed a very smooth process which, at least for not too high velocity of the antikink, can be explained, with a very good precision, within linear perturbation theory, i.e., the underlying spectral structure.  The soliton behaves as a elastic particle which can transmit a part of its energy into oscillations of the lump, which again is a kink-antikink state. Notably, there is extremely little radiation produced during the scattering. It is plausible that this behavior is a generic feature of the scattering of the particles in the BPS sector and should be observed for other potentials and BPS preserving defects. 

There are many directions in which the present work can be continued. First of all, owing to its similarity to the Abelian Higgs model with the partially BPS-preserving impurity \cite{Tong:2013iqa}, it can be used as a guidance for a systematic investigation of the scattering of vortices in the presence of such an impurity. Of course, the higher dimensionality will result in new phenomena (as for example the $90^\circ$ scattering \cite{Ashcroft:2018gkp}). However, one may expect a profound asymmetry in the vortex-impurity and antivortex-impurity interaction, where only one scattering process is a close-to BPS phenomenon. One may also try to apply the Bradlow law to get an analytical insight into the static vortex-impurity BPS state (in the spirt of the section \ref{sec-composite}).

Furthermore, as the impurity can be viewed as a sort of frozen soliton \cite{Tong:2013iqa} (here, a kink or antikink), our results may be also applied for the analysis of interactions of solitons in multi-field scalar theories in (1+1) dimensions (see for example \cite{Ferreira:2018ntx}, \cite{Gani:2016fqq}). Then, the half-BPS impurity model provides a first approximation to a scattering process with the assumption that the kink in one field is a dynamical object while the other kink in the second field is unaffected by the interaction. However, a rigorous derivation of such a frozen soliton limit is still required. 

Last but not least, one may ask about possible experimental realizations of such a partially BPS-preserving impurity. As the form of the impurity is arbitrary, it is conceivable that the deformation of the $\phi^4$ theory investigated in this paper can indeed be realised in a condensed matter system.

%%%%%%%%%%%%%%%%%%%%%%%%%%%%%%%%%%%%%%%%%
\section*{Acknowledgements}
%%%%%%%%%%%%%%%%%%%%%%%%%%%%%%%%%%%%%%%%%
The authors acknowledge financial support from the Ministry of Education, Culture, and Sports, Spain (Grant No. FPA2017-83814-P), the Xunta de Galicia (Grant No. INCITE09.296.035PR and Conselleria de Educacion), the Spanish Consolider-Ingenio 2010 Programme CPAN (CSD2007-00042), Maria de Maeztu Unit of Excellence MDM-2016-0692, and FEDER. 

We would like to thank Hubert Weigel and Yakov Shnir for their discussion and valuable remarks.

%%%%%%%%%%%%%%%%%%%%%%%%%%%
\appendix
%%%%%%%%%%%%%%%

%\nocite{*}
\bibliographystyle{JHEP}
\bibliography{refs}

\end{document}